\documentclass[a4paper,11pt]{article}

\usepackage{./auxi/jheppub}
\usepackage[utf8]{inputenc}
\usepackage[english]{babel}
\usepackage[T1]{fontenc}
\usepackage[titletoc]{appendix}
\usepackage{titletoc}
\usepackage{graphicx}
\usepackage{subcaption} 
\usepackage{xcolor}
\usepackage{amsmath,amssymb,amsthm}
\usepackage{bm}
\usepackage{float}
\usepackage[format=plain,labelfont={bf,it},textfont=it]{caption}
\usepackage{multirow}
\usepackage{slashed} 
\usepackage{mathtools} 
\usepackage{tabularx}
\usepackage{bbold}
\usepackage[shortlabels]{enumitem}
\usepackage{tikz}
\usetikzlibrary{positioning,arrows,calc}
\usetikzlibrary{arrows.meta}
\usetikzlibrary{decorations.pathmorphing}
\usetikzlibrary{decorations.markings}
\usetikzlibrary{shapes.geometric}
\usetikzlibrary{patterns}
\tikzset{
  snake it/.style={
    decorate, 
    decoration=snake,
    segment length=3
  }
}
\usepackage{xparse}
\usepackage{pdfpages} 
\usepackage{titlesec} 
\usepackage{fancyhdr} 
\usepackage{emptypage} 
\usepackage{chngcntr} 
\usepackage{etoolbox} 
\usepackage{dsfont} 
\definecolor{DarkBlueGrey}{RGB}{76,94,107}
\definecolor{MediumBlueGrey}{RGB}{110,135,153}
\definecolor{LightBlueGrey}{RGB}{134,163,184}
\definecolor{VeryLightBlueGrey}{RGB}{242,249,255}
\definecolor{WCOrange}{RGB}{242,146,29}
\definecolor{VeryLightOrange}{RGB}{255,245,233}
\definecolor{SCRed}{RGB}{179,48,48}
\definecolor{VeryLightRed}{RGB}{255,239,239}
\definecolor{VertexColor}{RGB}{242,146,29}
\definecolor{GluonColor}{RGB}{255,172,172}
\definecolor{SEColor}{RGB}{134,163,184}

\def\shuffle{\sqcup\mathchoice{\mkern-7mu}{\mkern-7mu}{\mkern-3.2mu}{\mkern-3.8mu}\sqcup}

\newcommand{\pd}{\partial}
\newcommand{\spd}{\slashed{\partial}}
\newcommand{\sx}{\slashed{x}}

\newcommand{\Li}{{\normalfont\text{Li}}}

\newcommand{\vev}[1]{\langle\, #1 \, \rangle}

\newcommand{\ub}{\bar{u}}

\def\veps{\varepsilon}

\newcommand{\Wl}{\mathcal{W}_\ell}

\newcommand{\Op}{\mathcal{O}}

\def\Am{{\mathcal{A}}}

\def\Cm{{\mathcal{C}}}

\def\Km{{\mathcal{K}}}

\def\Nm{{\mathcal{N}}}
\def\Om{{\mathcal{O}}}
\def\Pm{{\mathcal{P}}}

\def\Sm{{\mathcal{S}}}

\def\Wm{{\mathcal{W}}}

\def\Fds{{\mathds{F}}}

\def\Ids{{\mathds{I}}}

\usepackage{bbm}

\def\veps{\varepsilon}

\newcommand\psib{{\bar{\psi}}}
\newcommand\cb{{\bar{c}}}

\newcommand\zb{{\bar{z}}}

\newif\ifstartcompletesineup
\newif\ifendcompletesineup
\pgfkeys{
    /pgf/decoration/.cd,
    start up/.is if=startcompletesineup,
    start up=true,
    start up/.default=true,
    start down/.style={/pgf/decoration/start up=false},
    end up/.is if=endcompletesineup,
    end up=true,
    end up/.default=true,
    end down/.style={/pgf/decoration/end up=false}
}
\pgfdeclaredecoration{complete sines}{initial}
{
    \state{initial}[
        width=+0pt,
        next state=upsine,
        persistent precomputation={
            \ifstartcompletesineup
                \pgfkeys{/pgf/decoration automaton/next state=upsine}
                \ifendcompletesineup
                    \pgfmathsetmacro\matchinglength{
                        0.5*\pgfdecoratedinputsegmentlength / (ceil(0.5* \pgfdecoratedinputsegmentlength / \pgfdecorationsegmentlength) )
                    }
                \else
                    \pgfmathsetmacro\matchinglength{
                        0.5 * \pgfdecoratedinputsegmentlength / (ceil(0.5 * \pgfdecoratedinputsegmentlength / \pgfdecorationsegmentlength ) - 0.499)
                    }
                \fi
            \else
                \pgfkeys{/pgf/decoration automaton/next state=downsine}
                \ifendcompletesineup
                    \pgfmathsetmacro\matchinglength{
                        0.5* \pgfdecoratedinputsegmentlength / (ceil(0.5 * \pgfdecoratedinputsegmentlength / \pgfdecorationsegmentlength ) - 0.4999)
                    }
                \else
                    \pgfmathsetmacro\matchinglength{
                        0.5 * \pgfdecoratedinputsegmentlength / (ceil(0.5 * \pgfdecoratedinputsegmentlength / \pgfdecorationsegmentlength ) )
                    }
                \fi
            \fi
            \setlength{\pgfdecorationsegmentlength}{\matchinglength pt}
        }] {}
    \state{downsine}[width=\pgfdecorationsegmentlength,next state=upsine]{
        \pgfpathsine{\pgfpoint{0.5\pgfdecorationsegmentlength}{0.5\pgfdecorationsegmentamplitude}}
        \pgfpathcosine{\pgfpoint{0.5\pgfdecorationsegmentlength}{-0.5\pgfdecorationsegmentamplitude}}
    }
    \state{upsine}[width=\pgfdecorationsegmentlength,next state=downsine]{
        \pgfpathsine{\pgfpoint{0.5\pgfdecorationsegmentlength}{-0.5\pgfdecorationsegmentamplitude}}
        \pgfpathcosine{\pgfpoint{0.5\pgfdecorationsegmentlength}{0.5\pgfdecorationsegmentamplitude}}
}
    \state{final}{}
}

\tikzset{
corner/.style={line width=1pt,dashed,draw=black,dash pattern=on 6pt off 4pt},
scalar/.style={line width=1pt,draw=black},
gluon/.style={line width=1pt,decorate, draw=GluonColor,
    decoration={complete sines,aspect=0,amplitude=1.25mm,segment length=1.5mm,start up,end up}},
gluontwo/.style={line width=1pt,decorate, draw=GluonColor,
    decoration={complete sines,aspect=0,amplitude=.7mm,segment length=1mm,start up,end up}},
ghost/.style={line width=1pt,loosely dotted,draw=black},
wilson/.style={line width=2pt,draw=black},
 }
\NewDocumentCommand\semiloop{O{black}mmmO{}O{above}}
{%
\draw[#1] let \p1 = ($(#3)-(#2)$) in (#3) arc (#4:({#4+180}):({0.5*veclen(\x1,\y1)})node[midway, #6] {#5};)
}

\newcommand\x{.75}

\newcommand\vertexsize{.07}
\newcommand\scaleCC{1.3}

\newcommand{\DiagramIntro}{\begin{tikzpicture}[baseline={([yshift=0ex]current bounding box.center)}]
\draw[-Latex] (-.65,0)--(-.65,-2+.35);
\draw[-Latex] (.65,0)--(.65,-2+.35);
\draw[-Latex] (-.65,-2)--(-.65,-4+.35);
\draw[-Latex] (.65,-2)--(.65,-4+.35);
\draw[-Latex] (0,-4)--(0,-6+.35);
\draw[-Latex] (-5.25+1,0)--(-1,0);
\draw[-Latex] (-5.25+1,-4)--(-1,-4);
\draw[-Latex] (-5.25+1,-6)--(-1,-6);
\draw[-Latex] (5.25,-4)--(5.25,-2+.75-.35);
\draw[] (-5.25,-2+.75) -- (0-1.65-.25,-2+.75);
\draw[-Latex] (0-1.65-.25,-2+.75)--(0-1.65-.25,-2+.35);
\draw[] (-5.25,-2-.75) -- (0+1.65,-2-.75);
\draw[-Latex] (0+1.65,-2-.75)--(0+1.65,-2-.35);
\draw[] (5.25,-2+.75) -- (0-1.65+.25,-2+.75);
\draw[-Latex] (0-1.65+.25,-2+.75)--(0-1.65+.25,-2+.35);
\draw[-Latex] (5.25+1,0)--(1,0);
\draw[-Latex] (5.25,-4)--(1,-4);
\draw[-Latex] (5.25-1,-4+.05)--(2,-2.35);
\draw[-Latex] (5.25-1,-4-.05)--(1,-6);
\draw[rounded corners,fill=VeryLightBlueGrey] (0-1, 0-.35) rectangle (0+1,0+.35) {};
\node[] at (0,0) {$\vev{1111}$};
\draw[rounded corners,fill=VeryLightBlueGrey] (0-1.5-1.65, -2-.35) rectangle (0+1.5-1.65,-2+.35) {};
\node[] at (0-1.65,-2) {$\vev{1122}$};
\draw[rounded corners,fill=VeryLightBlueGrey] (0-1.5+1.65, -2-.35) rectangle (0+1.5+1.65,-2+.35) {};
\node[] at (0+1.65,-2) {$\vev{1212}$};
\draw[rounded corners,fill=VeryLightBlueGrey] (0-1, -4-.35) rectangle (0+1,-4+.35) {};
\node[] at (0,-4) {$\vev{11112}$};
\draw[rounded corners,fill=VeryLightBlueGrey] (0-1, -6-.35) rectangle (0+1,-6+.35) {};
\node[] at (0,-6) {$\vev{111111}$};
\draw[rounded corners,fill=VeryLightOrange] (-5.25-1, 0-.35) rectangle (-5.25+1,0+.35) {};
\node[] at (-5.25,0) {$\Fds_{1111}$};
\draw[rounded corners,fill=VeryLightOrange] (-5.25-1,-2-.35+.75) rectangle (-5.25+1,-2+.35+.75) {};
\node[] at (-5.25,-2+.75) {$\Fds_{1122}$};
\draw[rounded corners,fill=VeryLightOrange] (-5.25-1,-2-.35-.75) rectangle (-5.25+1,-2+.35-.75) {};
\node[] at (-5.25,-2-.75) {$\Fds_{1212}$};
\draw[rounded corners,fill=VeryLightOrange] (-5.25-1, -4-.35) rectangle (-5.25+1,-4+.35) {};
\node[] at (-5.25,-4) {$\Fds_{11112}$};
\draw[rounded corners,fill=VeryLightOrange] (-5.25-1, -6-.35) rectangle (-5.25+1,-6+.35) {};
\node[] at (-5.25,-6) {$\Fds_{111111}$};
\draw[rounded corners,fill=VeryLightRed] (5.25-1.5,-.35) rectangle (5.25+1.5,.35) {};
\node[] at (5.25,0) {1 $R$-channel};
\draw[rounded corners,fill=VeryLightRed] (5.25-1.5,-2+.75-.35) rectangle (5.25+1.5,-2+.75+.35) {};
\node[] at (5.25,-2+.75) {1 $R$-channel};
\draw[rounded corners,fill=VeryLightRed] (5.25-1.8, -4-.35) rectangle (5.25+1.8,-4+.35) {};
\node[] at (5.25,-4) {Train track integral};
\end{tikzpicture}}
\newcommand{\MethodFivePoint}{\begin{tikzpicture}[baseline={([yshift=0ex]current bounding box.center)}]
\draw[-Latex] (0,0)--(0,-1.5+.35);
\draw[-Latex] (0,-1.5)--(0,-3+.35);
\draw[-Latex] (0,-3)--(0,-4.5+.35);
\draw[-Latex] (0,-4.5)--(0,-6+.35);
\draw[-Latex] (0,-6)--(0,-7.5+.35);
\draw[-Latex] (-5.25+1,-1.5)--(-1.5,-1.5);
\draw[-Latex] (-5.25+1,-3)--(-1.5,-3);
\draw[-Latex] (5.25-1,-4.5)--(1.5,-4.5);
\draw[-Latex] (5.25-1,-6)--(1.5,-6);
\draw[-Latex] (-5.25-1,-7.5)--(-1.5,-7.5);
\draw[rounded corners,fill=VeryLightBlueGrey] (0-1.5, 0-.35) rectangle (0+1.5,0+.35) {};
\node[] at (0,0) {$6$ $R$-channels};
\draw[rounded corners,fill=VeryLightBlueGrey] (0-1.5, -1.5-.35) rectangle (0+1.5,-1.5+.35) {};
\node[] at (0,-1.5) {$2$ $R$-channels};
\draw[rounded corners,fill=VeryLightBlueGrey] (0-1.5, -3-.35) rectangle (0+1.5,-3+.35) {};
\node[] at (0,-3) {$1$ $R$-channel};
\draw[rounded corners,fill=VeryLightBlueGrey] (0-1.5, -4.5-.35) rectangle (0+1.5,-4.5+.35) {};
\node[] at (0,-4.5) {$231$ coefficients};
\draw[rounded corners,fill=VeryLightBlueGrey] (0-1.5, -6-.35) rectangle (0+1.5,-6+.35) {};
\node[] at (0,-6) {$20$ coefficients};
\draw[rounded corners,fill=VeryLightBlueGrey] (0-1.5, -7.5-.35) rectangle (0+1.5,-7.5+.35) {};
\node[] at (0,-7.5) {fully fixed};
\draw[rounded corners,fill=VeryLightOrange] (-5.25-1.5, -1.5-.35) rectangle (-5.25+1.5,-1.5+.35) {};
\node[] at (-5.25,-1.5) {SCWI$\: + \:\Fds_{11112}$};
\draw[rounded corners,fill=VeryLightOrange] (-5.25-1.5, -3-.35) rectangle (-5.25+1.5,-3+.35) {};
\node[] at (-5.25,-3) {Crossing};
\draw[rounded corners,fill=VeryLightRed] (5.25-1.9, -4.5-.35) rectangle (5.25+1.9,-4.5+.35) {};
\node[] at (5.25,-4.5) {Ansatz};
\draw[rounded corners,fill=VeryLightRed] (5.25-1.9, -6-.35) rectangle (5.25+1.9,-6+.35) {};
\node[] at (5.25,-6) {Train track integral};
\draw[rounded corners,fill=VeryLightOrange] (-5.25-2.5, -7.5-.35) rectangle (-5.25+2.5,-7.5+.35) {};
\node[] at (-5.25,-7.5) {Pinching to $\vev{1212}$, $\vev{1122}$};
\end{tikzpicture}}
\newcommand{\MethodSixPoint}{\begin{tikzpicture}[baseline={([yshift=0ex]current bounding box.center)}]
\draw[-Latex] (0,0)--(0,-1.5+.35);
\draw[-Latex] (0,-1.5)--(0,-3+.35);
\draw[-Latex] (0,-3)--(0,-4.5+.35);
\draw[-Latex] (0,-4.5)--(0,-6+.35);
\draw[-Latex] (0,-6)--(0,-7.5+.35);
\draw[-Latex] (0,-7.5)--(0,-9+.35);
\draw[-Latex] (-5.25+1,-1.5)--(-1.5,-1.5);
\draw[-Latex] (-5.25+1,-3)--(-1.5,-3);
\draw[-Latex] (5.25-1,-4.5)--(1.5,-4.5);
\draw[-Latex] (5.25-1,-6)--(1.5,-6);
\draw[-Latex] (5.25-1,-7.5)--(1.5,-7.5);
\draw[-Latex] (-5.25-1,-9)--(-1.5,-9);
\draw[rounded corners,fill=VeryLightBlueGrey] (0-1.5, 0-.35) rectangle (0+1.5,0+.35) {};
\node[] at (0,0) {$15$ $R$-channels};
\draw[rounded corners,fill=VeryLightBlueGrey] (0-1.5, -1.5-.35) rectangle (0+1.5,-1.5+.35) {};
\node[] at (0,-1.5) {$4$ $R$-channels};
\draw[rounded corners,fill=VeryLightBlueGrey] (0-1.5, -3-.35) rectangle (0+1.5,-3+.35) {};
\node[] at (0,-3) {$2$ $R$-channels};
\draw[rounded corners,fill=VeryLightBlueGrey] (0-1.5, -4.5-.35) rectangle (0+1.5,-4.5+.35) {};
\node[] at (0,-4.5) {$1$ $R$-channel};
\draw[rounded corners,fill=VeryLightBlueGrey] (0-1.5, -6-.35) rectangle (0+1.5,-6+.35) {};
\node[] at (0,-6) {$1406$ coefficients};
\draw[rounded corners,fill=VeryLightBlueGrey] (0-1.5, -7.5-.35) rectangle (0+1.5,-7.5+.35) {};
\node[] at (0,-7.5) {$21$ coefficients};
\draw[rounded corners,fill=VeryLightBlueGrey] (0-1.5, -9-.35) rectangle (0+1.5,-9+.35) {};
\node[] at (0,-9) {fully fixed};
\draw[rounded corners,fill=VeryLightOrange] (-5.25-1.5, -1.5-.35) rectangle (-5.25+1.5,-1.5+.35) {};
\node[] at (-5.25,-1.5) {SCWI $+ \Fds_{111111}$};
\draw[rounded corners,fill=VeryLightOrange] (-5.25-1.5, -3-.35) rectangle (-5.25+1.5,-3+.35) {};
\node[] at (-5.25,-3) {Crossing};
\draw[rounded corners,fill=VeryLightRed] (5.25-1.9, -4.5-.35) rectangle (5.25+1.9,-4.5+.35) {};
\node[] at (5.25,-4.5) {$F_{14} = 0$};
\draw[rounded corners,fill=VeryLightRed] (5.25-1.9, -6-.35) rectangle (5.25+1.9,-6+.35) {};
\node[] at (5.25,-6) {Ansatz};
\draw[rounded corners,fill=VeryLightRed] (5.25-1.9, -7.5-.35) rectangle (5.25+1.9,-7.5+.35) {};
\node[] at (5.25,-7.5) {Train track integral};
\draw[rounded corners,fill=VeryLightOrange] (-5.25-1.9, -9-.35) rectangle (-5.25+1.9,-9+.35) {};
\node[] at (-5.25,-9) {Pinching to $\vev{11112}$};
\end{tikzpicture}}

\newcommand{\MultipointCorrelators}{\begin{tikzpicture}[baseline={([yshift=-2ex]current bounding box.center)}]
\draw[wilson] (-5,0) -- (3,0);
\draw[fill=white] (-4,0) circle (.1);
\draw[fill=white] (-2.5,0) circle (.1);
\draw[fill=white] (0.5,0) circle (.1);
\draw[fill=white] (2,0) circle (.1);
\node[] at (-1,.4) {$\dots$};
\node[] at (-4,.4) {$\tau_1$};
\node[] at (-2.5,.4) {$\tau_2$};
\node[] at (0.5,.4) {$\tau_{n-1}$};
\node[] at (2,.4) {$\tau_{n}$};
\node[] at (-4,-.6) {$\Op_{\Delta_1}$};
\node[] at (-2.5,-.6)  {$\Op_{\Delta_2}$};
\node[] at (0.5,-.6)  {$\Op_{\Delta_{n-1}}$};
\node[] at (2,-.6)  {$\Op_{\Delta_n}$};
\node[] at (3.5,0) {$\Wl$};
\end{tikzpicture}}

\newcommand{\ScalarPropagator}{\begin{tikzpicture}[baseline={([yshift=-.85ex]current bounding box.center)}]
\draw[scalar] (0,.5) -- (1,.5);
\draw[fill=white] (0,.5) circle (.075);
\draw[fill=white] (1,.5) circle (.075);
\node[] at (0,.1) {\footnotesize \makebox[\widthof{$\smash{\mu,a}$}][c]{$\smash{I,a}$}};
\node[] at (0,.9) {\footnotesize $1$};
\node[] at (1,.1) {\footnotesize \makebox[\widthof{$\smash{\mu,a}$}][c]{$\smash{J,b}$}};
\node[] at (1,.9) {\footnotesize $2$};
\end{tikzpicture}}
\newcommand{\GluonPropagator}{\begin{tikzpicture}[baseline={([yshift=-.85ex]current bounding box.center)}]
\draw[gluon] (0,.5) -- (1,.5);
\draw[fill=white] (0,.5) circle (.075);
\draw[fill=white] (1,.5) circle (.075);
\node[] at (0,.1) {\footnotesize \makebox[\widthof{$\mu,a$}][c]{$\smash{\mu,a}$}};
\node[] at (0,.9) {\footnotesize $1$};
\node[] at (1,.1) {\footnotesize \makebox[\widthof{$\nu,b$}][c]{$\smash{\nu,b}$}};
\node[] at (1,.9) {\footnotesize $2$};
\end{tikzpicture}}
\newcommand{\FermionPropagator}{\begin{tikzpicture}[baseline={([yshift=-.85ex]current bounding box.center)}]
\draw[scalar] (0,.5) -- (1,.5);
\draw[fill=white] (0,.5) circle (.075);
\draw[fill=white] (1,.5) circle (.075);
\draw[-Latex] (.62,.5)--(.63,.5);
\node[] at (0,.1) {\footnotesize \makebox[\widthof{$\smash{\mu,a}$}][c]{$\smash{a}$}};
\node[] at (0,.9) {\footnotesize $1$};
\node[] at (1,.1) {\footnotesize \makebox[\widthof{$\smash{\nu,b}$}][c]{$\smash{b}$}};
\node[] at (1,.9) {\footnotesize $2$};
\end{tikzpicture}}
\newcommand{\GhostPropagator}{\begin{tikzpicture}[baseline={([yshift=-.85ex]current bounding box.center)}]
\draw[ghost] (0,.5) -- (1,.5);
\draw[fill=white] (0,.5) circle (.075);
\draw[fill=white] (1,.5) circle (.075);
\node[] at (0,.1) {\footnotesize \makebox[\widthof{$\smash{\mu,a}$}][c]{$\smash{a}$}};
\node[] at (0,.9) {\footnotesize $1$};
\node[] at (1,.1) {\footnotesize \makebox[\widthof{$\smash{\nu,b}$}][c]{$\smash{b}$}};
\node[] at (1,.9) {\footnotesize $2$};
\end{tikzpicture}}

\newcommand{\VertexScalarScalarGluon}{\begin{tikzpicture}[baseline={([yshift=-.35ex]current bounding box.center)}]
\draw[gluon] (0,.5) -- (.75,.5);
\draw[scalar] (.75,.5) -- (1.25,1);
\draw[scalar] (.75,.5) -- (1.25,0);
\draw[fill=VertexColor,VertexColor] (.75,.5) circle (\vertexsize);
\draw[fill=white] (0,.5) circle (.075);
\draw[fill=white] (1.25,1) circle (.075);
\draw[fill=white] (1.25,0) circle (.075);
\node[anchor=west] at (1.3,1) {\footnotesize $I,a$};
\node[] at (1.25,1.3) {\footnotesize $1$};
\node[anchor=west] at (1.3,0) {\footnotesize $J,b$};
\node[] at (1.25,-.3) {\footnotesize $2$};
\node[] at (0,.1) {\footnotesize $\mu,c$};
\node[] at (0,.9) {\footnotesize $3$};
\end{tikzpicture}}
\newcommand{\VertexFermionFermionScalar}{\begin{tikzpicture}[baseline={([yshift=-.45ex]current bounding box.center)}]
\draw[scalar] (0,.5) -- (.75,.5);
\draw[scalar] (.75,.5) -- (1.25,1);
\draw[-Latex] (1.05,.8)--(.95,.7);
\draw[scalar] (.75,.5) -- (1.25,0);
\draw[-Latex] (1,.25)--(1.1,.15);
\draw[fill=VertexColor,VertexColor] (.75,.5) circle (\vertexsize);
\draw[fill=white] (0,.5) circle (.075);
\draw[fill=white] (1.25,1) circle (.075);
\draw[fill=white] (1.25,0) circle (.075);
\node[anchor=west] at (1.3,1) {\footnotesize $a$};
\node[] at (1.25,1.3) {\footnotesize $1$};
\node[anchor=west] at (1.3,0) {\footnotesize $b$};
\node[] at (1.25,-.3) {\footnotesize $2$};
\node[] at (0,.1) {\footnotesize $I,c$};
\node[] at (0,.9) {\footnotesize $3$};
\end{tikzpicture}}
\newcommand{\VertexFermionFermionGluon}{\begin{tikzpicture}[baseline={([yshift=-.35ex]current bounding box.center)}]
\draw[gluon] (0,.5) -- (.75,.5);
\draw[scalar] (.75,.5) -- (1.25,1);
\draw[-Latex] (1.05,.8)--(.95,.7);
\draw[scalar] (.75,.5) -- (1.25,0);
\draw[-Latex] (1,.25)--(1.1,.15);
\draw[fill=VertexColor,VertexColor] (.75,.5) circle (\vertexsize);
\draw[fill=white] (0,.5) circle (.075);
\draw[fill=white] (1.25,1) circle (.075);
\draw[fill=white] (1.25,0) circle (.075);
\end{tikzpicture}}
\newcommand{\VertexGhostGhostGluon}{\begin{tikzpicture}[baseline={([yshift=-.35ex]current bounding box.center)}]
\draw[gluon] (0,.5) -- (.75,.5);
\draw[ghost] (.75,.5) -- (1.25,1);
\draw[ghost] (.75,.5) -- (1.25,0);
\draw[fill=VertexColor,VertexColor] (.75,.5) circle (\vertexsize);
\draw[fill=white] (0,.5) circle (.075);
\draw[fill=white] (1.25,1) circle (.075);
\draw[fill=white] (1.25,0) circle (.075);
\end{tikzpicture}}
\newcommand{\VertexGluonGluonGluon}{\begin{tikzpicture}[baseline={([yshift=-.35ex]current bounding box.center)}]
\draw[gluon] (0,.5) -- (.75,.5);
\draw[gluon] (.75,.5) -- (1.25,1);
\draw[gluon] (.75,.5) -- (1.25,0);
\draw[fill=VertexColor,VertexColor] (.75,.5) circle (\vertexsize);
\draw[fill=white] (0,.5) circle (.075);
\draw[fill=white] (1.25,1) circle (.075);
\draw[fill=white] (1.25,0) circle (.075);
\end{tikzpicture}}
\newcommand{\VertexFourScalars}{\begin{tikzpicture}[baseline={([yshift=-.35ex]current bounding box.center)}]
\draw[scalar] (0,0) -- (1.25,1);
\draw[scalar] (0,1) -- (1.25,0);
\draw[fill=VertexColor,VertexColor] (.625,.5) circle (\vertexsize);
\draw[fill=white] (0,0) circle (.075);
\draw[fill=white] (0,1) circle (.075);
\draw[fill=white] (1.25,1) circle (.075);
\draw[fill=white] (1.25,0) circle (.075);
\node[anchor=west] at (1.3,1) {\footnotesize $I,a$};
\node[] at (1.25,1.3) {\footnotesize $1$};
\node[anchor=west] at (1.3,0) {\footnotesize $J,b$};
\node[] at (1.25,-.3) {\footnotesize $2$};
\node[anchor=east] at (-0.05,1) {\footnotesize $K,c$};
\node[] at (0,1.3) {\footnotesize $3$};
\node[anchor=east] at (-0.05,0) {\footnotesize $L,d$};
\node[] at (0,-.3) {\footnotesize $4$};
\end{tikzpicture}}

\newcommand{\VertexScalarScalarGluonGluon}{\begin{tikzpicture}[baseline={([yshift=-.35ex]current bounding box.center)}]
\draw[scalar] (.625,.5) -- (1.25,1);
\draw[scalar] (.625,.5) -- (1.25,0);
\draw[gluon] (0,0) -- (.625,.5);
\draw[gluon] (0,1) -- (.625,.5);
\draw[fill=VertexColor,VertexColor] (.625,.5) circle (\vertexsize);
\draw[fill=white] (0,0) circle (.075);
\draw[fill=white] (0,1) circle (.075);
\draw[fill=white] (1.25,1) circle (.075);
\draw[fill=white] (1.25,0) circle (.075);
\end{tikzpicture}}
\newcommand{\VertexGluonGluonGluonGluon}{\begin{tikzpicture}[baseline={([yshift=-.35ex]current bounding box.center)}]
\draw[gluon] (.625,.5) -- (1.25,1);
\draw[gluon] (.625,.5) -- (1.25,0);
\draw[gluon] (0,0) -- (.625,.5);
\draw[gluon] (0,1) -- (.625,.5);
\draw[fill=VertexColor,VertexColor] (.625,.5) circle (\vertexsize);
\draw[fill=white] (0,0) circle (.075);
\draw[fill=white] (0,1) circle (.075);
\draw[fill=white] (1.25,1) circle (.075);
\draw[fill=white] (1.25,0) circle (.075);
\end{tikzpicture}}

\newcommand{\SelfEnergyNoText}{\begin{tikzpicture}[baseline={([yshift=-.35ex]current bounding box.center)}]
\draw[scalar] (0,.5) -- (1,.5);
\draw[fill=white] (0,.5) circle (.075);
\draw[fill=white] (1,.5) circle (.075);
\draw[fill=SEColor,SEColor] (.5,.5) circle (.15);
\end{tikzpicture}}
\newcommand{\SelfEnergyDiagramOne}{\begin{tikzpicture}[baseline={([yshift=-.4ex]current bounding box.center)}]
\draw[scalar] (0,.5) -- (2,.5);
\draw[gluon] (.5,.5) arc (180:-20:.5);
\draw[fill=white] (0,.5) circle (.075);
\draw[fill=white] (2,.5) circle (.075);
\draw[fill=VertexColor,VertexColor] (.5,.5) circle (\vertexsize);
\draw[fill=VertexColor,VertexColor] (1.5,.5) circle (\vertexsize);
\end{tikzpicture}}
\newcommand{\SelfEnergyDiagramTwo}{\begin{tikzpicture}[baseline={([yshift=-.5ex]current bounding box.center)}]
\draw[scalar] (0,.5) -- (.5,.5);
\draw[scalar] (1.5,.5) -- (2,.5);
\draw[scalar,postaction={decorate}] (.5,.5) arc (180:0:.5);
\draw[-Latex] (1.1,1)--(1.11,1);
\draw[scalar,postaction={decorate}] (.5,.5) arc (-180:0:.5);
\draw[-Latex] (.9,0)--(.89,0);
\draw[fill=white] (0,.5) circle (.075);
\draw[fill=white] (2,.5) circle (.075);
\draw[fill=VertexColor,VertexColor] (.5,.5) circle (\vertexsize);
\draw[fill=VertexColor,VertexColor] (1.5,.5) circle (\vertexsize);
\end{tikzpicture}}
\newcommand{\SelfEnergyDiagramThree}{\begin{tikzpicture}[baseline={([yshift=-.75ex]current bounding box.center)}]
\draw[scalar] (0,.5) -- (2,.5);
\draw[scalar] (1,1) circle (.5);
\draw[fill=white] (0,.5) circle (.075);
\draw[fill=white] (2,.5) circle (.075);
\draw[fill=VertexColor,VertexColor] (1,.5) circle (\vertexsize);
\end{tikzpicture}}
\newcommand{\SelfEnergyDiagramFour}{\begin{tikzpicture}[baseline={([yshift=-1.05ex]current bounding box.center)}]
\draw[scalar] (0,.5) -- (2,.5);
\draw[gluon] (1,1) circle (.5);
\draw[fill=white] (0,.5) circle (.075);
\draw[fill=white] (2,.5) circle (.075);
\draw[fill=VertexColor,VertexColor] (1,.5) circle (\vertexsize);
\end{tikzpicture}}

\newcommand{\DefectVertexOnePointScalar}{\begin{tikzpicture}[baseline={([yshift=-3ex]current bounding box.center)}]
\draw[wilson] (0,0) -- (1.5,0);
\draw[scalar] (.75,0) -- (.75,1);
\draw[fill=white] (.75,1) circle (.075);
\draw[fill=white] (.3,0) circle (.075);
\draw[fill=white] (1.2,0) circle (.075);
\draw[fill=VertexColor,VertexColor] (.75,0) circle (\vertexsize);
\end{tikzpicture}}
\newcommand{\DefectVertexOnePointGluon}{\begin{tikzpicture}[baseline={([yshift=-1ex]current bounding box.center)}]
\draw[wilson] (0,0) -- (1.5,0);
\draw[gluon] (.75,0) -- (.75,1);
\draw[fill=white] (.75,1) circle (.075);
\draw[fill=white] (.3,0) circle (.075);
\draw[fill=white] (1.2,0) circle (.075);
\draw[fill=VertexColor,VertexColor] (.75,0) circle (\vertexsize);
\node[anchor=east] at (.7,1) {\footnotesize $\mu$};
\node[] at (1.05,1) {\footnotesize $1$};
\node[] at (.3,-.3) {\footnotesize $2$};
\node[] at (1.2,-.3) {\footnotesize $3$};
\end{tikzpicture}}

\newcommand{\TrainTrack}{\begin{tikzpicture}[baseline={([yshift=-.5ex]current bounding box.center)}]
\draw[scalar] (0,0) -- (2.25,0);
\draw[scalar] (.75,-.75) -- (.75,.75);
\draw[scalar] (1.5,-.75) -- (1.5,.75);
\draw[fill=white] (0,0) circle (.075);
\draw[fill=white] (2.25,0) circle (.075);
\draw[fill=white] (.75,-.75) circle (.075);
\draw[fill=white] (.75,.75) circle (.075);
\draw[fill=white] (1.5,-.75) circle (.075);
\draw[fill=white] (1.5,.75) circle (.075);
\draw[fill=VertexColor,VertexColor] (.75,0) circle (\vertexsize);
\draw[fill=VertexColor,VertexColor] (1.5,0) circle (\vertexsize);
\node[] at (.75,1.1) {\footnotesize $1$};
\node[] at (-.25,0) {\footnotesize $2$};
\node[] at (.75,-1.1) {\footnotesize $3$};
\node[] at (1.5,-1.1) {\footnotesize $4$};
\node[] at (2.5,0) {\footnotesize $5$};
\node[] at (1.5,1.1) {\footnotesize $6$};
\end{tikzpicture}}
\newcommand{\Hintegral}{\begin{tikzpicture}[baseline={([yshift=-.5ex]current bounding box.center)}]
\draw[scalar] (.75,0) -- (1.5,0);
\draw[scalar] (.75,-.75) -- (.75,.75);
\draw[scalar] (1.5,-.75) -- (1.5,.75);
\draw[fill=white] (.75,-.75) circle (.075);
\draw[fill=white] (.75,.75) circle (.075);
\draw[fill=white] (1.5,-.75) circle (.075);
\draw[fill=white] (1.5,.75) circle (.075);
\draw[fill=VertexColor,VertexColor] (.75,0) circle (\vertexsize);
\draw[fill=VertexColor,VertexColor] (1.5,0) circle (\vertexsize);
\node[] at (.75,1.1) {\footnotesize $1$};
\node[] at (.75,-1.1) {\footnotesize $2$};
\node[] at (1.5,-1.1) {\footnotesize $3$};
\node[] at (1.5,1.1) {\footnotesize $4$};
\end{tikzpicture}}

\newcommand{\NLOX}{\begin{tikzpicture}[baseline={([yshift=-2ex]current bounding box.center)}]
\draw[wilson] (0,0) -- (2,0);
\draw[fill=white] (.25,0) circle (.075);
\draw[fill=white] (.75,0) circle (.075);
\draw[fill=white] (1.25,0) circle (.075);
\draw[fill=white] (1.75,0) circle (.075);
\draw[fill=VertexColor,VertexColor] (1,.43) circle (\vertexsize);
\path[clip] (0,0) rectangle (2,1);
\draw[scalar] (.75,0) circle (.5);
\draw[scalar] (1.25,0) circle (.5);
\draw[wilson] (0,0) -- (2,0);
\draw[fill=white] (.25,0) circle (.075);
\draw[fill=white] (.75,0) circle (.075);
\draw[fill=white] (1.25,0) circle (.075);
\draw[fill=white] (1.75,0) circle (.075);
\draw[fill=VertexColor,VertexColor] (1,.43) circle (\vertexsize);
\end{tikzpicture}}
\newcommand{\NNLOXXOne}{\begin{tikzpicture}[baseline={([yshift=-2ex]current bounding box.center)}]
\draw[wilson] (0,0) -- (2,0);
\draw[fill=white] (.25,0) circle (.075);
\draw[fill=white] (.75,0) circle (.075);
\draw[fill=white] (1.25,0) circle (.075);
\draw[fill=white] (1.75,0) circle (.075);
\draw[fill=VertexColor,VertexColor] (1,.43) circle (\vertexsize);
\path[clip] (0,0) rectangle (2,1);
\draw[scalar] (.75,0) circle (.5);
\draw[scalar] (1.25,0) circle (.5);
\draw[wilson] (0,0) -- (2,0);
\draw[scalar] (1.25,0) .. controls (1.25-.65,.3) and (1.5,.4) .. (1.75,0);
\draw[fill=white] (.25,0) circle (.075);
\draw[fill=white] (.75,0) circle (.075);
\draw[fill=white] (1.25,0) circle (.075);
\draw[fill=white] (1.75,0) circle (.075);
\draw[fill=VertexColor,VertexColor] (1,.43) circle (\vertexsize);
\draw[fill=VertexColor,VertexColor] (1.25-0.075,.25) circle (\vertexsize);
\end{tikzpicture}}
\newcommand{\NNLOXXTwo}{\begin{tikzpicture}[baseline={([yshift=-2ex]current bounding box.center)}]
\draw[wilson] (0,0) -- (2,0);
\draw[fill=white] (.25,0) circle (.075);
\draw[fill=white] (.75,0) circle (.075);
\draw[fill=white] (1.25,0) circle (.075);
\draw[fill=white] (1.75,0) circle (.075);
\draw[fill=VertexColor,VertexColor] (1,.43) circle (\vertexsize);
\path[clip] (0,0) rectangle (2,1);
\draw[scalar] (.75,0) circle (.5);
\draw[scalar] (1.25,0) circle (.5);
\draw[wilson] (0,0) -- (2,0);
\draw[scalar] (1.75,0) .. controls (1.75+.65,.3) and (1.5,.4) .. (1.25,0);
\draw[fill=white] (.25,0) circle (.075);
\draw[fill=white] (.75,0) circle (.075);
\draw[fill=white] (1.25,0) circle (.075);
\draw[fill=white] (1.75,0) circle (.075);
\draw[fill=VertexColor,VertexColor] (1,.43) circle (\vertexsize);
\draw[fill=VertexColor,VertexColor] (1.75-0.075,.25) circle (\vertexsize);
\end{tikzpicture}}
\newcommand{\NNLOXHOne}{\begin{tikzpicture}[baseline={([yshift=-2ex]current bounding box.center)}]
\draw[wilson] (0,0) -- (2,0);
\draw[scalar] (1.75,0) arc (20:160:.265);
\draw[gluon] (1.125,.35) .. controls (1.25,.35) and (1.35,.3) .. (1.5,.15);
\draw[fill=white] (.25,0) circle (.075);
\draw[fill=white] (.75,0) circle (.075);
\draw[fill=white] (1.25,0) circle (.075);
\draw[fill=white] (1.75,0) circle (.075);
\draw[fill=VertexColor,VertexColor] (1,.43) circle (\vertexsize);
\path[clip] (0,0) rectangle (2,1);
\draw[scalar] (.75,0) circle (.5);
\draw[scalar] (1.25,0) circle (.5);
\draw[wilson] (0,0) -- (2,0);
\draw[fill=white] (.25,0) circle (.075);
\draw[fill=white] (.75,0) circle (.075);
\draw[fill=white] (1.25,0) circle (.075);
\draw[fill=white] (1.75,0) circle (.075);
\draw[fill=VertexColor,VertexColor] (1,.43) circle (\vertexsize);
\draw[fill=VertexColor,VertexColor] (1.175,.275) circle (\vertexsize);
\draw[fill=VertexColor,VertexColor] (1.5,.15) circle (\vertexsize);
\end{tikzpicture}}
\newcommand{\NNLOXHTwo}{\begin{tikzpicture}[baseline={([yshift=-2ex]current bounding box.center)}]
\draw[wilson] (0,0) -- (2,0);
\draw[scalar] (1.75,0) arc (20:160:.265);
\draw[gluon] (1.25,.505) -- (1.5,.15);
\draw[fill=white] (.25,0) circle (.075);
\draw[fill=white] (.75,0) circle (.075);
\draw[fill=white] (1.25,0) circle (.075);
\draw[fill=white] (1.75,0) circle (.075);
\draw[fill=VertexColor,VertexColor] (1,.43) circle (\vertexsize);
\path[clip] (0,0) rectangle (2,1);
\draw[scalar] (.75,0) circle (.5);
\draw[scalar] (1.25,0) circle (.5);
\draw[wilson] (0,0) -- (2,0);
\draw[fill=white] (.25,0) circle (.075);
\draw[fill=white] (.75,0) circle (.075);
\draw[fill=white] (1.25,0) circle (.075);
\draw[fill=white] (1.75,0) circle (.075);
\draw[fill=VertexColor,VertexColor] (1,.43) circle (\vertexsize);
\draw[fill=VertexColor,VertexColor] (1.225,.5) circle (\vertexsize);
\draw[fill=VertexColor,VertexColor] (1.5,.15) circle (\vertexsize);
\end{tikzpicture}}
\newcommand{\NNLOSubtractFactorOne}{\begin{tikzpicture}[baseline={([yshift=.3ex]current bounding box.center)}]
\draw[wilson] (1.1,0) -- (1.9,0);
\draw[scalar] (1.75,0) arc (20:160:.265);
\draw[fill=white] (1.25,0) circle (.075);
\draw[fill=white] (1.75,0) circle (.075);
\end{tikzpicture}}
\newcommand{\NNLOSubtractFactorTwo}{\begin{tikzpicture}[baseline={([yshift=-.6ex]current bounding box.center)}]
\draw[wilson] (.8,0) -- (2.2,0);
\draw[gluon] (1.5,.175) -- (1.5,.5);
\draw[scalar] (1.75,0) arc (20:160:.265);
\draw[fill=white] (1.25,0) circle (.075);
\draw[fill=white] (1.75,0) circle (.075);
\draw[fill=VertexColor,VertexColor] (1,0) circle (\vertexsize);
\draw[fill=VertexColor,VertexColor] (2,0) circle (\vertexsize);
\draw[fill=VertexColor,VertexColor] (1.5,.175) circle (\vertexsize);
\end{tikzpicture}}
\newcommand{\NNLOXYSubtractOne}{\begin{tikzpicture}[baseline={([yshift=-2ex]current bounding box.center)}]
\draw[wilson] (0,0) -- (2,0);
\draw[gluon] (1.125,.35) .. controls (1.25,.35) and (1.35,.3) .. (1.5,0);
\draw[fill=white] (.25,0) circle (.075);
\draw[fill=white] (.75,0) circle (.075);
\draw[fill=white] (1.25,0) circle (.075);
\draw[fill=white] (1.75,0) circle (.075);
\draw[fill=VertexColor,VertexColor] (1,.43) circle (\vertexsize);
\draw[fill=VertexColor,VertexColor] (1.5,0) circle (\vertexsize);
\path[clip] (0,0) rectangle (2,1);
\draw[scalar] (.75,0) circle (.5);
\draw[scalar] (1.25,0) circle (.5);
\draw[wilson] (0,0) -- (2,0);
\draw[fill=white] (.25,0) circle (.075);
\draw[fill=white] (.75,0) circle (.075);
\draw[fill=white] (1.25,0) circle (.075);
\draw[fill=white] (1.75,0) circle (.075);
\draw[fill=VertexColor,VertexColor] (1,.43) circle (\vertexsize);
\draw[fill=VertexColor,VertexColor] (1.175,.275) circle (\vertexsize);
\draw[fill=VertexColor,VertexColor] (1.5,0) circle (\vertexsize);
\end{tikzpicture}}
\newcommand{\NNLOXYSubtractTwo}{\begin{tikzpicture}[baseline={([yshift=-2ex]current bounding box.center)}]
\draw[wilson] (0,0) -- (2,0);
\draw[gluon] (1.25,.505) -- (1.5,0);
\draw[fill=white] (.25,0) circle (.075);
\draw[fill=white] (.75,0) circle (.075);
\draw[fill=white] (1.25,0) circle (.075);
\draw[fill=white] (1.75,0) circle (.075);
\draw[fill=VertexColor,VertexColor] (1,.43) circle (\vertexsize);
\draw[fill=VertexColor,VertexColor] (1.5,0) circle (\vertexsize);
\path[clip] (0,0) rectangle (2,1);
\draw[scalar] (.75,0) circle (.5);
\draw[scalar] (1.25,0) circle (.5);
\draw[wilson] (0,0) -- (2,0);
\draw[fill=white] (.25,0) circle (.075);
\draw[fill=white] (.75,0) circle (.075);
\draw[fill=white] (1.25,0) circle (.075);
\draw[fill=white] (1.75,0) circle (.075);
\draw[fill=VertexColor,VertexColor] (1,.43) circle (\vertexsize);\draw[fill=VertexColor,VertexColor] (1.225,.5) circle (\vertexsize);
\draw[fill=VertexColor,VertexColor] (1.5,0) circle (\vertexsize);
\end{tikzpicture}}
\newcommand{\NNLOFourPtTrainTrack}{\begin{tikzpicture}[baseline={([yshift=-2ex]current bounding box.center)}]
\draw[wilson] (0,0) -- (2,0);
\draw[scalar] (1.75,0) arc (35:150:.6);
\draw[fill=white] (.25,0) circle (.075);
\draw[fill=white] (.75,0) circle (.075);
\draw[fill=white] (1.25,0) circle (.075);
\draw[fill=white] (1.75,0) circle (.075);
\draw[fill=VertexColor,VertexColor] (1,.43) circle (\vertexsize);
\path[clip] (0,0) rectangle (2,1);
\draw[scalar] (.75,0) circle (.5);
\draw[scalar] (1.25,0) circle (.5);
\draw[wilson] (0,0) -- (2,0);
\draw[fill=white] (.25,0) circle (.075);
\draw[fill=white] (.75,0) circle (.075);
\draw[fill=white] (1.25,0) circle (.075);
\draw[fill=white] (1.75,0) circle (.075);
\draw[fill=VertexColor,VertexColor] (1,.43) circle (\vertexsize);
\draw[fill=VertexColor,VertexColor] (1.25-0.075,.25) circle (\vertexsize);
\end{tikzpicture}}

\newcommand{\NNLOFivePt}{\begin{tikzpicture}[baseline={([yshift=-2ex]current bounding box.center)}]
\draw[wilson] (0,0) -- (2.5,0);
\draw[scalar] (1.75,0) arc (10:170:.775);
\draw[scalar] (2.25,0) arc (10:170:.775);
\draw[scalar] (2.25,0) arc (10:170:.5);
\draw[fill=white] (.25,0) circle (.075);
\draw[fill=white] (.75,0) circle (.075);
\draw[fill=white] (1.25,0) circle (.075);
\draw[fill=white] (1.75,0) circle (.075);
\draw[fill=white] (2.25,0) circle (.075);
\draw[fill=VertexColor,VertexColor] (1.25,.58) circle (\vertexsize);
\draw[fill=VertexColor,VertexColor] (1.57,.375) circle (\vertexsize);
\end{tikzpicture}}

\newcommand{\NNLOSixPt}{\begin{tikzpicture}[baseline={([yshift=-2ex]current bounding box.center)}]
\draw[wilson] (0,0) -- (3,0);
\draw[scalar] (2.25,0) arc (10:170:1.015);
\draw[scalar] (2.75,0) arc (10:170:.775);
\draw[scalar] (1.75,0) arc (10:170:.5);
\draw[fill=white] (.25,0) circle (.075);
\draw[fill=white] (.75,0) circle (.075);
\draw[fill=white] (1.25,0) circle (.075);
\draw[fill=white] (1.75,0) circle (.075);
\draw[fill=white] (2.25,0) circle (.075);
\draw[fill=white] (2.75,0) circle (.075);
\draw[fill=VertexColor,VertexColor] (1.42,.37) circle (\vertexsize);
\draw[fill=VertexColor,VertexColor] (1.86,.625) circle (\vertexsize);
\end{tikzpicture}}

\newcommand{\NNNLOEightPt}{\begin{tikzpicture}[baseline={([yshift=-3ex]current bounding box.center)}]
\draw[wilson] (0,0) -- (4,0);
\draw[scalar] (2.25,0) arc (10:170:1.015);
\draw[scalar] (2.75,0) arc (10:170:.5);
\draw[scalar] (3.25,0) arc (10:170:1.015);
\draw[scalar] (3.75,0) arc (10:170:1.525);
\draw[fill=white] (.25,0) circle (.075);
\draw[fill=white] (.75,0) circle (.075);
\draw[fill=white] (1.25,0) circle (.075);
\draw[fill=white] (1.75,0) circle (.075);
\draw[fill=white] (2.25,0) circle (.075);
\draw[fill=white] (2.75,0) circle (.075);
\draw[fill=white] (3.25,0) circle (.075);
\draw[fill=white] (3.75,0) circle (.075);
\draw[fill=VertexColor,VertexColor] (2.1,.375) circle (\vertexsize);
\draw[fill=VertexColor,VertexColor] (1.75,.7) circle (\vertexsize);
\draw[fill=VertexColor,VertexColor] (1.2,.83) circle (\vertexsize);
\end{tikzpicture}}

\newcommand{\CombChannel}{\scalebox{\scaleCC}{\begin{tikzpicture}[baseline={([yshift=-0ex]current bounding box.center)}]
\draw[scalar] (-4.5,0) -- (0,0);
\draw[scalar] (-3.75,0) -- (-3.75,0.7);
\draw[scalar] (-2.75,0) -- (-2.75,0.7);
\draw[scalar] (-1.75,0) -- (-1.75,0.7);
\draw[scalar] (-0.75,0) -- (-0.75,0.7);
\node[] at (-3.25,.2) {\tiny $\mathds{1}$};
\node[] at (-2.25,.2) {\tiny $\Op_1$};
\node[] at (-1.25,.2) {\tiny $\phi^6$};
\draw[MediumBlueGrey,fill=MediumBlueGrey] (-3.75,0) circle (.05);
\draw[MediumBlueGrey,fill=MediumBlueGrey] (-2.75,0) circle (.05);
\draw[MediumBlueGrey,fill=MediumBlueGrey] (-1.75,0) circle (.05);
\draw[MediumBlueGrey,fill=MediumBlueGrey] (-0.75,0) circle (.05);
\end{tikzpicture}}}
\newcommand{\SnowflakeChannel}{\scalebox{\scaleCC}{\begin{tikzpicture}[baseline={([yshift=-0ex]current bounding box.center)}]
\draw[scalar] (0,0) -- (0,.7);
\draw[scalar] (0,.7) -- (.7,1.15);
\draw[scalar] (0,.7) -- (-.7,1.15);
\draw[scalar] (0,0) -- (-.7,-.45);
\draw[scalar] (-.7,-.45) -- (-1.4,0);
\draw[scalar] (-.7,-.45) -- (-.7,-1.15);
\draw[scalar] (0,0) -- (.7,-.45);
\draw[scalar] (.7,-.45) -- (1.4,0);
\draw[scalar] (.7,-.45) -- (.7,-1.15);
\node[] at (-.5,0) {\tiny $\mathds{1}$};
\node[] at (.2,-.4) {\tiny $\phi^6$};
\node[] at (.25,.35) {\tiny $\phi^6$};
\draw[MediumBlueGrey,fill=MediumBlueGrey] (0,0) circle (.05);
\draw[MediumBlueGrey,fill=MediumBlueGrey] (0,.7) circle (.05);
\draw[MediumBlueGrey,fill=MediumBlueGrey] (-.7,-.45) circle (.05);
\draw[MediumBlueGrey,fill=MediumBlueGrey] (.7,-.45) circle (.05);
\end{tikzpicture}}}

\newcommand{\DefectSSSSTwoLoopsSelfEnergyOne}{\begin{tikzpicture}[baseline={([yshift=-2ex]current bounding box.center)}]
\draw[wilson] (0,0) -- (2,0);
\draw[fill=white] (.25,0) circle (.075);
\draw[fill=white] (.75,0) circle (.075);
\draw[fill=white] (1.25,0) circle (.075);
\draw[fill=white] (1.75,0) circle (.075);
\draw[fill=VertexColor,VertexColor] (1,.43) circle (\vertexsize);
\path[clip] (0,0) rectangle (2,1);
\draw[scalar] (.75,0) circle (.5);
\draw[scalar] (1.25,0) circle (.5);
\draw[wilson] (0,0) -- (2,0);
\draw[fill=white] (.25,0) circle (.075);
\draw[fill=white] (.75,0) circle (.075);
\draw[fill=white] (1.25,0) circle (.075);
\draw[fill=white] (1.75,0) circle (.075);
\draw[fill=VertexColor,VertexColor] (1,.43) circle (\vertexsize);
\draw[fill=SEColor,SEColor] (.5,.4) circle (.1);
\end{tikzpicture}}
\newcommand{\DefectSSSSTwoLoopsSelfEnergyTwo}{\begin{tikzpicture}[baseline={([yshift=-2ex]current bounding box.center)}]
\draw[wilson] (0,0) -- (2,0);
\draw[fill=white] (.25,0) circle (.075);
\draw[fill=white] (.75,0) circle (.075);
\draw[fill=white] (1.25,0) circle (.075);
\draw[fill=white] (1.75,0) circle (.075);
\draw[fill=VertexColor,VertexColor] (1,.43) circle (\vertexsize);
\path[clip] (0,0) rectangle (2,1);
\draw[scalar] (.75,0) circle (.5);
\draw[scalar] (1.25,0) circle (.5);
\draw[wilson] (0,0) -- (2,0);
\draw[fill=white] (.25,0) circle (.075);
\draw[fill=white] (.75,0) circle (.075);
\draw[fill=white] (1.25,0) circle (.075);
\draw[fill=white] (1.75,0) circle (.075);
\draw[fill=VertexColor,VertexColor] (1,.43) circle (\vertexsize);
\draw[fill=SEColor,SEColor] (.82,.24) circle (.1);
\end{tikzpicture}}
\newcommand{\DefectSSSSTwoLoopsSelfEnergyThree}{\begin{tikzpicture}[baseline={([yshift=-2ex]current bounding box.center)}]
\draw[wilson] (0,0) -- (2,0);
\draw[fill=white] (.25,0) circle (.075);
\draw[fill=white] (.75,0) circle (.075);
\draw[fill=white] (1.25,0) circle (.075);
\draw[fill=white] (1.75,0) circle (.075);
\draw[fill=VertexColor,VertexColor] (1,.43) circle (\vertexsize);
\path[clip] (0,0) rectangle (2,1);
\draw[scalar] (.75,0) circle (.5);
\draw[scalar] (1.25,0) circle (.5);
\draw[wilson] (0,0) -- (2,0);
\draw[fill=white] (.25,0) circle (.075);
\draw[fill=white] (.75,0) circle (.075);
\draw[fill=white] (1.25,0) circle (.075);
\draw[fill=white] (1.75,0) circle (.075);
\draw[fill=VertexColor,VertexColor] (1,.43) circle (\vertexsize);
\draw[fill=SEColor,SEColor] (1.18,.24) circle (.1);
\end{tikzpicture}}
\newcommand{\DefectSSSSTwoLoopsSelfEnergyFour}{\begin{tikzpicture}[baseline={([yshift=-2ex]current bounding box.center)}]
\draw[wilson] (0,0) -- (2,0);
\draw[fill=white] (.25,0) circle (.075);
\draw[fill=white] (.75,0) circle (.075);
\draw[fill=white] (1.25,0) circle (.075);
\draw[fill=white] (1.75,0) circle (.075);
\draw[fill=VertexColor,VertexColor] (1,.43) circle (\vertexsize);
\path[clip] (0,0) rectangle (2,1);
\draw[scalar] (.75,0) circle (.5);
\draw[scalar] (1.25,0) circle (.5);
\draw[wilson] (0,0) -- (2,0);
\draw[fill=white] (.25,0) circle (.075);
\draw[fill=white] (.75,0) circle (.075);
\draw[fill=white] (1.25,0) circle (.075);
\draw[fill=white] (1.75,0) circle (.075);
\draw[fill=VertexColor,VertexColor] (1,.43) circle (\vertexsize);
\draw[fill=SEColor,SEColor] (1.5,.4) circle (.1);
\end{tikzpicture}}

\newcommand{\DefectSSSSTwoLoopsXXOne}{\begin{tikzpicture}[baseline={([yshift=-2ex]current bounding box.center)}]
\draw[scalar] (.5,.365) -- (1.5,.365);
\draw[wilson] (0,0) -- (2,0);
\draw[fill=white] (.25,0) circle (.075);
\draw[fill=white] (.75,0) circle (.075);
\draw[fill=white] (1.25,0) circle (.075);
\draw[fill=white] (1.75,0) circle (.075);
\draw[fill=VertexColor,VertexColor] (.5,.365) circle (\vertexsize);
\draw[fill=VertexColor,VertexColor] (1.5,.365) circle (\vertexsize);
\path[clip] (0,0) rectangle (2,1);
\draw[scalar] (.5,.085) circle (.28);
\draw[scalar] (1.5,.085) circle (.28);
\draw[wilson] (0,0) -- (2,0);
\draw[fill=white] (.25,0) circle (.075);
\draw[fill=white] (.75,0) circle (.075);
\draw[fill=white] (1.25,0) circle (.075);
\draw[fill=white] (1.75,0) circle (.075);
\draw[fill=VertexColor,VertexColor] (.5,.365) circle (\vertexsize);
\draw[fill=VertexColor,VertexColor] (1.5,.365) circle (\vertexsize);
\path[clip] (0,.365) rectangle (2,1);
\draw[scalar] (1,0) circle (.62);
\draw[fill=VertexColor,VertexColor] (.5,.365) circle (\vertexsize);
\draw[fill=VertexColor,VertexColor] (1.5,.365) circle (\vertexsize);
\end{tikzpicture}}
\newcommand{\DefectSSSSTwoLoopsXXTwo}{\begin{tikzpicture}[baseline={([yshift=-2ex]current bounding box.center)}]
\draw[wilson] (0,0) -- (2,0);
\draw[fill=white] (.25,0) circle (.075);
\draw[fill=white] (.75,0) circle (.075);
\draw[fill=white] (1.25,0) circle (.075);
\draw[fill=white] (1.75,0) circle (.075);
\draw[fill=VertexColor,VertexColor] (1,.6) circle (\vertexsize);
\draw[fill=VertexColor,VertexColor] (1,.245) circle (\vertexsize);
\path[clip] (0,0) rectangle (2,1);
\draw[scalar] (1,-.2) circle (.8);
\draw[scalar] (1,-.025) circle (.27);
\draw[wilson] (0,0) -- (2,0);
\draw[fill=white] (.25,0) circle (.075);
\draw[fill=white] (.75,0) circle (.075);
\draw[fill=white] (1.25,0) circle (.075);
\draw[fill=white] (1.75,0) circle (.075);
\draw[fill=VertexColor,VertexColor] (1,.6) circle (\vertexsize);
\draw[fill=VertexColor,VertexColor] (1,.245) circle (\vertexsize);
\path[clip] (.8,.245) rectangle (1.2,.6);
\draw[scalar] (.8,.4225) circle (.27);
\draw[scalar] (1.2,.4225) circle (.27);
\draw[fill=VertexColor,VertexColor] (1,.6) circle (\vertexsize);
\draw[fill=VertexColor,VertexColor] (1,.245) circle (\vertexsize);
\end{tikzpicture}}

\newcommand{\DefectSSSSTwoLoopsXHOne}{\begin{tikzpicture}[baseline={([yshift=-2ex]current bounding box.center)}]
\draw[gluontwo] (.4,.36) -- (.81,.225);
\draw[wilson] (0,0) -- (2,0);
\draw[fill=white] (.25,0) circle (.075);
\draw[fill=white] (.75,0) circle (.075);
\draw[fill=white] (1.25,0) circle (.075);
\draw[fill=white] (1.75,0) circle (.075);
\draw[fill=VertexColor,VertexColor] (1,.43) circle (\vertexsize);
\draw[fill=VertexColor,VertexColor] (.4,.36) circle (\vertexsize);
\draw[fill=VertexColor,VertexColor] (.81,.225) circle (\vertexsize);
\path[clip] (0,0) rectangle (2,1);
\draw[scalar] (.75,0) circle (.5);
\draw[scalar] (1.25,0) circle (.5);
\draw[wilson] (0,0) -- (2,0);
\draw[fill=white] (.25,0) circle (.075);
\draw[fill=white] (.75,0) circle (.075);
\draw[fill=white] (1.25,0) circle (.075);
\draw[fill=white] (1.75,0) circle (.075);
\draw[fill=VertexColor,VertexColor] (1,.43) circle (\vertexsize);
\draw[fill=VertexColor,VertexColor] (.4,.36) circle (\vertexsize);
\draw[fill=VertexColor,VertexColor] (.81,.225) circle (\vertexsize);
\end{tikzpicture}}
\newcommand{\DefectSSSSTwoLoopsXHTwo}{\begin{tikzpicture}[baseline={([yshift=-2ex]current bounding box.center)}]
\draw[gluontwo] (.81,.225) -- (1.19,.225);
\draw[wilson] (0,0) -- (2,0);
\draw[fill=white] (.25,0) circle (.075);
\draw[fill=white] (.75,0) circle (.075);
\draw[fill=white] (1.25,0) circle (.075);
\draw[fill=white] (1.75,0) circle (.075);
\draw[fill=VertexColor,VertexColor] (1,.43) circle (\vertexsize);
\draw[fill=VertexColor,VertexColor] (.81,.225) circle (\vertexsize);
\draw[fill=VertexColor,VertexColor] (1.19,.225) circle (\vertexsize);
\path[clip] (0,0) rectangle (2,1);
\draw[scalar] (.75,0) circle (.5);
\draw[scalar] (1.25,0) circle (.5);
\draw[wilson] (0,0) -- (2,0);
\draw[fill=white] (.25,0) circle (.075);
\draw[fill=white] (.75,0) circle (.075);
\draw[fill=white] (1.25,0) circle (.075);
\draw[fill=white] (1.75,0) circle (.075);
\draw[fill=VertexColor,VertexColor] (1,.43) circle (\vertexsize);
\draw[fill=VertexColor,VertexColor] (.81,.225) circle (\vertexsize);
\draw[fill=VertexColor,VertexColor] (1.19,.225) circle (\vertexsize);
\end{tikzpicture}}
\newcommand{\DefectSSSSTwoLoopsXHThree}{\begin{tikzpicture}[baseline={([yshift=-2ex]current bounding box.center)}]
\draw[gluontwo] (1.6,.36) -- (1.19,.225);
\draw[wilson] (0,0) -- (2,0);
\draw[fill=white] (.25,0) circle (.075);
\draw[fill=white] (.75,0) circle (.075);
\draw[fill=white] (1.25,0) circle (.075);
\draw[fill=white] (1.75,0) circle (.075);
\draw[fill=VertexColor,VertexColor] (1,.43) circle (\vertexsize);
\draw[fill=VertexColor,VertexColor] (1.6,.36) circle (\vertexsize);
\draw[fill=VertexColor,VertexColor] (1.19,.225) circle (\vertexsize);
\path[clip] (0,0) rectangle (2,1);
\draw[scalar] (.75,0) circle (.5);
\draw[scalar] (1.25,0) circle (.5);
\draw[wilson] (0,0) -- (2,0);
\draw[fill=white] (.25,0) circle (.075);
\draw[fill=white] (.75,0) circle (.075);
\draw[fill=white] (1.25,0) circle (.075);
\draw[fill=white] (1.75,0) circle (.075);
\draw[fill=VertexColor,VertexColor] (1,.43) circle (\vertexsize);
\draw[fill=VertexColor,VertexColor] (1.6,.36) circle (\vertexsize);
\draw[fill=VertexColor,VertexColor] (1.19,.225) circle (\vertexsize);
\end{tikzpicture}}
\newcommand{\DefectSSSSTwoLoopsXHFour}{\begin{tikzpicture}[baseline={([yshift=-2ex]current bounding box.center)}]
\draw[wilson] (0,0) -- (2,0);
\draw[fill=white] (.25,0) circle (.075);
\draw[fill=white] (.75,0) circle (.075);
\draw[fill=white] (1.25,0) circle (.075);
\draw[fill=white] (1.75,0) circle (.075);
\draw[fill=VertexColor,VertexColor] (1,.43) circle (\vertexsize);
\draw[fill=VertexColor,VertexColor] (.4,.36) circle (\vertexsize);
\draw[fill=VertexColor,VertexColor] (1.6,.36) circle (\vertexsize);
\path[clip] (0,0) rectangle (2,1);
\draw[scalar] (.75,0) circle (.5);
\draw[scalar] (1.25,0) circle (.5);
\draw[wilson] (0,0) -- (2,0);
\draw[fill=white] (.25,0) circle (.075);
\draw[fill=white] (.75,0) circle (.075);
\draw[fill=white] (1.25,0) circle (.075);
\draw[fill=white] (1.75,0) circle (.075);
\draw[fill=VertexColor,VertexColor] (.4,.36) circle (\vertexsize);
\draw[fill=VertexColor,VertexColor] (1.6,.36) circle (\vertexsize);
\path[clip] (0,.36) rectangle (2,1);
\draw[gluontwo] (1,.15) circle (.65);
\draw[fill=VertexColor,VertexColor] (1,.43) circle (\vertexsize);
\draw[fill=VertexColor,VertexColor] (.4,.36) circle (\vertexsize);
\draw[fill=VertexColor,VertexColor] (1.6,.36) circle (\vertexsize);
\end{tikzpicture}}

\newcommand{\DefectSSSSTwoLoopsSpider}{\begin{tikzpicture}[baseline={([yshift=-2ex]current bounding box.center)}]
\draw[wilson] (0,0) -- (2,0);
\draw[fill=white] (.25,0) circle (.075);
\draw[fill=white] (.75,0) circle (.075);
\draw[fill=white] (1.25,0) circle (.075);
\draw[fill=white] (1.75,0) circle (.075);
\draw[fill=VertexColor,VertexColor] (1,.43) circle (\vertexsize);
\path[clip] (0,0) rectangle (2,1);
\draw[scalar] (.75,.1) circle (.52);
\draw[scalar] (1.25,.1) circle (.52);
\draw[wilson] (0,0) -- (2,0);
\draw[fill=white] (.25,0) circle (.075);
\draw[fill=white] (.75,0) circle (.075);
\draw[fill=white] (1.25,0) circle (.075);
\draw[fill=white] (1.75,0) circle (.075);
\draw[scalar, fill=white] (1,.45) circle (.275);
\draw[-Latex] (1,.725)--(1.1,.725);
\draw[-Latex] (1,.18)--(.9,.18);
\draw[-Latex] (.725,.43)--(.725,.53);
\draw[-Latex] (1.275,.44)--(1.275,.34);
\draw[fill=VertexColor,VertexColor] (.775,.615) circle (\vertexsize);
\draw[fill=VertexColor,VertexColor] (.785,.25) circle (\vertexsize);
\draw[fill=VertexColor,VertexColor] (1.225,.615) circle (\vertexsize);
\draw[fill=VertexColor,VertexColor] (1.215,.25) circle (\vertexsize);
\end{tikzpicture}}

\newcommand{\DefectSSSSTwoLoopsXYOne}{\begin{tikzpicture}[baseline={([yshift=-2ex]current bounding box.center)}]
\draw[wilson] (-.2,0) -- (2.2,0);
\draw[gluon] (.4,.36) -- (.1,.66);
\draw[fill=white] (.25,0) circle (.075);
\draw[fill=white] (.75,0) circle (.075);
\draw[fill=white] (1.25,0) circle (.075);
\draw[fill=white] (1.75,0) circle (.075);
\draw[fill=VertexColor,VertexColor] (1,.43) circle (\vertexsize);
\draw[fill=VertexColor,VertexColor] (.4,.36) circle (\vertexsize);
\draw[fill=VertexColor,VertexColor] (0,0) circle (\vertexsize);
\draw[fill=VertexColor,VertexColor] (.5,0) circle (\vertexsize);
\draw[fill=VertexColor,VertexColor] (2,0) circle (\vertexsize);
\path[clip] (0,0) rectangle (2,1);
\draw[scalar] (.75,0) circle (.5);
\draw[scalar] (1.25,0) circle (.5);
\draw[wilson] (0,0) -- (2,0);
\draw[fill=white] (.25,0) circle (.075);
\draw[fill=white] (.75,0) circle (.075);
\draw[fill=white] (1.25,0) circle (.075);
\draw[fill=white] (1.75,0) circle (.075);
\draw[fill=VertexColor,VertexColor] (1,.43) circle (\vertexsize);
\draw[fill=VertexColor,VertexColor] (.4,.36) circle (\vertexsize);
\draw[fill=VertexColor,VertexColor] (0,0) circle (\vertexsize);
\draw[fill=VertexColor,VertexColor] (.5,0) circle (\vertexsize);
\draw[fill=VertexColor,VertexColor] (2,0) circle (\vertexsize);
\end{tikzpicture}}

\newcommand{\DefectSSSSTwoLoopsXYTwo}{\begin{tikzpicture}[baseline={([yshift=-2ex]current bounding box.center)}]
\draw[wilson] (0,0) -- (2,0);
\draw[gluon] (.81,.225) -- (.46,.225);
\draw[fill=white] (.25,0) circle (.075);
\draw[fill=white] (.75,0) circle (.075);
\draw[fill=white] (1.25,0) circle (.075);
\draw[fill=white] (1.75,0) circle (.075);
\draw[fill=VertexColor,VertexColor] (1,.43) circle (\vertexsize);
\draw[fill=VertexColor,VertexColor] (.81,.225) circle (\vertexsize);
\draw[fill=VertexColor,VertexColor] (.5,0) circle (\vertexsize);
\draw[fill=VertexColor,VertexColor] (1,0) circle (\vertexsize);
\path[clip] (0,0) rectangle (2,1);
\draw[scalar] (.75,0) circle (.5);
\draw[scalar] (1.25,0) circle (.5);
\draw[wilson] (0,0) -- (2,0);
\draw[fill=white] (.25,0) circle (.075);
\draw[fill=white] (.75,0) circle (.075);
\draw[fill=white] (1.25,0) circle (.075);
\draw[fill=white] (1.75,0) circle (.075);
\draw[fill=VertexColor,VertexColor] (1,.43) circle (\vertexsize);
\draw[fill=VertexColor,VertexColor] (.81,.225) circle (\vertexsize);
\draw[fill=VertexColor,VertexColor] (.5,0) circle (\vertexsize);
\draw[fill=VertexColor,VertexColor] (1,0) circle (\vertexsize);
\end{tikzpicture}}

\newcommand{\DefectSSSSTwoLoopsXYThree}{\begin{tikzpicture}[baseline={([yshift=-2ex]current bounding box.center)}]
\draw[wilson] (0,0) -- (2,0);
\draw[gluon] (1.19,.225) -- (1.54,.225);
\draw[fill=white] (.25,0) circle (.075);
\draw[fill=white] (.75,0) circle (.075);
\draw[fill=white] (1.25,0) circle (.075);
\draw[fill=white] (1.75,0) circle (.075);
\draw[fill=VertexColor,VertexColor] (1,.43) circle (\vertexsize);
\draw[fill=VertexColor,VertexColor] (1.19,.225) circle (\vertexsize);
\draw[fill=VertexColor,VertexColor] (1.5,0) circle (\vertexsize);
\draw[fill=VertexColor,VertexColor] (1,0) circle (\vertexsize);
\path[clip] (0,0) rectangle (2,1);
\draw[scalar] (.75,0) circle (.5);
\draw[scalar] (1.25,0) circle (.5);
\draw[wilson] (0,0) -- (2,0);
\draw[fill=white] (.25,0) circle (.075);
\draw[fill=white] (.75,0) circle (.075);
\draw[fill=white] (1.25,0) circle (.075);
\draw[fill=white] (1.75,0) circle (.075);
\draw[fill=VertexColor,VertexColor] (1,.43) circle (\vertexsize);
\draw[fill=VertexColor,VertexColor] (1.19,.225) circle (\vertexsize);
\draw[fill=VertexColor,VertexColor] (1.5,0) circle (\vertexsize);
\draw[fill=VertexColor,VertexColor] (1,0) circle (\vertexsize);
\end{tikzpicture}}

\newcommand{\DefectSSSSTwoLoopsXYFour}{\begin{tikzpicture}[baseline={([yshift=-2ex]current bounding box.center)}]
\draw[wilson] (-.2,0) -- (2.2,0);
\draw[gluon] (1.6,.36) -- (1.9,.66);
\draw[fill=white] (.25,0) circle (.075);
\draw[fill=white] (.75,0) circle (.075);
\draw[fill=white] (1.25,0) circle (.075);
\draw[fill=white] (1.75,0) circle (.075);
\draw[fill=VertexColor,VertexColor] (1,.43) circle (\vertexsize);
\draw[fill=VertexColor,VertexColor] (1.6,.36) circle (\vertexsize);
\draw[fill=VertexColor,VertexColor] (0,0) circle (\vertexsize);
\draw[fill=VertexColor,VertexColor] (1.5,0) circle (\vertexsize);
\draw[fill=VertexColor,VertexColor] (2,0) circle (\vertexsize);
\path[clip] (0,0) rectangle (2,1);
\draw[scalar] (.75,0) circle (.5);
\draw[scalar] (1.25,0) circle (.5);
\draw[wilson] (0,0) -- (2,0);
\draw[fill=white] (.25,0) circle (.075);
\draw[fill=white] (.75,0) circle (.075);
\draw[fill=white] (1.25,0) circle (.075);
\draw[fill=white] (1.75,0) circle (.075);
\draw[fill=VertexColor,VertexColor] (1,.43) circle (\vertexsize);
\draw[fill=VertexColor,VertexColor] (1.6,.36) circle (\vertexsize);
\draw[fill=VertexColor,VertexColor] (0,0) circle (\vertexsize);
\draw[fill=VertexColor,VertexColor] (1.5,0) circle (\vertexsize);
\draw[fill=VertexColor,VertexColor] (2,0) circle (\vertexsize);
\end{tikzpicture}}

\newcommand{\XIntegral}{\begin{tikzpicture}[baseline={([yshift=-.35ex]current bounding box.center)}]
\draw[scalar] (0,0) -- (1.25,1);
\draw[scalar] (0,1) -- (1.25,0);
\draw[fill=VertexColor,VertexColor] (.625,.5) circle (\vertexsize);
\draw[fill=white] (0,0) circle (.075);
\draw[fill=white] (0,1) circle (.075);
\draw[fill=white] (1.25,1) circle (.075);
\draw[fill=white] (1.25,0) circle (.075);
\end{tikzpicture}}
\newcommand{\YIntegral}{\begin{tikzpicture}[baseline={([yshift=-.45ex]current bounding box.center)}]
\draw[scalar] (0,.5) -- (.75,.5);
\draw[scalar] (.75,.5) -- (1.25,1);
\draw[scalar] (.75,.5) -- (1.25,0);
\draw[fill=VertexColor,VertexColor] (.75,.5) circle (\vertexsize);
\draw[fill=white] (0,.5) circle (.075);
\draw[fill=white] (1.25,1) circle (.075);
\draw[fill=white] (1.25,0) circle (.075);
\end{tikzpicture}}

\newcommand{\IntegralYOneTwoTwo}{\scalebox{\x}{\begin{tikzpicture}[baseline={([yshift=-.55ex]current bounding box.center)}]
\draw[scalar] (0,.5) -- (1,.5);
\draw[scalar] (1.5,.5) circle (.5);
\draw[fill=white] (0,.5) circle (.075);
\draw[fill=white] (2,.5) circle (.075);
\draw[fill=VertexColor,VertexColor] (1,.5) circle (\vertexsize);
\end{tikzpicture}}}
\newcommand{\IntegralXOneTwoThreeThree}{\scalebox{\x}{\begin{tikzpicture}[baseline={([yshift=-.55ex]current bounding box.center)}]
\draw[scalar] (.5,1) -- (1,.5);
\draw[scalar] (.5,0) -- (1,.5);
\draw[scalar] (1.5,.5) circle (.5);
\draw[fill=white] (.5,1) circle (.075);
\draw[fill=white] (.5,0) circle (.075);
\draw[fill=white] (2,.5) circle (.075);
\draw[fill=VertexColor,VertexColor] (1,.5) circle (\vertexsize);
\end{tikzpicture}}}
\newcommand{\IntegralXOneOneTwoTwo}{\scalebox{\x}{\begin{tikzpicture}[baseline={([yshift=-.55ex]current bounding box.center)}]
\draw[scalar] (.5,.5) circle (.5);
\draw[scalar] (1.5,.5) circle (.5);
\draw[fill=white] (0,.5) circle (.075);
\draw[fill=white] (2,.5) circle (.075);
\draw[fill=VertexColor,VertexColor] (1,.5) circle (\vertexsize);
\end{tikzpicture}}}
\newcommand{\IntegralFOneThreeTwoThree}{\scalebox{\x}{\begin{tikzpicture}[baseline={([yshift=-.55ex]current bounding box.center)}]
\draw[scalar] (0,0) -- (.625,0);
\draw[scalar] (0,1) -- (.625,1);
\draw[scalar] (.625,0) arc (270:450:.5);
\draw[scalar] (.625,0) -- (.625,1);
\draw[fill=VertexColor,VertexColor] (.625,0) circle (\vertexsize);
\draw[fill=VertexColor,VertexColor] (.625,1) circle (\vertexsize);
\draw[fill=white] (0,0) circle (.075);
\draw[fill=white] (0,1) circle (.075);
\draw[fill=white] (1.125,.5) circle (.075);
\end{tikzpicture}}}

\newcommand{\KiteIntegral}{\begin{tikzpicture}[baseline={([yshift=-.35ex]current bounding box.center)}]
\draw[scalar] (0,0) -- (2,0);
\draw[scalar] (.5,0) -- (1,1);
\draw[scalar] (.5,0) -- (1,-1);
\draw[scalar] (1.5,0) -- (1,1);
\draw[scalar] (1.5,0) -- (1,-1);
\draw[ghost] (1,-1) -- (1,1);
\draw[fill=VertexColor,VertexColor] (.5,0) circle (\vertexsize);
\draw[fill=VertexColor,VertexColor] (1.5,0) circle (\vertexsize);
\draw[fill=white] (0,0) circle (.075);
\draw[fill=white] (2,0) circle (.075);
\draw[fill=white] (1,1) circle (.075);
\draw[fill=white] (1,-1) circle (.075);
\node[] at (-.3,0) {{\footnotesize $2$}};
\node[] at (2.3,0) {{\footnotesize $4$}};
\node[] at (1,1.3) {{\footnotesize $1$}};
\node[] at (1,-1.3) {{\footnotesize $3$}};
\end{tikzpicture}}

\usepackage{simplewick}
\usepackage{dsfont}
\usepackage{float}
\usepackage{pgfplots}
\pgfplotsset{compat=1.14}

\let\oldbfseries=\bfseries
\let\oldmdseries=\mdseries
\let\oldnormalfont=\normalfont
\renewcommand{\bfseries}{\oldbfseries\boldmath}
\renewcommand{\mdseries}{\oldmdseries\unboldmath}
\renewcommand{\normalfont}{\oldnormalfont\unboldmath}

\makeatletter
\newlength{\apb@width}
\newcommand{\autoparbox}[2][c]{\settowidth{\apb@width}{#2}\parbox[#1]{\apb@width}{#2}}

\makeatother

\DeclareMathOperator{\tr}{tr}

\def\Am{{\mathcal{A}}}

\def\Cm{{\mathcal{C}}}

\def\Km{{\mathcal{K}}}

\def\Nm{{\mathcal{N}}}
\def\Om{{\mathcal{O}}}
\def\Pm{{\mathcal{P}}}

\def\Sm{{\mathcal{S}}}

\def\zb{{\bar{z}}}

\def\veps{\varepsilon}

\def\pd{\partial}

\newcommand{\beq}{\begin{equation}}
\newcommand{\eeq}{\end{equation}}

\definecolor{nicegreen}{rgb}{0.1,0.6,0.1}

\newmuskip\pFqskip
\mathchardef\pFcomma=\mathcode`,

\makeatletter
\renewcommand*\env@matrix[1][\arraystretch]{%
  \edef\arraystretch{#1}%
  \hskip -\arraycolsep
  \let\@ifnextchar\new@ifnextchar
  \array{*\c@MaxMatrixCols c}}
\makeatother

\graphicspath{{./figures/}}

\begin{document} 

\thispagestyle{empty}

\vspace*{-.6in}
\begin{flushright}
    HU-EP-24/27-RTG |  DESY-24-165
\end{flushright}

\vspace{1cm}

{\large
\begin{center}
    {\Large \bf Perturbative bootstrap of the Wilson-line defect CFT: Multipoint correlators
    }\\
\end{center}}

\vspace{0.5cm}

\begin{center}
    Daniele Artico,$^{a,}\footnote{daniele.artico@physik.hu-berlin.de}$
    Julien Barrat,$^{b,}\footnote{julien.barrat@desy.de}$ and
    Giulia Peveri$^{c,}\footnote{giulia.peveri@fht.org}$\\[0.5cm] 
    { \small
    $^{a}$ Institut f\"ur Physik und IRIS Adlershof, Humboldt-Universit{\"a}t zu Berlin, Zum Gro{\ss}en Windkanal 2, 12489 Berlin, Germany\\
    \vspace{0.3em}
    $^{b}$ Deutsches Elektronen-Synchrotron DESY, Notkestr. 85, 22607 Hamburg, Germany\\
    \vspace{0.3em}
    $^{c}$ Human Technopole, V.le Rita Levi-Montalcini, 1, 20157 Milano, Italy\\
    }
    \vspace{1cm} 

    \bf Abstract
\end{center}

\begin{abstract}
    \noindent We study the defect CFT associated with the half-BPS Wilson line in $\Nm=4$ Super Yang-Mills theory in four dimensions.
    Using a perturbative bootstrap approach, we derive new analytical results for multipoint correlators of protected defect operators at large $N$ and weak coupling.
    At next-to-next-to-leading order, we demonstrate that the simplest five- and six-point functions are fully determined by non-perturbative constraints — which include superconformal symmetry, crossing symmetry, and the pinching of operators to lower-point functions — as well as by a single integral, known as the train track integral.
    Additionally, we present new analytical results for the four-point functions $\vev{1122}$ and $\vev{1212}$.
\end{abstract}

\newpage

\setcounter{tocdepth}{2}
\tableofcontents
\thispagestyle{empty}

\newpage

\setcounter{page}{1}

\section{Introduction}
\label{sec:introduction}

\setcounter{footnote}{0}

Defects play a crucial role as observables in physics, with applications spanning from condensed-matter systems to high-energy physics.
In condensed matter, line defects typically correspond to point-like impurities in atomic lattices, while in gauge theories Wilson lines are essential in probing confinement.
In quantum chromodynamics (QCD), the expectation value of the Wilson line serves as an order parameter for confinement \cite{Polyakov:1978vu,Witten:1998zw}.
In conformal field theories (CFTs), there exists a specialized class of defects, known as conformal defects, which break conformal symmetry in a controlled manner \cite{Billo:2016cpy,Lauria:2020emq}.
For line defects, this symmetry breaking preserves a one-dimensional CFT.
Conformal line defects have been studied in a wide range of critical systems, from the $\mathrm{O}(N)$ model (and related theories) in various dimensions \cite{Cuomo:2021kfm,Gimenez-Grau:2022czc,Gimenez-Grau:2022ebb,Bianchi:2022sbz,Aharony:2023amq,Cuomo:2024psk} to supersymmetric theories \cite{Drukker:1999zq,Drukker:2000rr,Semenoff:2001xp,Drukker:2011za,Bianchi:2020hsz}, employing powerful techniques such as the conformal bootstrap, integrability, and supersymmetric localization. 

Four-dimensional $\Nm=4$ Super Yang-Mills (SYM) occupies a special position in the space of quantum field theories, due to its rich structure: it is a conformal field theory, believed to be integrable \cite{Beisert:2003tq,Beisert:2010jr}, and it has a well-studied holographic dual in the context of the AdS/CFT correspondence \cite{Maldacena:1997re,Witten:1998qj}.
A particularly notable conformal defect in this theory is the supersymmetric Maldacena-Wilson loop \cite{Maldacena:1998im}, defined in Euclidean space along a path $\Cm$ as
\begin{equation}
    \Wm_{\Cm} = \frac{1}{N} \tr \Pm \exp \oint_{\Cm} d\tau (i \dot{x}_\mu A_\mu (\tau) + |\dot{x}|\, \theta^I \phi^I (\tau))\,,
    \label{eq:MaldacenaWilsonLoop}
\end{equation}
where $\theta^{I=1, \ldots, 6}$ is a polarization vector that defines which of the $\mathfrak{so}(6)_R$ scalar fields couple to the defect.
For a circular geometry, the operator becomes half-BPS, and the exact expectation value of this operator is given by a Bessel function \cite{Erickson:2000af,Drukker:2000rr,Pestun:2007rz,Pestun:2009nn}.
For an infinite straight line, it reduces to the simple value of $1$.
In recent years, the Wilson-line defect CFT in $\Nm=4$ SYM has garnered significant attention and it has been explored through approaches such as the conformal bootstrap \cite{Liendo:2016ymz,Liendo:2018ukf,Ferrero:2021bsb,Barrat:2021yvp,Barrat:2022psm,Ferrero:2023znz,Ferrero:2023gnu,Bonomi:2024lky,Carmi:2024tmp}, integrability \cite{Giombi:2009ds,Giombi:2018qox,Giombi:2018hsx}, and a combination of both techniques, known as bootstrability \cite{Cavaglia:2021bnz,Cavaglia:2022qpg,Cavaglia:2022yvv,Cavaglia:2023mmu}.
Perturbative calculations, both at weak \cite{Barrat:2021tpn,Barrat:2022eim,Bianchi:2022ppi,Artico:2024wnt} and strong coupling \cite{Giombi:2017cqn,Gimenez-Grau:2023fcy,Giombi:2023zte,Artico:2024wnt}, have yielded a remarkable amount of data.

In this work, we focus on multipoint correlation functions of defect half-BPS operators.
Several motivations drive this direction of study.
Higher-point correlation functions are expected to become a central focus of conformal bootstrap studies in the near future, as they encode vast amounts of CFT data and provide an alternative framework to the conventional analysis of multiple four-point functions.
Early progress in this area has been made, e.g., with studies of five-point functions in the Ising model \cite{Poland:2023vpn,Poland:2023bny} and six-point functions in one-dimensional systems \cite{Antunes:2023kyz,Harris:2024nmr}.\footnote{See \cite{Bercini:2020msp,Antunes:2021kmm,Buric:2021ywo,Buric:2021ttm,Buric:2021kgy,Kaviraj:2022wbw,Bargheer:2024hfx} for various works in this direction.}
The Wilson-line defect CFT is particularly well-suited for advancing these techniques, and a fascinating direction would be to merge the techniques of \cite{Antunes:2023kyz} and \cite{Cavaglia:2021bnz} to do a multipoint bootstrability study.
Moreover, we aim to demonstrate how symmetry considerations, combined with appropriately constructed Ans\"atze, can lead to novel analytical results in the weak-coupling regime.
In particular, multipoint superconformal Ward identities have been conjectured in \cite{Barrat:2021tpn}, before being confirmed and extended in \cite{Bliard:2024und,Barrat:2024ta}.

\begin{figure}
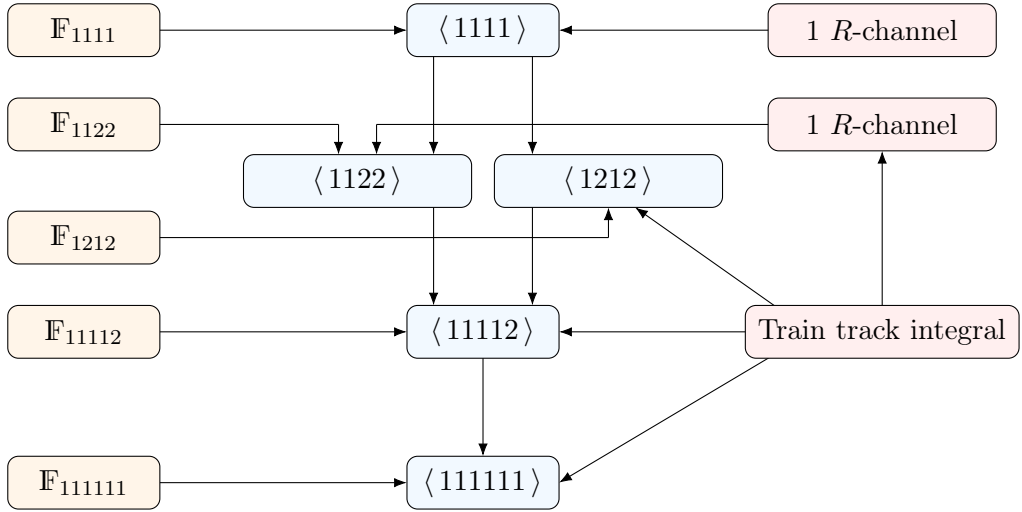

    \centering
    \DiagramIntro
    \caption{Illustration of the strategy used for calculating multipoint correlation functions. The center column consists of the correlators that are computed throughout this paper. The left column refers to the data accessible non-perturbatively, while the right column refers to the perturbative input. The correlators $\vev{1111}$ and $\vev{1122}$ require the input of one $R$-channel each, while the other correlators necessitate the topological data $\Fds$ and a single six-point integral, known as the train track.}
    \label{fig:DiagramIntro}
\end{figure}

We introduce a novel bootstrap approach for the perturbative analysis of multipoint correlators, leading to new results for the five-point function $\vev{11112}$ and the six-point function $\vev{111111}$.
Our method leverages non-perturbative constraints such as superconformal symmetry, crossing symmetry, and the pinching behavior of correlators to lower-point functions.
Remarkably, we show that the correlators are governed only by either one or two functions of the spacetime cross-ratios, significantly simplifying the computations.
This reduction allows us to focus on a selected class of diagrams, from which we can determine the correlators at next-to-next-to-leading order.
We specifically demonstrate that the six-point train track integral in the collinear limit controls the correlators.
By inputting the result for this integral along with the non-perturbative constraints, we fully determine the correlators, providing new analytical expressions.
Each correlator is systematically constructed by building upon lower-point results, as illustrated in Figure \ref{fig:DiagramIntro}.

The structure of the paper is as follows.
In Section \ref{sec:Preliminaries}, we provide a brief overview of the Wilson-line defect CFT, exploring it from both the group-theoretical and perturbative perspectives.
Section \ref{sec:NonPerturbativeConstraints} outlines the non-perturbative constraints that form the foundation for our calculations.
Given the dependence of our results on lower-point functions, Section \ref{sec:FourPointFunctions} presents explicit calculations for the correlators $\vev{1111}$, $\vev{1122}$, and $\vev{1212}$ up to next-to-next-to-leading order.
Our main results — the correlators $\vev{11112}$ and $\vev{111111}$ at next-to-next-to-leading order — are detailed in Section \ref{sec:HigherPointFunctions}.
Section \ref{sec:conclusions} summarizes our findings and proposes directions for future exploration.
The paper concludes with four appendices: Appendix \ref{app:SymbolsAndGoncharovPolylogarithms} reviews symbols and Goncharov polylogarithms; Appendix \ref{app:Integrals} provides the integrals necessary throughout this work; and Appendices \ref{app:FeynmanDiagramsOf1111} and \ref{app:FeynmanDiagramsOf1122} summarize the Feynman diagram computations for the correlators $\vev{1111}$ and $\vev{1122}$, respectively.

\section{Preliminaries}
\label{sec:Preliminaries}

This section provides the foundational material essential for the discussion in this paper.
We begin with a concise overview of the Wilson-line defect CFT, then we introduce the correlation functions that are the focus of subsequent sections.
Finally, we discuss a special conformal integral (the six-point train track) that turns out to be essential for the calculation of the correlators.

\subsection{The Wilson-line defect CFT}
\label{subsec:TheWilsonLineDefectCFT}

We begin by outlining the Wilson-line defect CFT, approached from both perturbative and group-theoretical perspectives.

\subsubsection{The bulk action}
\label{subsubsec:TheBulkAction}

The four-dimensional $\Nm=4$ SYM theory consists of six scalar fields, four Weyl fermions, one gauge field, and one ghost field.
The corresponding action, which we refer to as the \textit{bulk} action, is given by
\begin{equation}
    \begin{split}
    S
    =\ &
    \frac{1}{g^2} \tr \int d^4 x\, \biggl( 
    \frac{1}{2} F_{\mu\nu} F_{\mu\nu}
    + D_\mu \phi^I D_\mu \phi^I
    - \frac{1}{2} [ \phi^I, \phi^J ] [ \phi^I, \phi^J ] \\
    &+
    i\, \psib \slashed{D} \psi
    + \psib \Gamma^I [ \phi^I\,, \psi ]
    + \pd_\mu \cb D_\mu c
    + \xi ( \pd_\mu A_\mu )^2
    \biggr)\,,
    \end{split}
    \label{eq:BulkAction}
\end{equation}
where $\mu=0, \ldots, 3$ denote spacetime directions, and $I=1, \ldots, 6$ are indices of the $\mathfrak{so}(6)_R$ $R$-symmetry. We work in the Feynman gauge $\xi = 1$.
The Weyl fermions are combined into a single $16$-component Majorana fermion.
All fields are in the adjoint representation of the gauge group, and carry a generator in their definition:
\begin{equation}
    \phi = T^a \phi^a\,, \quad \psi = T^a \psi^a\,, \quad A_\mu = T^a A_\mu^a\,,
    \label{eq:FieldsAndGenerators}
\end{equation}
where $a = 1, \ldots, N$ is the color index associated with the $\mathfrak{su}(N)$ algebra.
In this paper, we set $N \to \infty$ while keeping the coupling constant $\lambda = g^2 N$ small.

\subsubsection{The Maldacena-Wilson line}
\label{subsubsec:TheMaldacenaWilsonLine}

The Wilson-line defect CFT considered in this paper is defined by the Maldacena-Wilson loop \eqref{eq:MaldacenaWilsonLoop}, which we orient along the temporal direction as a straight line: 
\begin{equation}
    \Wl
    =
    \frac{1}{N}
    \tr \Pm \exp \int_{-\infty}^\infty d\tau\,
    (i A_0 (\tau) + \phi^6 (\tau))\,,
    \label{eq:WilsonLine}
\end{equation}
where, without loss of generality, we choose $\phi^6$ to be the scalar field coupling to the line defect by setting $\theta$ to be
\begin{equation}
	\theta
	=
	(0,0,0,0,0,1)\,.
	\label{eq:theta}
\end{equation}

\noindent This operator breaks the (Euclidean) conformal group as follows:
\begin{equation}
    \mathfrak{so} (5,1) \to \mathfrak{so} (2,1) \times \mathfrak{so} (3)\,.
    \label{eq:SymmetryBreaking}
\end{equation}
Here, $\mathfrak{so}(2,1)$ corresponds to the conformal group of the one-dimensional CFT associated with the line defect, with the quantum number being the scaling dimension $\Delta$.
The $\mathfrak{so}(3)$ factor represents the rotational symmetry around the defect, which can be interpreted as an internal symmetry with quantum number $s$, referred to as the \textit{transverse spin}.
The defect also breaks the $R$-symmetry group $\mathfrak{so}(6)_R$ down to $\mathfrak{so}(5)_R$, with $[a,b]$ denoting the corresponding quantum numbers of the irreducible representation of the operators living on the defect.
Altogether, the full supersymmetric algebra $\mathfrak{psu}(2,2|4)$ of $\Nm=4$ SYM is broken down to the defect algebra $\mathfrak{osp}(4^*|4)$.

Among the representations of the defect algebra, a special class consists of the scalar half-BPS operators $\Op_\Delta$, which have protected scaling dimensions and satisfy $[a,b] = [0, \Delta ]$ and $s=0$.
We focus on the operators that are inserted along the trace of the Wilson line.\footnote{In principle, higher-trace operators exist.
However, they do not play a significant role in the large $N$ limit for the one-dimensional theory.
They do, however, become relevant in the study of correlation functions that involve bulk operators in the presence of a defect, as discussed in Appendix A of \cite{Giombi:2018hsx} or in the microbootstrap section of \cite{Barrat:2024nod}, even if $N$ is taken to be infinity.}
These operators are defined by
\begin{equation}
    \Op_\Delta (u,\tau)
    =
    \frac{1}{\sqrt{n_\Delta}}
    \Wl [ (u \cdot \phi(\tau))^\Delta ]\,,
    \label{eq:HalfBPSOperator}
\end{equation}
where $u^2 = 0$ and $u \cdot \theta = 0$ to ensure that the representation remains symmetric traceless and decoupled from the field $\phi^6$ present in the Wilson line.
Here, $\Wl[\ldots]$ indicates that the fields are \textit{inserted} along the Wilson line, meaning
\begin{equation}
    \Wl [ \Op ]
    =
    \frac{1}{N} \tr
    \Pm [\, \Op\, \exp \int d\tau\,
    (i A_0 (\tau) + \phi^6 (\tau)) ]\,.
    \label{eq:Insertion}
\end{equation}

The normalization constants $n_\Delta$ depend on the coupling $\lambda$ only.
For half-BPS operators, these constants can be computed using the methods outlined in \cite{Giombi:2018qox}.
Below are the expressions for the operators relevant to this paper:
\begin{align}
    n_1
    &=
    \frac{\sqrt{\lambda}}{2\pi^2} \frac{\Ids_1}{\Ids_2}\,, \label{eq:n1} \\
    n_2
    &=
    \frac{1}{4\pi^4} (3 \lambda - (\Ids_1 - 2)(\Ids_1 + 10))\,, \label{eq:n2} \\
    n_3
    &=
    \frac{3}{8 \pi^6} \biggl(
    \frac{(5 \lambda + 72) \Ids_1 \Ids_2}{\sqrt{\lambda}}
    - \frac{\lambda (26 \Ids_1 + 3\lambda - 32) + 288(\Ids_1 - 1)}{\Ids_1 -2}
    \biggr)\,,
    \label{eq:n3}
\end{align}
where we define the function
\begin{equation}
    \Ids_a
    =
    \frac{\sqrt{\lambda}\, I_0 (\sqrt{\lambda})}{I_a (\sqrt{\lambda})}\,.
    \label{eq:Ids}
\end{equation}
Note that $n_1$ has a direct \textit{physical} interpretation, being related to the Bremsstrahlung function, as it describes the emission of soft particles from the line defect \cite{Alday:2007hr}.

\subsubsection{Feynman rules}
\label{subsubsec:FeynmanRules}

In this section, we collect the Feynman rules derived from the bulk action \eqref{eq:BulkAction}, supplemented by the inclusion of the Wilson-line defect.

\paragraph{Bulk Feynman rules.}
We begin by listing the relevant propagators and vertices for four-dimensional $\Nm=4$ SYM.
The (free) propagators are as follows:
\begin{equation}
    \begin{split}
    \text{Scalars:} \qquad 
    & \ScalarPropagator = g^2 \delta^{IJ} \delta^{ab}\, I_{12}\,, \\
    \text{Gluons:} \qquad 
    & \GluonPropagator = g^2 \delta_{\mu\nu} \delta^{ab}\, I_{12}\,, \\
    \text{Fermions:} \qquad 
    & \FermionPropagator = i g^2 \delta^{ab} \slashed{\pd}_1 I_{12}\,, \\
    \text{Ghosts:} \qquad 
    & \GhostPropagator = g^2 \delta^{ab} I_{12}\,,
    \end{split}
    \label{eq:Propagators}
\end{equation}
where $I_{ij}$ is the $4d$ scalar propagator, given by
\begin{equation}
    I_{ij} = \frac{1}{4\pi^2 x_{ij}^2}\,.
    \label{eq:PropagatorFunction4d}
\end{equation}
In our conventions, the free propagators include the (dimensionless) coupling $g^2$.

The two following cubic vertices play an important role in our calculations:
\begin{align}
    \VertexScalarScalarGluon
    &=
    - g^4 f^{abc} \delta^{IJ} (\pd_1 - \pd_2)_\mu Y_{123}\,, \label{eq:VertexScalarScalarGluon} \\
    \VertexFermionFermionScalar
    &=
    - g^4 f^{abc} \Gamma^I \spd_1 \spd_2 Y_{123}\,,
    \label{eq:VertexFermionFermionScalar}
\end{align}
as well as the quartic coupling
\begin{align}
    \VertexFourScalars
    &=
    - g^6
    \left\lbrace f^{abe}f^{cde} \left( \delta^{IK}\delta^{JL}
    -
    \delta_{IL}\delta_{JK} \right)
    +
    f^{ace}f^{bde} \left( \delta_{IJ}\delta_{KL}
    -
    \delta_{IL}\delta_{JK} \right) \right. \notag \\[-1.5em]
    &\phantom{=\ }
    \left. + f^{ade}f^{bce} \left( \delta_{IJ}\delta_{KL}
    -
    \delta_{IK}\delta_{JL} \right) \right\rbrace X_{1234}\,.
    \label{eq:VertexFourScalars}
\end{align}
Here, $X_{1234}$ and $Y_{123}$ are massless conformal integrals defined in \eqref{eq:X1234} and \eqref{eq:Y123}, respectively.
For completeness, we list without details the remaining vertices, which primarily contribute to self-energy diagrams:
\begin{equation}
    \VertexGluonGluonGluon \quad
    \VertexFermionFermionGluon \quad
    \VertexGhostGhostGluon\ \quad
    \VertexScalarScalarGluonGluon \quad
    \VertexGluonGluonGluonGluon\ .
    \label{eq:MoreVertices}
\end{equation}
The explicit forms of these vertices can be found for instance in \cite{Beisert:2002bb} and \cite{Drukker:2009sf} in the form of insertion rules.
Note that the vertices carry a factor $1/g^2$, which is compensated in the expressions above by the powers of $g$ carried by the external propagators.

For later purposes, we provide the insertion rule for the one-loop correction to the scalar propagator as reported in \cite{Erickson:2000af}:
\begin{equation}
    \begin{split}
    \SelfEnergyNoText &=
    \SelfEnergyDiagramOne
    +
    \SelfEnergyDiagramTwo
    +
    \SelfEnergyDiagramThree
    +
    \SelfEnergyDiagramFour \\
    &=
    - 2 g^4 N \delta^{ab} \delta^{IJ}\, Y_{112}\,,
    \end{split}
    \label{eq:SelfEnergy}
\end{equation}
where $Y_{112}$ is a logarithmically divergent integral, which is given explicitly in \eqref{eq:Y112}.

\paragraph{Defect Feynman rules.}
The presence of the Maldacena-Wilson line introduces additional vertices into the theory.
One crucial vertex arises from the coupling of the Wilson line to the gluon field, expressed as
\begin{equation}
    \DefectVertexOnePointGluon
    \sim
    \frac{1}{N} \delta^{\mu 0} \int_{\tau_2}^{\tau_3} d\tau_4\, I_{14}\,.
    \label{eq:DefectVertexOnePointGluon}
\end{equation}
Here, the contribution of the generators of the gauge group depends on the number of insertions, which is determined by the structure of the bulk action and of the correlator of interest.
Additionally, a scalar vertex exists, though it does not play a role in our specific calculations:
\begin{equation}
    \DefectVertexOnePointScalar\,.
    \label{eq:DefectVertexOnePointScalar}
\end{equation}
Note however that this scalar vertex is essential for ensuring that the Wilson line operator \eqref{eq:WilsonLine} maintains a finite expectation value without requiring renormalization.

\subsection{Correlation functions}
\label{subsec:CorrelationFunctions}

\begin{figure}
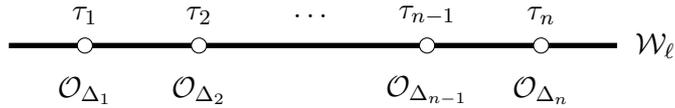

    \centering
    \MultipointCorrelators
    \caption{Illustration of the multipoint correlators defined in \eqref{eq:DefinitionCorrelators}.
    The half-BPS defect operators are ordered along the Wilson line.}
    \label{fig:MultipointCorrelators}
\end{figure}

We now focus on the correlation functions within the Wilson-line defect CFT.
The $1d$ CFT is described by correlators defined as
\begin{equation}
    \vev{\Op_1 (u_1, \tau_1) \ldots \Op_n (u_n, \tau_n)}_{1d}
    =
    \vev{\Wl [ \Op_1 (u_1, \tau_1) \ldots \Op_n (u_n, \tau_n) ]}_{4d}\,,
    \label{eq:DefinitionCorrelators}
\end{equation}
where $\Wl[\ldots]$ refers to the Wilson-line insertion as defined in \eqref{eq:Insertion}.
We assume that the $\tau_i$'s are ordered as $\tau_1 < \tau_2 < \ldots < \tau_n$.
The subscript on the left-hand side signifies a correlator in the $1d$ CFT, while the one on the right-hand side represents its definition within the four-dimensional $\Nm=4$ SYM theory.
Since we are consistently dealing with $1d$ correlators throughout this work, the subscripts will remain implicit from this point forward.

An illustration of the multipoint correlators is provided in Figure \ref{fig:MultipointCorrelators}.
We restrict our analysis to correlators involving half-BPS operators.
For notational convenience, we introduce the shorthand notation
\begin{equation}
    \vev{\Delta_1 \ldots \Delta_n}
    =
    \vev{\Op_1 (u_1, \tau_1) \ldots \Op_n (u_n, \tau_n)}_{1d}\,.
    \label{eq:ShorthandCorrelators}
\end{equation}
In the following, we introduce the kinematic structure of these correlators, with a focus on cases involving up to six operators.

\subsubsection{Two- and three-point functions}
\label{subsubsec:TwoAndThreePointFunctions}

\paragraph{Two-point functions.}
The two-point functions in the defect CFT are constrained by conformal symmetry.
For half-BPS operators, they take the following form:
\begin{equation}
    \vev{\Delta_1 \Delta_2}
    =
    \delta_{\Delta_1 \Delta_2} (12)^{\Delta_1}\,,
    \label{eq:TwoPointFunctions}
\end{equation}
where the shorthand notation
\begin{equation}
    (ij) = \frac{u_i \cdot u_j}{\tau_{ij}^2}
    \label{eq:SuperPropagator}
\end{equation}
has been introduced.
The normalization of the two-point functions in \eqref{eq:TwoPointFunctions} has been chosen to ensure that they are unit-normalized.
The scaling dimensions $\Delta_i$ of half-BPS operators are protected by supersymmetry and integer-valued.

\paragraph{Three-point functions.}
Similarly, the three-point functions of scalar operators are constrained by conformal symmetry and take the form
\begin{equation}
    \vev{\Delta_1 \Delta_2 \Delta_3}
    =
    \lambda_{\Delta_1 \Delta_2 \Delta_3}
    (12)^{\Delta_{123}} (23)^{\Delta_{231}} (31)^{\Delta_{312}}\,,
    \label{eq:ThreePointFunctions}
\end{equation}
where the exponents are given by
\begin{equation}
    \Delta_{ijk}
    =
    \Delta_i + \Delta_j - \Delta_k\,.
    \label{eq:ShorthandThreePoint}
\end{equation}
The coefficients $\lambda_{\Delta_1 \Delta_2 \Delta_3}$, known as OPE coefficients, depend only on the coupling constant.
These coefficients can be computed exactly using integrability-based methods \cite{Giombi:2018qox}.
For the purposes of our calculations, we will make use of the following OPE coefficients:
\begin{align}
    \lambda_{112} &= \frac{-2 \sqrt{\lambda}
   \Ids_1 (7 + \Ids_1) \Ids_2 \Ids_3 - (-32 + 14 \Ids_1 + \Ids_1^2) \Ids_2^2 \Ids_3 + \lambda (3 \Ids_1 \Ids_2^2 - \Ids_1^2 \Ids_3 + 9 \Ids_2^2 \Ids_3)}{4 \Ids_1 \sqrt{\lambda (3 \lambda - (-2 + \Ids_1) (10 + \Ids_1))}
  \Ids_2 \Ids_3} \,, \label{eq:lambda112} \\
    \lambda_{123} &= \frac{\sqrt{3} \Ids_2 \sqrt{\sqrt{\lambda} (\lambda  (-26 \Ids_1-3 \lambda +32)-288 (\Ids_1-1))+(\Ids_1-2) \Ids_1 \Ids_2 (5
   \lambda +72)}}{\sqrt{\lambda } \sqrt{(\Ids_1-2) \Ids_1 \Ids_2 (3 \lambda -(\Ids_1-2) (\Ids_1+10))}} \,, \label{eq:lambda123} \\
    \lambda_{222} &= \frac{6 (\Ids_1-17) \lambda +2 (\Ids_1-2) (\Ids_1 (\Ids_1+14)+148)}{(3 \lambda -(\Ids_1-2) (\Ids_1+10))^{3/2}}\,,\label{eq:lambda222}
\end{align}
where the functions $\Ids_a$ are defined in \eqref{eq:Ids}.

\subsubsection{Four-point functions}
\label{subsubsec:Correlators_FourPointFunctions}

We now turn our attention to the lowest correlators of half-BPS operators with non-trivial kinematics: the four-point functions.
A reduced correlator can be defined as
\begin{equation}
    \vev{\Delta_1 \Delta_2 \Delta_3 \Delta_4}
    =
    \Km_{\Delta_1 \Delta_2 \Delta_3 \Delta_4} \Am_{\Delta_1 \Delta_2 \Delta_3 \Delta_4} (x; r,s)\,,
    \label{eq:4Pt_Correlator}
\end{equation}
where $\Km_{\Delta_1 \Delta_2 \Delta_3 \Delta_4}$ is a (super)conformal prefactor, and $\Am_{\Delta_1 \Delta_2 \Delta_3 \Delta_4} (x; r,s)$ is a dimensionless function depending on one spacetime cross-ratio $x$ and two $R$-symmetry variables $r$ and $s$.
The spacetime cross-ratio is defined as
\begin{equation}
    x
    =
    \frac{\tau_{12} \tau_{34}}{\tau_{13} \tau_{24}}\,,
    \label{eq:4Pt_CrossRatio_x}
\end{equation}
while the $R$-symmetry cross-ratios are given by
\begin{equation}
    r
    =
    \frac{(u_1 \cdot u_2)(u_3 \cdot u_4)}{(u_1 \cdot u_3)(u_2 \cdot u_4)}\,,
    \qquad
    s
    =
    \frac{(u_1 \cdot u_4)(u_2 \cdot u_3)}{(u_1 \cdot u_3)(u_2 \cdot u_4)}\,.
    \label{eq:4Pt_CrossRatio_rs}
\end{equation}

The reduced correlator can be decomposed into $R$-symmetry channels (or $R$-channels):
\begin{equation}
    \Am_{\Delta_1 \Delta_2 \Delta_3 \Delta_4}
    =
    \sum_{i=1}^{r} R_i\, F_i (x)\,,
    \label{eq:4Pt_RSymmetryChannels}
\end{equation}
where $r = r(\Delta_1, \ldots, \Delta_4)$ is the number of channels, and $R_i$ represents the basis elements, which remain unspecified for now but are linearly independent.
The number of channels $r$ can be determined using a recursion relation for $n$-point functions independently of the chosen basis: 
\begin{equation}
    r (\Delta_1, \ldots, \Delta_n)
    =
    r (\Delta_1 - 1, \Delta_2 - 1, \ldots, \Delta_n)
    + \ldots
    +
    r (\Delta_1 - 1, \Delta_2, \ldots, \Delta_n - 1)\,,
    \label{eq:Recursion_RSymmetryChannels}
\end{equation}
with the initial conditions
\begin{equation}
    \begin{split}
        r (\Delta_1, \ldots, \Delta_i, 0, \Delta_{i+2}, \ldots, \Delta_n) &= r (\Delta_1, \ldots, \Delta_i, \Delta_{i+2}, \ldots, \Delta_n)\,, \\
        r (\Delta) &= 0\,, \\
        r (\Delta_1, \Delta_2) &= \delta_{\Delta_1 \Delta_2}\,.
    \end{split}
    \label{eq:Recursion_StartingValues}
\end{equation}
For external operators all having dimension $\Delta = 1$, the number of channels is given by the closed-form expression
\begin{equation}
    r(1,1, \ldots, 1)
    =
    (n-1)!!\,.
    \label{eq:Recursion_SpecialCase}
\end{equation}

In this work, we focus on three specific cases, all of which involve exactly three $R$-symmetry channels:\footnote{It is straightforward to extend this analysis to the more general correlators $\vev{11kk}$ and $\vev{1k1k}$, which also contain three $R$-symmetry channels.}
\begin{align}
        \vev{1111} &= (13)(24) \Am_{1111} (x;r,s)\,, \label{eq:4Pt_1111} \\
        \vev{1122} &= (13)(24)(34) \Am_{1122} (x;r,s)\,, \label{eq:4Pt_1122} \\
        \vev{1212} &= (13)(24)^2 \Am_{1212} (x;r,s)\,. \label{eq:4Pt_1212}
\end{align}
A natural choice for the basis of $R$-symmetry channels is
\begin{equation}
    \Am (x;r,s)
    =
    F_1 (x)
    +
    \frac{r}{x^2} F_2 (x)
    +
    \frac{s}{(1-x)^2} F_3 (x)\,.
    \label{eq:4Pt_NaturalBasis}
\end{equation}
The functions $F_j (x)$ are different for each of the correlators $\vev{1111}$, $\vev{1122}$, and $\vev{1212}$, but we omit additional subscripts to streamline the notation.
The context will clarify which correlator the respective $F_j$ functions refer to.

\subsubsection{Five-point functions}
\label{subsubsec:Correlators_FivePointFunctions}

We now focus on five-point functions, specifically on the case of $\vev{11112}$.
While more general configurations can be studied using the techniques presented here, we restrict ourselves to this case for clarity.
The reduced correlator is defined as
\begin{equation}
    \vev{11112}
    =
    \Km_{11112} \Am_{11112} ( \lbrace x; r, s, t \rbrace )\,,
    \label{eq:11112_Correlator}
\end{equation}
where the kinematic dependence should be understood as
\begin{equation}
    \lbrace x; r,s,t \rbrace
    =
    ( x_1, x_2; r_1, s_1, r_2, s_2, t_{12} )\,.
    \label{eq:11112_Shorthand}
\end{equation}
Five-point functions depend on two spacetime cross-ratios:
\begin{equation}
    x_1
    =
    \frac{\tau_{12} \tau_{45}}{\tau_{14} \tau_{25}}\,,
    \qquad
    x_2
    =
    \frac{\tau_{13} \tau_{45}}{\tau_{14} \tau_{35}}\,,
    \label{eq:11112_CrossRatios_x}
\end{equation}
and five $R$-symmetry variables:
\begin{gather}
        r_1
        =
        \frac{(u_1 \cdot u_2) (u_4 \cdot u_5)}{(u_1 \cdot u_4) (u_2 \cdot u_5)}\,,
        \qquad
        s_1
        =
        \frac{(u_1 \cdot u_5) (u_2 \cdot u_4)}{(u_1 \cdot u_4) (u_2 \cdot u_5)}\,, \notag \\
        r_2
        =
        \frac{(u_1 \cdot u_3) (u_4 \cdot u_5)}{(u_1 \cdot u_4) (u_3 \cdot u_5)}\,,
        \qquad
        s_2
        =
        \frac{(u_1 \cdot u_5) (u_3 \cdot u_4)}{(u_1 \cdot u_4) (u_3 \cdot u_5)}\,, \label{eq:11112_CrossRatio_rst} \\
        t_{12}
        =
        \frac{(u_1 \cdot u_5) (u_2 \cdot u_3) (u_4 \cdot u_5)}{(u_1 \cdot u_4) (u_2 \cdot u_5) (u_3 \cdot u_5)}\,. \notag
\end{gather}
The (super)conformal prefactor $\Km_{11112}$ is chosen as
\begin{equation}
    \Km_{11112}
    =
    (14)(25)(35)\,.
    \label{eq:11112_Prefactor}
\end{equation}
The reduced correlator can be decomposed into six $R$-symmetry channels:
\begin{equation}
    \Am_{11112}
    =
    \sum_{i=1}^6 R_i F_i (x_1, x_2)\,,
    \label{eq:11112_RSymmetryChannels}
\end{equation}
for which the natural choice of basis is
\begin{equation}
    \lbrace R_i \rbrace
    =
    \biggl\lbrace
    1, \frac{r_1}{x_1^2}, \frac{s_1}{(1-x_1)^2}, \frac{r_2}{x_2^2}, \frac{s_2}{(1-x_2)^2}, \frac{t_{12}}{x_{12}^2}
    \biggr\rbrace\,.
    \label{eq:11112_NaturalBasis}
\end{equation}
\subsubsection{Six-point functions}
\label{subsubsec:Correlators_SixPointFunctions}

We now examine the six-point function of elementary insertions $\Op_1$.
Analogous to the lower-point functions discussed previously, this correlator can be expressed as
\begin{equation}
    \vev{111111}
    =
    \Km_{111111} \Am_{111111} ( \lbrace x; r, s, t \rbrace )\,.
    \label{eq:111111_Correlator}
\end{equation}
The function $\Am_{111111}$ depends on three spacetime cross-ratios and nine $R$-symmetry variables.\footnote{In fact, it depends on eight variables only.
For convenience, we keep the ninth variable throughout the paper, though it should be kept in mind that it can be expressed in terms of the other eight cross-ratios.
We thank Pietro Ferrero for bringing this to our attention.}
The spacetime cross-ratios are defined as
\begin{equation}
    x_1
    =
    \frac{\tau_{12}\tau_{56}}{\tau_{15}\tau_{26}}\,,
    \quad
    x_2\
    =
    \frac{\tau_{13}\tau_{56}}{\tau_{15}\tau_{36}}\,,
    \quad
    x_3\
    =
    \frac{\tau_{14}\tau_{56}}{\tau_{15}\tau_{46}}\,,
    \label{eq:111111_CrossRatio_x}
\end{equation}
while the $R$-symmetry variables are
\begin{gather}
    r_1
    =
    \frac{(u_1 \cdot u_2)(u_5 \cdot u_6)}{(u_1 \cdot u_5)(u_2 \cdot u_6)}\,,
    \quad
    r_2
    =
    \frac{(u_1 \cdot u_3)(u_5 \cdot u_6)}{(u_1 \cdot u_5)(u_3 \cdot u_6)}\,,
    \quad
    r_3
    =
    \frac{(u_1 \cdot u_4)(u_5 \cdot u_6)}{(u_1 \cdot u_5)(u_4 \cdot u_6)}\,, \notag \\
    s_1
    =
    \frac{(u_1 \cdot u_6)(u_2 \cdot u_5)}{(u_1 \cdot u_5)(u_2 \cdot u_6)}\,,
    \quad
    s_2
    =
    \frac{(u_1 \cdot u_6)(u_3 \cdot u_5)}{(u_1 \cdot u_5)(u_3 \cdot u_6)}\,,
    \quad
    s_3
    =
    \frac{(u_1 \cdot u_6)(u_4 \cdot u_5)}{(u_1 \cdot u_5)(u_4 \cdot u_6)}\,, \notag \\
    t_{12}
    =
    \frac{(u_1 \cdot u_6)(u_2 \cdot u_3)(u_5 \cdot u_6)}{(u_1 \cdot u_5)(u_2 \cdot u_6)(u_3 \cdot u_6)}\,, \label{eq:111111_CrossRatio_rst}
    \quad
    t_{13}
    =
    \frac{(u_1 \cdot u_6)(u_2 \cdot u_4)(u_5 \cdot u_6)}{(u_1 \cdot u_5)(u_2 \cdot u_6)(u_4 \cdot u_6)}\,, \notag \\
    t_{23}
    =
    \frac{(u_1 \cdot u_6)(u_3 \cdot u_4)(u_5 \cdot u_6)}{(u_1 \cdot u_5)(u_3 \cdot u_6)(u_4 \cdot u_6)}\,.
\end{gather}
The (super)conformal prefactor is chosen as
\begin{equation}
    \Km_{111111}
    =
    \frac{(15)^2 (26) (36) (46)}{(16) (56)}\,.
    \label{eq:111111_Prefactor}
\end{equation}
This choice is motivated by the natural basis of $R$-symmetry channels which, with this choice of prefactor, is
\begin{equation}
    \begin{split}
    \lbrace R_i \rbrace
    &=
    \biggl\lbrace
    \frac{t_{23}}{x_{23}^2}, \frac{t_{23} r_1}{x_{23}^2 x_1^2}, \frac{t_{23} s_1}{x_{23}^2 (1-x_1)^2},
    \frac{t_{13}}{x_{13}^2}, \frac{t_{13} r_2}{x_{13}^2 x_2^2}, \frac{t_{13} s_2}{x_{13}^2 (1-x_2)^2},
    \frac{t_{12}}{x_{12}^2}, \frac{t_{12} r_3}{x_{12}^2 x_3^2}, \frac{t_{12} s_3}{x_{12}^2 (1-x_3)^2}, \\
    & \phantom{=\ }
    \phantom{\biggl\lbrace}
    \frac{r_1 s_2}{x_1^2 (1-x_2)^2}, \frac{r_1 s_3}{x_1^2 (1-x_3)^2}, \frac{r_2 s_1}{x_2^2 (1-x_1)^2}, \frac{r_2 s_3}{x_2^2 (1-x_3)^2}, \frac{r_3 s_1}{x_3^2 (1-x_1)^2}, \frac{r_3 s_2}{x_3^2 (1-x_2)^2}
    \biggr\rbrace\,,
    \end{split}
    \label{eq:111111_NaturalBasis}
\end{equation}
i.e., $\Am_{111111}$ is a polynomial in the $R$-symmetry variables \eqref{eq:111111_CrossRatio_rst}.

\subsection{One integral to rule them all}
\label{subsec:OneIntegralToRuleThemAll}

A central part of this work is to demonstrate that higher-point correlators at next-to-next-to-leading order can be computed by imposing symmetry constraints, provided we know \textit{one} integral.
This section introduces this integral -- the six-point train track -- and discusses some important limits used throughout the work.

\subsubsection{The train track}
\label{subsubsec:TheTrainTrack}

The six-point train track integral is a conformal two-loop integral defined as
\begin{equation}
    B_{123,456}
    =
    \TrainTrack\
    =
    \int d^4 x_7\, I_{15} I_{25} I_{35} X_{4567}\,,
    \label{eq:TrainTrack_Definition}
\end{equation}
which is expected to be elliptic when the external points are four-dimensional \cite{Bourjaily:2017bsb,Bourjaily:2018ycu,Ananthanarayan:2020ncn,Loebbert:2020glj,Kristensson:2021ani,McLeod:2023qdf}.
The integral $X_{4567}$ is defined in \eqref{eq:X1234}.
This integral can be expressed in terms of polylogarithms in the collinear limit, i.e., when all external points are aligned.
This configuration was studied in \cite{Rodrigues:2024znq}, and the result can be expressed in terms of Goncharov polylogarithms as
\begin{equation}
    B_{123,456}
    =
    \frac{I_{15} I_{24} I_{36}}{128 \pi^{4}} b_{123,456} (x_1, x_2, x_3) \,,
    \label{eq:TrainTrack_Result}
\end{equation}
where
\begingroup
\allowdisplaybreaks
\begin{align}
    b_{123,456}
    &=
    \frac{x_{13}^2}{x_1 x_2 (1-x_3) x_{12}}
    \bigl(
    -G(1,x_1) G(1,x_2) G(1,x_3)+G(1,x_2) G(x_3,x_1) G(1,x_3) \notag \\
    &\phantom{=\ } -G(1,x_1) G(x_3,x_2)
    G(1,x_3)+G(x_3,x_1) G(x_3,x_2) G(1,x_3) \notag \\
    &\phantom{=\ }-2 G(1,0,x_1) G(1,x_3)+2 G(1,x_2,x_1)
     G(1,x_3)+2 G(x_3,0,x_1) G(1,x_3) \notag \\
    &\phantom{=\ }-2 G(x_3,x_2,x_1) G(1,x_3)+G(x_3,x_1)
    G(0,1,x_2)-G(1,x_1) G(0,x_3,x_2) \notag \\
    &\phantom{=\ }+G(x_3,x_1) G(1,0,x_2)+G(0,x_3) (-G(1,x_2)
    G(x_3,x_1)+G(1,x_1) (G(1,x_2) \notag \\
    &\phantom{=\ }+G(x_3,x_2))+2 G(1,0,x_1)-2 G(1,x_2,x_1))-G(x_3,x_2)
    G(1,x_2,x_1) \notag \\
    &\phantom{=\ }+G(x_3,x_2) G(1,x_3,x_1)+G(1,x_1) G(1,x_3,x_2)-G(x_3,x_1)
    G(1,x_3,x_2) \notag \\
    &\phantom{=\ }-G(1,x_1) G(x_3,0,x_2)-G(1,x_2) G(x_3,1,x_1)+G(0,x_2) (-2 G(0,x_3)
    G(1,x_1) \notag \\
    &\phantom{=\ }+2 G(1,x_3) G(1,x_1)-2 G(1,x_3)
    G(x_3,x_1)-G(1,x_3,x_1)+G(x_3,1,x_1)) \notag \\
    &\phantom{=\ }+G(1,x_1) G(x_3,1,x_2)-G(x_3,x_1)
    G(x_3,1,x_2)+G(1,x_2)
    G(x_3,x_2,x_1) \notag \\
    &\phantom{=\ }+G(1,0,x_3,x_1)-G(1,x_2,x_3,x_1)+G(1,x_3,0,x_1)-G(1,x_3,x_2,x_1) \notag \\
    &\phantom{=\ }-
    G(x_3,0,1,x_1)-G(x_3,1,0,x_1)+G(x_3,1,x_2,x_1)+G(x_3,x_2,1,x_1)
    \bigr)\,.
\end{align}
\endgroup
The variables $x_1, x_2, x_3$ correspond to the six-point spacetime cross-ratios defined in \eqref{eq:111111_CrossRatio_x}.
Note that, in the collinear limit, the result of the integral depends on the ordering of the external points, i.e., the subscripts on $B$ are not generally commutative.
The results for different orderings can be found in \cite{Rodrigues:2024znq}.
We provide a review of Goncharov polylogarithms in Appendix \ref{app:SymbolsAndGoncharovPolylogarithms}.

Interestingly, the well-known conformal kite integral \cite{Usyukina:1994iw,Drummond:2006rz} can be obtained as a special pinching limit of the train track:
\begin{align}
    K_{13,24}
    =
    \KiteIntegral
    =
    I_{13}^{-1} \lim\limits_{6 \to 1, 5 \to 3}
    B_{123,546}
    =
    \frac{I_{13} I_{24}}{512 \pi^{6}} \Phi^{(2)} (x)\,,
    \label{eq:KiteIntegral}
\end{align}
where
\begin{equation}
    \begin{split}
    \Phi^{(2)} (x) &=
    \frac{1}{x (1-x)}
    \bigl(
    G(0,0,1,x)+G(1,1,0,x)
    +G(0,1,0,x)+G(1,0,1,x) \\
    &\phantom{=\ }
    -2 (G(1,0,0,x) + G(0,1,1,x))
    \bigr)\,,
    \end{split}
\end{equation}
with $x$ being the four-point cross-ratio defined in \eqref{eq:4Pt_CrossRatio_x}.

\subsubsection{Non-conformal integrals from conformal integrals}
\label{subsubsec:NonConformalIntegralsFromConformalIntegrals}

The train track integral can also be used to access \textit{non-conformal} integrals.
For example, the $H$-integral, defined as
\begin{equation}
    H_{12,34}
    =
    \Hintegral\
    =
    \int d^4 x_5\, I_{15} I_{25}\, Y_{345}\,,
    \label{eq:H1234_Definition}
\end{equation}
is non-conformal and remains unsolved to the best of our knowledge.
It can be derived from the conformal train track integral as follows:
\begin{equation}
    H_{12,34}
    =
    \lim\limits_{\tau_5 \to \infty} I_{25}^{-1} I_{35}^{-1} B_{125,345}\,.
    \label{eq:H_From_B}
\end{equation}
This expression is valid even when the points are not aligned, suggesting that the $H$-integral is likely elliptic in the general case.
In the collinear limit, the $H$-integral can be expressed in Goncharov polylogarithms and depends on two variables (up to a prefactor): 
\begin{equation}
    H_{12,34} =
    \frac{h(u,v)}{8192 \pi^{10} \tau_{12} \tau_{34}}\,,
   \label{eq:H1234_1d}
\end{equation}
with
\begin{equation}
    \begin{split}
    h (u,v) &=
    G (\ub,v)
    \left( G (0,1,u)+G (1,0,u) \right)
    +
    2 (G (0,u) G (1,0,v) - G (1,u) G (1,0,v)) \\
    &\phantom{=\ }
    - G(1,v) \left(G (0,1,u)+G (1,0,u)-2 G (1,1,u) \right)
    +
    G (1,u) G (1,1-u,v) \\
    &\phantom{=\ }
    -
    2 G (0,u) G (1-u,0,v)-G (1,0,1-u,v) - G (1,1-u,0,v) \\
    &\phantom{=\ }
    +
    G (1-u,0,1,v)+G (1-u,1,0,v)\,,
   \end{split}
   \label{eq:H1234_1d_Result}
\end{equation}
where we defined the variables
\begin{equation}
    u = \frac{\tau_{12}}{\tau_{14}}\,, \qquad
    v = \frac{\tau_{34}}{\tau_{14}}\,.
    \label{eq:H_Variables}
\end{equation}
This integral will play an important role in the upcoming sections.

\section{Non-perturbative constraints}
\label{sec:NonPerturbativeConstraints}

In this section, we present the non-perturbative constraints that are instrumental for deriving the correlators in Sections \ref{sec:FourPointFunctions} and \ref{sec:HigherPointFunctions}.
We begin by discussing the constraints imposed by superconformal symmetry, which are then further refined by the requirement of crossing symmetry.
A key conclusion of this analysis is that all the correlators under consideration depend on one or two functions of the spacetime cross-ratios.
We then explore additional constraints on this function emerging from the pinching limits.

\subsection{Superconformal symmetry}
\label{subsec:SuperconformalSymmetry}

The constraints arising from superconformal symmetry can be expressed in the form of superconformal Ward identities (SCWI).
We study how these constraints can be applied to the correlators presented in Section \ref{subsec:CorrelationFunctions}.\footnote{Note that some of the content presented in this section overlaps with \cite{Barrat:2024ta}.}
To begin, we review the case of four-point functions, which was previously addressed in \cite{Liendo:2018ukf}.
Building on this approach, we then extend the analysis to higher-point functions and provide solutions to the SCWI for the correlators $\vev{11112}$ and $\vev{111111}$.

\subsubsection{Four-point functions}
\label{subsubsec:WI_FourPointFunctions}

Superconformal symmetry imposes stringent constraints on the four-point correlators \eqref{eq:4Pt_Correlator}.
These constraints are embodied in the differential constraint
\begin{equation}
    \left.
    \biggl(
    \frac{1}{2} \pd_x
    +
    \alpha \pd_r
    - (1-\alpha) \pd_s
    \biggr)
    \Am_{\Delta_1 \Delta_2 \Delta_3 \Delta_4} (x;r,s)
    \right|_{r \to \alpha x, s \to (1-\alpha) (1-x)}
    =
    0\,,
    \label{eq:4Pt_SCWI}
\end{equation}
for $\alpha \in \mathbb{•}{R}$, and which were originally derived and solved in \cite{Liendo:2016ymz} using superspace techniques.

A noteworthy consequence from \eqref{eq:4Pt_SCWI} is that all four-point functions exhibit a \textit{topological sector}, meaning that the dependence on kinematic variables disappears when the $R$-symmetry variables are aligned with the spacetime ones.
Specifically,
\begin{equation}
    \Am_{\Delta_1 \Delta_2 \Delta_3 \Delta_4} (x;x^2,(1-x)^2)
    =
    \Fds_{\Delta_1 \Delta_2 \Delta_3 \Delta_4}\,,
    \label{eq:4Pt_Topological}
\end{equation}
where $\Fds_{\Delta_1 \Delta_2 \Delta_3 \Delta_4}$ is a function of the coupling $\lambda$ alone.
These functions can be determined using localization techniques \cite{Giombi:2018qox}.
For our cases of interest, we have
\begin{align}
    \Fds_{1111} &= \frac{3 \Ids_2^2}{\lambda \Ids_1^2} (\lambda + 8 - 4 \Ids_1)\,, \label{eq:1111_Fds} \\
    \Fds_{1122} &= \frac{1}{{8 \Ids_1 \Ids_2^2 \Ids_3 \Ids_4 \sqrt{\lambda } (3 \lambda -(\Ids_1-2)
   (\Ids_1+10))}} \bigl(
   15 \Ids_1 \Ids_3 \Ids_2^3 \lambda ^{3/2} \notag \\
   &\phantom{=\ }
   +
   \Ids_4 \bigl(\Ids_3 \bigl(\Ids_1^3 \lambda ^{3/2}+3 (\Ids_1+10)
   \Ids_2 \Ids_1^2 \lambda +3 \Ids_2^2 \Ids_1 \sqrt{\lambda } (\Ids_1 (\Ids_1+20)+14 \lambda +256) \notag \\
   &\phantom{=\ }
   +\Ids_2^3 (9 (3
   \Ids_1-56) \lambda +(\Ids_1-2) (\Ids_1 (\Ids_1+32)+832))\bigr) \notag \\
   &\phantom{=\ }
   -
   6 \Ids_1 \Ids_2^2 \lambda  \left(\Ids_1 \sqrt{\lambda
   }+(\Ids_1+28) \Ids_2\right)\bigr)
   \bigr)\,.
   \label{eq:1122And1212_Fds} \\
   \Fds_{1212} &= \Fds_{1122}
\end{align}

As discussed in Section \ref{subsubsec:Correlators_FourPointFunctions}, the correlators $\vev{1111}$, $\vev{1122}$, and $\vev{1212}$ depend on three distinct $R$-symmetry channels.
Applying the superconformal Ward identities to the natural basis \eqref{eq:4Pt_NaturalBasis} leads to relations between these channels:
\begin{equation}
    \begin{split}
        F_1' (x) = - \frac{1}{1-x} F_3'(x)\,, \\
        F_2' (x) = - \frac{x}{1-x} F_3'(x)\,,
    \end{split}
    \label{eq:4Pt_NaiveSCWI}
\end{equation}
where $F'(x)$ denotes the derivative with respect to $x$.
This basis, however, is cumbersome as it requires integrating one channel to obtain another.
This issue becomes even more challenging for higher-point correlators, where the $R$-symmetry channels depend on multiple spacetime cross-ratios.
We expect however that there exists a basis in which two functions can be eliminated, reducing the solution of the Ward identity to a dependence on a single function, following the methods of \cite{Liendo:2016ymz}.
Below, we outline how to construct such a change of basis.

We first define a new $R$-symmetry basis
\begin{equation}
    \Am (x;r,s)
    =
    \sum_{i=1}^{3} \tilde{R}_i (x;r,s) G_i(x)\,,
    \label{eq:4Pt_NewBasis}
\end{equation}
where $\Am$ may refer to $\Am_{1111}$, $\Am_{1212}$ or $\Am_{1122}$.
To circumvent the issue mentioned below \eqref{eq:4Pt_NaiveSCWI}, we impose the following conditions:
\begin{enumerate}
    \item One of the functions, $G_1(x)$, corresponds to the topological sector:
    \begin{equation}
        G_1 (x) = \Fds\,,
        \label{eq:4Pt_TopologicalCondition}
    \end{equation}
    where $\Fds$ represents $\Fds_{1111}$, $\Fds_{1212}$ or $\Fds_{1122}$.
    \item The derivative of $G_2(x)$ should not appear in the Ward identities, which leads to the condition
    \begin{equation}
        \left. \tilde{R}_2 \right|_{r \to \alpha x, s \to (1-\alpha) (1-x)}
        =
        0\,.
        \label{eq:4Pt_DerivativeCondition}
    \end{equation}
    \item We demand that $G_3 (x)$ does not appear when applying the Ward identity.
    This can be obtained by demanding that the corresponding basis element $\tilde{R}_3(x)$ is an invariant of the Ward identity:
    \begin{equation}
        \left.
        \biggl(
        \frac{1}{2} \pd_x
        +
        \alpha \pd_r
        - (1-\alpha) \pd_s
        \biggr)
        \tilde{R}_3
        \right|_{r \to \alpha x, s \to (1-\alpha) (1-x)}
        =
        0\,.
        \label{eq:4Pt_SCWICondition}
    \end{equation}
\end{enumerate}
The remaining coefficients are arbitrary as long as the basis elements are linearly independent.
We normalize the solution by imposing the following additional conditions:
\begin{enumerate}
    \item[4.] At weak coupling and large $N$, in the natural basis \eqref{eq:4Pt_NaturalBasis}, $F_1(x)$ is simpler than $F_2(x)$ and $F_3(x)$, so we relate $G_2(x)$ to $F_1(x)$ directly:
    \begin{equation}
        G_2 (x) \sim F_1 (x)\,,
        \label{eq:4Pt_PlanarityCondition}
    \end{equation}
    allowing a proportionality function of $x$.
    \item[5.] Anticipating the results of Section \ref{subsec:CrossingSymmetry}, we demand that $G_2(x)$ is anti-self-crossing for $\vev{1111}$ and $\vev{1212}$:
    \begin{equation}
        G_2 (x) + G_2 (1-x)
        =
        0\,,
        \label{eq:4Pt_CrossingCondition}
    \end{equation}
    which implies that $G_2(x)$ vanishes at leading order:
    \begin{equation}
        G_2 (x)
        =
        0 + \Op(\lambda)\,.
        \label{eq:4Pt_LeadingOrderCondition}
    \end{equation}
    Note that the condition \eqref{eq:4Pt_CrossingCondition} does not hold for $\vev{1122}$.
    This is due to the fact that this correlator does not cross to itself (except in a trivial way), as commented in Section \ref{subsubsec:Crossing_FourPointFunctions}.
\end{enumerate}

Based on these conditions, we can define an appropriate change of basis between the natural basis \eqref{eq:4Pt_NaturalBasis} and $\tilde{R}_j$ to apply the Ward identities.
For instance, the following dictionary provides a convenient change of basis:
\begin{equation}
    \begin{split}
        G_1 (x) &= F_1 (x) + F_2 (x) + F_3 (x) = \Fds\,, \\
        G_2 (x) &= F_1 (x)\,, \\
        G_3 (x) &= \frac{1}{2} ((2x-1) F_1 (x)+F_2 (x)-F_3 (x))\,.
    \end{split}
    \label{eq:4Pt_ChangeOfBasis}
\end{equation}
Finally, relabeling $G_3(x)$ as $f(x)$, the solution can be written succinctly as
\begin{equation}
    \Am (x;r,s)
    =
    \frac{1}{2} \biggl( \frac{r}{x^2} + \frac{s}{(1-x)^2} \biggr) \Fds
    +
    \pd_x ( \xi f(x) )\,,
    \label{eq:4Pt_SolutionSCWI}
\end{equation}
where the auxiliary function $\xi$ is defined as
\begin{equation}
    \xi
    =
    1 - \frac{r}{x} - \frac{s}{1-x}\,.
    \label{eq:4Pt_HelpFunction}
\end{equation}
The function $f(x)$ is closely related to the solution to the SCWI in \cite{Liendo:2018ukf}, which is obtained using a different change of $R-$symmetry basis. The choice here was made by requesting \eqref{eq:4Pt_AntiSelfCrossing} for $\vev{1111}$ and $\vev{1212}$, which disallows the presence of a constant term in $f(x)$.\footnote{The results in Section \ref{sec:FourPointFunctions} seem to include a constant, but this is an artefact of the chosen fibration basis.
A series expansions around $x \sim 0$ or $x \sim 1$ shows no constant.}
The following relation is useful to keep in mind:
\begin{equation}
    f'(x)
    =
    F_1(x)\,,
    \label{eq:fFromF1}
\end{equation}
which follows directly from the choice of basis above.
In Section \ref{sec:FourPointFunctions}, we provide the function $f(x)$ for the correlators $\vev{1111}$, $\vev{1122}$, and $\vev{1212}$ up to next-to-next-to-leading order.
In the following sections, we extend this method to solve the SCWI for higher-point functions, such as $\vev{11112}$ and $\vev{111111}$.

\subsubsection{Five-point functions}
\label{subsubsec:WI_FivePointFunctions}

The superconformal Ward identities discussed for four-point functions in \eqref{eq:4Pt_SCWI} have a natural extension to multipoint correlators.
It was conjectured in \cite{Barrat:2021tpn} that for five-point functions, the SCWI take the following form:
\begin{equation}
    \sum_{i = 1,2}
    \biggl(
    \frac{1}{2} \partial_{x_i}
    +
    \alpha_i \partial_{r_i}
    -
    (1-\alpha)_i \partial_{s_i}
    \biggr)
    \Am_{\Delta_1 \ldots \Delta_5} \bigr|_{r_i \to \alpha_i x_i, s_i \to (1-\alpha)_i (1-x_i), t_{ij} \to \alpha_{ij} x_{ij}} = 0\,,
    \label{eq:5Pt_SCWI_Conjecture}
\end{equation}
where $\alpha_i \in \mathbb{R}$.
Later work \cite{Bliard:2024und,Barrat:2024ta} found that these constraints are part of a more general set of Ward identities:
\begin{equation}
    \sum_{i \neq j}^2 \beta_i
    \biggl(
    \frac{1}{2} \partial_{x_i}
    +
    \alpha_i \partial_{r_i}
    -
    (1-\alpha)_i \partial_{s_i}
    +
    \alpha_{ij} \partial_{t_{12}}
    \biggr)
    \Am_{\Delta_1 \ldots \Delta_5} \bigr|_{r_i \to \alpha_i x_i, s_i \to (1-\alpha)_i (1-x_i), t_{ij} \to \alpha_{ij} x_{ij}} = 0\,,
    \label{eq:5Pt_SCWI}
\end{equation}
with $\alpha_i, \beta_i \in \mathbb{R}$.
This equation encodes the full constraints imposed by superconformal symmetry.

Similarly to the case of four-point functions, the SCWI suggest the existence of a topological subsector for five-point functions.
For instance, in the case of $\vev{11112}$, an exact evaluation of the correlator yields
\begin{equation}
    \Fds_{11112}
    =
    \frac{6 \Ids_2^2}{\lambda \Ids_1^2} \frac{2(\Ids_1 - 2)(\Ids_1 + 28) + \lambda (2 \Ids_1 - 19)}{\sqrt{3 \lambda - (\Ids_1 - 2)(\Ids_1 + 10)}}\,.
    \label{eq:11112_Fds}
\end{equation}

The differential constraints in \eqref{eq:5Pt_SCWI} can be solved similarly to those for four-point functions.
Below, we outline the steps for transforming the natural basis of $\vev{11112}$ into a form analogous to \eqref{eq:4Pt_SolutionSCWI}.

We introduce a new basis for the correlator:
\begin{equation}
    \Am_{11112}
    =
    \sum_{j=1}^6 \tilde{R}_j G_j (x_1, x_2)\,.
    \label{eq:11112_NewBasis}
\end{equation}
In analogy to the four-point case, we impose the following conditions on the functions $G_j$:
\begin{enumerate}
    \item The topological limit is given by
    \begin{equation}
        G_1 (x_1, x_2)
        =
        \Fds_{11112}\,,
        \label{eq:11112_TopologicalCondition}
    \end{equation}
    meaning that all other basis functions vanish when $r_i \to x_i^2, s_i \to (1-x_i)^2, t_{ij} \to x_{ij}^2$.
    \item The derivatives of $G_{2,3,4} (x_1, x_2)$ do not appear after applying the Ward identities,\footnote{The number of functions that can be eliminated in this way is not arbitrary.
    Starting with the Ansatz
    \begin{equation}
        \tilde{R}_{i=2, \ldots, i_\text{max}+1}
        =
        a_1^{(i)} + a_2^{(i)} \frac{r_1}{x_1} + a_3^{(i)} \frac{s_1}{1-x_1} + a_4^{(i)} \frac{r_2}{x_2} + a_5^{(i)} \frac{s_2}{1-x_2} + a_6^{(i)} \frac{t_{12}}{x_{12}}\,,
        \label{eq:11112_DerivativeCondition}
    \end{equation}
    we find that $i_\text{max}=3$ to ensure linear independence.} leading to the condition
    \begin{equation}
        \left. \tilde{R}_{2,3,4} \right|_{r_i \to \alpha_i x_i, s_i \to (1-\alpha)_i (1-x_i), t_{ij} \to \alpha_{ij} x_{ij}}
        =
        0\,.
        \label{eq:11112_DerivativeConditionExplained}
    \end{equation}
    \item The remaining basis elements, $\tilde{R}_{j=5,6}$, are chosen to satisfy the Ward identities \eqref{eq:5Pt_SCWI}.
    \item We demand that the functions $G_{j=2,3,4} (x_1, x_2)$ are directly related to the simplest channels at weak coupling, i.e.,
    \begin{equation}
        \begin{split}
        G_2 (x_1, x_2) &\sim F_1 (x_1, x_2)\,, \\
        G_3 (x_1, x_2) &\sim F_3 (x_1, x_2)\,, \\
        G_4 (x_1, x_2) &\sim F_4 (x_1, x_2)\,.
        \end{split}
        \label{eq:11112_PlanarityCondition}
    \end{equation}
    \item Anticipating Section \ref{subsec:CrossingSymmetry}, we select the functions $G_{j=5,6}$ such that they are related by crossing symmetry, i.e.,
    \begin{equation}
        G_5 (x)
        =
        G_6 (1-x)\,.
        \label{eq:11112_CrossingCondition}
    \end{equation}
\end{enumerate}

The transformation between the natural and new basis is not unique, but one possible choice that satisfies these conditions is
\begin{equation}
    \begin{split}
    G_1 (x_1, x_2) &= \sum_{i=1}^6 F_i (x_1, x_2) = \Fds_{11112}\,, \\
    G_2 (x_1, x_2) &= F_1 (x_1, x_2)\,, \\
    G_3 (x_1, x_2) &= \frac{F_3 (x_1, x_2)}{1-x_1}\,, \\
    G_4 (x_1, x_2) &= \frac{F_4 (x_1, x_2)}{x_2}\,, \\
    G_5 (x_1, x_2) &= (1-x_2) F_1 (x_1,x_2) + \frac{(1-x_2) F_3 (x_1,x_2)}{1-x_1} + F_5 (x_1,x_2)\,, \\
    G_6 (x_1, x_2) &= x_1 F_1 (x_1,x_2)+F_2 (x_1,x_2)+\frac{x_1 F_4 (x_1,x_2)}{x_2}\,.
    \end{split}
    \label{eq:11112_ChangeOfBasis}
\end{equation}
After applying the Ward identities and eliminating $G_{1,\ldots,4}$, the five-point function solution takes the elegant form
\begin{equation}
    \begin{split}
        \Am_{11112}
        &=
        \frac{t_{12}}{x_{12}^2} \Fds_{11112}
        + \rho f_1
        + \pd_{x_1} (\xi_1 f_1)
        + \pd_{x_2} (\xi_2 f_1)
        + \pd_{x_2} (\eta f_2)\,,
    \end{split}
    \label{eq:11112_SolutionSCWI}
\end{equation}
where we have relabeled $G_5 \to f_1$ and $G_6 \to f_2$.
The auxiliary functions are given by
\begin{align}
    \xi_1 &= \frac{1-x_1}{1-x_2} \biggl( 1 - \frac{r_1}{x_1} - \frac{s_1}{1-x_1} \biggr)\,, \label{eq:11112_HelpFunction1} \\
    \xi_2 &= 1 - \frac{r_1}{x_1} - \frac{s_2}{1-x_2} + \frac{t_{12}}{x_{12}}\,, \label{eq:11112_HelpFunction2} \\
    \eta &= \frac{x_2}{x_1} \biggl( \frac{r_1}{x_1} - \frac{r_2}{x_2} - \frac{t_{12}}{x_{12}} \biggr)\,, \label{eq:11112_HelpFunction3} \\
    \rho&= - \frac{1}{1-x_2}\biggl( 1 - \frac{r_1}{x_1^2} \biggr)\,. \label{eq:11112_HelpFunction4}
\end{align}
It is however important to note that the solution \eqref{eq:11112_SolutionSCWI} does not satisfy on its own the Ward identities.
The reason is that there exists one additional constraint on the derivatives of $f_1$ and $f_2$, which can be expressed as
\begin{equation}
    \biggl(
    \pd_{x_1} + \frac{x_2}{x_1} \pd_{x_2}
    \biggr) f_2
    +
    \biggl(
    \frac{1-x_1}{1-x_2} \pd_{x_1} + \pd_{x_2}
    \biggr) f_1
    =
    0\,.
    \label{eq:11112_ExtraConstraint}
\end{equation}
This constraint is not crucial for our purposes, as in the next section, we show that crossing symmetry allows us to eliminate $f_2$ altogether.
However, it should be taken into account when deriving superconformal blocks in the gist of \cite{Liendo:2016ymz}.

\subsubsection{Six-point functions}
\label{subsubsec:WI_SixPointFunctions}

The superconformal Ward identities for six-point functions can be written as
\begin{equation}
    \sum_{i \neq j}^3 \beta_i
    \biggl(
    \frac{1}{2} \partial_{x_i}
    +
    \alpha_i \partial_{r_i}
    -
    (1-\alpha)_i \partial_{s_i}
    +
    \alpha_{ij} \partial_{t_{ij}}
    \biggr)
    \Am_{\Delta_1 \ldots \Delta_6} \bigr|_{r_i \to \alpha_i x_i, s_i \to (1-\alpha)_i (1-x_i), t_{ij} \to \alpha_{ij} x_{ij}} = 0\,.
    \label{eq:6Pt_SCWI}
\end{equation}
As with lower-point functions, \eqref{eq:6Pt_SCWI} suggests the existence of a topological sector.
The topological limit of the six-point function $\vev{111111}$ is given by
\begin{equation}
    \Fds_{111111}
    =
    \frac{15 \Ids_2^3}{\lambda^{3/2} \Ids_1^3} ((\lambda + 24) \Ids_1 - 8 ( \lambda + 6))\,.
    \label{eq:111111_Fds}
\end{equation}
The Ward identities can be solved for $\vev{111111}$ in the same way as for the other correlators.
However the solution is quite intricate and too lengthy to be displayed explicitly here.
We provide the complete expression in an ancillary \textsc{Mathematica} notebook for those interested in the explicit expressions.

Importantly, the Ward identities allow us to eliminate \textit{ten} functions from the general solution.
Additionally, the remaining four functions are subject to non-trivial differential constraints, similar to those seen for the five-point correlator.
While solving these constraints in closed form is not critical for our current focus, it would be interesting to explore whether they admit direct solutions.

\subsection{Crossing symmetry}
\label{subsec:CrossingSymmetry}

We now turn our attention to the constraints imposed by crossing symmetry on the correlators of interest.

\subsubsection{Four-point functions}
\label{subsubsec:Crossing_FourPointFunctions}

For four-point functions, crossing symmetry imposes the relation
\begin{equation}
    \vev{\Op_{\Delta_1} (1) \Op_{\Delta_2} (2) \Op_{\Delta_3} (3) \Op_{\Delta_4} (4)}
    =
    \vev{\Op_{\Delta_1} (1) \Op_{\Delta_4} (4) \Op_{\Delta_3} (3) \Op_{\Delta_2} (2)}\,,
    \label{eq:4Pt_CrossingSymmetry}
\end{equation}
with $(i) = (u_i, \tau_i)$.
The implications of crossing symmetry for specific four-point correlators, such as $\vev{1111}$, $\vev{1122}$, and $\vev{1212}$, were investigated in \cite{Liendo:2018ukf}.
We review here those results.
For the correlators $\vev{1111}$ and $\vev{1212}$, the crossing relation \eqref{eq:4Pt_CrossingSymmetry} leads to the following conditions:
\begin{align}
    F_1 (x) &= F_1 (1-x)\,, \label{eq:4Pt_F1_Crossing} \\
    F_2 (x) &= F_3 (1-x)\,. \label{eq:4Pt_F2F3_Crossing}
\end{align}
These relations imply that one channel can be eliminated by crossing symmetry, while another channel is found to be self-crossing.
As mentioned in Section \ref{subsubsec:WI_FourPointFunctions}, this results in the function $f(x)$ of \eqref{eq:4Pt_SolutionSCWI} satisfying the following anti-self-crossing condition:
\begin{equation}
    f(x)
    =
    - f(1-x)\,.
    \label{eq:4Pt_AntiSelfCrossing}
\end{equation}
It is crucial to point out that the correlator $\vev{1122}$ is related to $\vev{1221}$ via crossing symmetry.
In this case, the channels themselves are not constrained, and the associated function $f(x)$ does \textit{not} obey the anti-self-crossing relation \eqref{eq:4Pt_AntiSelfCrossing}.

\subsubsection{Five-point functions}
\label{subsubsec:Crossing_FivePointFunctions}

The crossing symmetry relation \eqref{eq:4Pt_CrossingSymmetry} can be naturally extended to higher-point functions.
For the five-point correlator considered in \eqref{eq:11112_Correlator}, crossing symmetry takes the form
\begin{equation}
    \vev{\Op_{1} (1) \Op_{1} (2) \Op_{1} (3) \Op_{1} (4) \Op_{2} (5)}
    =
    \vev{\Op_{1} (4) \Op_{1} (3) \Op_{1} (2) \Op_{1} (1) \Op_{2} (5)}\,.
    \label{eq:5Pt_CrossingSymmetry}
\end{equation}
This crossing relation imposes constraints on the $R$-channels introduced in \eqref{eq:11112_NaturalBasis}.
Specifically, the following relations hold:
\begin{align}
    F_1 (x_1, x_2) = F_1 (1-x_2, 1-x_1)\,, \label{eq:11112_F1_Crossing} \\
    F_2 (x_1, x_2) = F_5 (1-x_2, 1-x_1)\,, \label{eq:11112_F2F5_Crossing} \\
    F_3 (x_1, x_2) = F_4 (1-x_2, 1-x_1)\,, \label{eq:11112_F3F4_Crossing} \\
    F_6 (x_1, x_2) = F_6 (1-x_2, 1-x_1)\,. \label{eq:11112_F6_Crossing}
\end{align}
The choice of basis made in Section \ref{subsubsec:WI_FivePointFunctions} for solving the Ward identities results in the simple crossing relation
\begin{equation}
    f_1 (x_1, x_2)
    =
    f_2 (1-x_2, 1-x_1)\,,
    \label{eq:11112_f1f2_Crossing}
\end{equation}
which implies that the correlator ultimately depends on a \textit{single function}, denoted in subsequent sections by $f(x_1,x_2) := f_1(x_1, x_2)$.
Determining this function at weak coupling is the subject of Section \ref{subsec:11112}.

\subsubsection{Six-point functions}
\label{subsubsec:Crossing_SixPointFunctions}

Next, we explore the crossing symmetry constraints for six-point functions, focusing on the correlator $\vev{111111}$.
Since all the operators are identical, the reduced correlators exhibit numerous relations.
These relations are summarized in Table \ref{tab:SixPoint_Crossing}, where redundant relations are ommitted.

The choice of basis for solving the Ward identities can be made such that, after imposing crossing symmetry, the correlator depends on two functions $f_1$ and $f_2$.
As we will see in Section \ref{subsec:111111}, our choice of basis results in $f_2=0$ up to next-to-next-to-leading order.
We thus focus on determining $f_1$ in subsequent sections.
The explicit relations used to relate $G_j$ to $f_{1,2}$ can be found in the ancillary notebook.

\begin{table}[]
    \centering
    \begin{tabular}{|c|ccccc|}
        \hline
        $(x_1, x_2, x_3)$ & $F_1$ & $F_2$ & $F_3$ & $F_4$ & $F_{14}$ \\[.25em]
        $(- \frac{x_{12}}{x_2}, - \frac{x_{13}}{x_3}, 1-x_1)$ & $F_7$ & $F_9$ & $F_8$ & $F_{12}$ & $F_{14}$ \\[.25em]
        $(\frac{x_1 x_{23}}{x_2 x_{13}}, \frac{x_1 (1-x_2)}{x_2 (1-x_1)}, \frac{x_1}{x_2})$ & $F_{10}$ & $F_2$ & $F_{11}$ & $F_{15}$ & $F_{14}$ \\[.25em]
        $(\frac{x_{12} (1-x_3)}{x_{13} (1-x_2)}, \frac{x_{12}}{x_{13}}, \frac{x_{12} x_3}{x_{13} x_2})$ & $F_6$ & $F_9$ & $F_3$ & $F_4$ & $F_{14}$ \\[.25em]
        $(- \frac{x_{23}}{1-x_2}, - \frac{x_{23}}{x_3 (1-x_2)}, \frac{x_{23} (1-x_1)}{x_{13} (1-x_2)})$ & $F_5$ & $F_2$ & $F_8$ & $F_{12}$ & $F_{14}$ \\
        $(1-x_3, \frac{1-x_3}{1-x_1}, \frac{1-x_3}{1-x_2})$ & $F_{13}$ & $F_9$ & $F_{11}$ & $F_{15}$ & $F_{14}$ \\[.25em]
        \hline
    \end{tabular}
    \caption{Summary of the relations between the $R$-channels in the natural basis \eqref{eq:111111_NaturalBasis}, induced by the crossing symmetry of $\vev{111111}$.
    The left column represents the arguments of the functions $F_j$.
    The equalities between the $R$-channels should be understood column-wise.}
    \label{tab:SixPoint_Crossing}
\end{table}

\subsection{Pinching}
\label{subsec:Pinching}

In this section, we describe the constraints that arise from lower-point functions, commonly referred to as \textit{pinching}.
This phenomenon occurs when higher-weight operators are formed by bringing together fields from distinct points.
Specifically, the operators defined via \eqref{eq:HalfBPSOperator} possess the interesting property that higher-weight operators can be constructed by pinching together operators of lower length.
For example,
\begin{equation}
    \Wl[\underbrace{\phi^{I_{n-1}} (\tau_{n-1}) \phi^{I_n} (\tau_n)}_{\text{two operators of length $1$}}]
    \overset{\tau_n \to \tau_{n-1}}{\longrightarrow}
    \Wl[\underbrace{\phi^{I_{n-1}} (\tau_{n-1}) \phi^{I_n} (\tau_{n-1})}_{\text{one operator of length $2$}}]\,.
    \label{eq:Pinching}
\end{equation}
This property is important because lower-point correlators can impose constraints on higher-point functions.
Note that in $\Nm=4$ SYM, the pinching of half-BPS operators does not close within the same class of operators; instead, it generates higher-trace operators.
On a broader level, one can interpret pinching as indicating that the correlators of scalar insertions are fully determined by the chain of fundamental insertions
\begin{equation}
    \vev{\phi^{I_1} (\tau_1) \ldots \phi^{I_n} (\tau_n)}\,.
    \label{eq:TheWorldDependsOnElementaryInsertions}
\end{equation}
Below, we enumerate the specific constraints that pinching imposes on the correlators of interest in this study.

\subsubsection{Four-point functions}
\label{subsubsec:Pinching_FourPointFunctions}

We now discuss the constraints on four-point functions arising from pinching, where two operators are combined to yield a three-point function.

\paragraph{$\vev{1111}$.}
For the correlator of elementary operators $\Op_1$, pinching the last two operators to obtain the three-point function $\vev{112}$ gives
\begin{equation}
    \lim_{4 \to 3} \vev{1111}
    =
    (13)(23) (F_1 (0) + F_3 (0))
    =
    \frac{\sqrt{n_2}}{n_1} \vev{112}\,,
    \label{eq:1111_PinchingTo112}
\end{equation}
with $n_1$ and $n_2$ the normalization constants given in \eqref{eq:n1} and \eqref{eq:n2}, respectively.
These normalization constants are required because the pinching operation \eqref{eq:Pinching} is applied at the level of the fields rather than the unit-normalized operators.

\paragraph{$\vev{1122}$.}
The correlator $\vev{1122}$ can be pinched in two distinct ways.
First, it can be pinched such that it collapses to $\vev{222}$:
\begin{equation}
    \lim_{2 \to 1} \vev{1122}
    =
    (13)(24)(34) (F_1 (0) + F_3 (0))
    =
    \frac{\sqrt{n_2}}{n_1} \vev{222}\,.
    \label{eq:1122_PinchingTo222}
\end{equation}
Second, one can bring the middle operators together in order to obtain
\begin{equation}
    \lim_{3 \to 2} \vev{1122}
    =
    (12)(24)^2 (F_1 (1) + F_2 (1))
    =
    \frac{\sqrt{n_3}}{\sqrt{n_1 n_2}} \vev{132}\,.
    \label{eq:1122_PinchingTo132}
\end{equation}

\paragraph{$\vev{1212}$.}
One can proceed similarly for $\vev{1212}$.
It can be collapsed to $\vev{123}$ or $\vev{132}$, which result into two constraints:
\begin{align}
    \lim_{4 \to 3} \vev{1212}
    &=
    (13)(23)^2 (F_1 (0) + F_3 (0))
    =
    \frac{\sqrt{n_3}}{\sqrt{n_1 n_2}} \vev{123}\,, \label{eq:1212_PinchingTo123} \\
    \lim_{3 \to 2} \vev{1212}
    &=
    (12)(24)^2 (F_1 (1) + F_2 (1))
    =
    \frac{\sqrt{n_3}}{\sqrt{n_1 n_2}} \vev{132}\,.
    \label{eq:1212_PinchingTo132}
\end{align}

\subsubsection{Five-point functions}
\label{subsubsec:Pinching_FivePointFunctions}

Five-point functions can be pinched to produce lower-point functions, such as four-point functions.
For the correlator $\vev{11112}$, two particularly useful pinching limits are
\begin{align}
    \lim\limits_{4 \to 3} \vev{11112} &= (13)(25)(35) \biggl( F_1(x,1) + F_4 (x,1) + \frac{r}{x} F_2(x,1) + \frac{s}{(1-x)} (F_3 (x,1) + F_6 (x,1)) \biggr) \notag \\
    &= \frac{\sqrt{n_2}}{n_1} \vev{1122}\,, \label{eq:11112_PinchingTo1122} \\
    \lim\limits_{3 \to 2} \vev{11112} &= (14)(25)^2 \biggl( F_1(x,x) + \frac{r}{x} (F_2(x,x) + F_4(x,x)) + \frac{s}{(1-x)} (F_3 (x,x) + F_5 (x,x)) \biggr) \notag \\
    &= \frac{\sqrt{n_2}}{n_1} \vev{1212}\,. \label{eq:11112_PinchingTo1212}
\end{align}
Pinching to $\vev{1113}$ is less informative because $\vev{1113}$ is an extremal correlator, yielding only a number, as discussed in \cite{Barrat:2021tpn}.
For our purposes, the relations \eqref{eq:11112_PinchingTo1122} and \eqref{eq:11112_PinchingTo1212} are sufficient to determine the correlators of interest up to next-to-next-to-leading order.

\subsubsection{Six-point functions}
\label{subsubsec:SixPointFunctions}

Six-point functions can be pinched to produce lower-point functions, such as five-point functions.
For the correlator $\vev{111111}$, a particularly useful pinching limit is
\begin{equation}
	\begin{split}
	\lim\limits_{6 \to 5} \vev{111111}
	&=
    (14)(25)(35)
    \lim_{\veps\to0} \biggl\lbrace
     \left( F_{14} (x_1\veps,x_2\veps,\veps) + F_{15}(x_1\veps,x_2\veps,\veps)\right)  \\ 
     & \phantom{(14)(25)(35) \lim_{\veps\to0} \lbrace} + \frac{r_1}{x_1}\left( F_{10} (x_1\veps,x_2\veps,\veps) + F_{11}(x_1\veps,x_2\veps,\veps)\right) \\ 
     & \phantom{(14)(25)(35) \lim_{\veps\to0} \lbrace} + \frac{s_1}{(1-x_1)} \left( F_{4} (x_1\veps,x_2\veps,\veps) + F_{6}(x_1\veps,x_2\veps,\veps)\right) \\
     & \phantom{(14)(25)(35) \lim_{\veps\to0} \lbrace} + \frac{r_2}{x_2}\left( F_{12} (x_1\veps,x_2\veps,\veps) + F_{13}(x_1\veps,x_2\veps,\veps)\right) \\ 
     & \phantom{(14)(25)(35) \lim_{\veps\to0} \lbrace} + \frac{s_2}{(1-x_2)}\left( F_{1} (x_1\veps,x_2\veps,\veps) + F_{3}(x_1\veps,x_2\veps,\veps)\right) \\
     & \phantom{(14)(25)(35) \lim_{\veps\to0} \lbrace} \left. + \frac{t_{12}}{x_{12}}\left( F_{7} (x_1\veps,x_2\veps,\veps) + F_{9}(x_1\veps,x_2\veps,\veps)\right) \right\rbrace \\
     &=
     \frac{\sqrt{n_2}}{n_1}\vev{11112} \,.
	\end{split}
	\label{eq:111111_PinchingTo11112}
\end{equation}
The limit $\veps\to 0$ means to take the limit of all three six-point cross-ratios to zero while keeping their ratios fixed to the two five-point cross-ratios.

\section{Four-point functions}
\label{sec:FourPointFunctions}

We now compute the four-point functions introduced in Section \ref{subsubsec:Correlators_FourPointFunctions}, employing the solution of the superconformal Ward identities \eqref{eq:4Pt_SolutionSCWI}, the topological sector \eqref{eq:1111_Fds}-\eqref{eq:1122And1212_Fds}, and the pinching relations \eqref{eq:1111_PinchingTo112}-\eqref{eq:1212_PinchingTo132}.
For all correlators, we focus on the channel $F_1$, from which the function $f(x)$ can be deduced.
Both functions will be expanded perturbatively as follows:
\begin{align}
    F_1 (x)
    &=
    \sum_{\ell=0}^\infty \lambda^\ell F_1^{(\ell)} (x)\,, \label{eq:F1_PerturbativeExpansion} \\
    f (x)
    &=
    \sum_{\ell=0}^\infty \lambda^\ell f^{(\ell)} (x)\,.
    \label{eq:f_PerturbativeExpansion}
\end{align}
In this section, we compute $F_1$ using Feynman diagrams.
These results will serve as the foundation for calculating higher-point functions in Section  \ref{sec:HigherPointFunctions}.

\subsection{$\vev{1111}$}
\label{subsec:1111}

We begin by revisiting the simplest four-point function, $\vev{1111}$.
This correlator was computed up to next-to-leading order in \cite{Kiryu:2018phb,Barrat:2021tpn} and extended to next-to-next-to-leading order in \cite{Cavaglia:2022qpg}.
In the latter case, a combination of the conformal bootstrap and integrability techniques was employed to derive the result.

In the following, we demonstrate how to achieve the same result using Feynman diagrams.
Interestingly, we find that only the two lowest orders in $x$ of $F_1$ are required to determine the full correlator, if using a transcendentality-based Ansatz.
Consequently, numerical integration alone would have sufficed to obtain the analytical result.
We show however that all diagrams can, in fact, be computed analytically in this case.

\subsubsection{Leading and next-to-leading orders}
\label{subsubsec:1111_LeadingAndNextToLeadingOrders}

The low orders in the coupling $\lambda$ are straightforward to compute by focusing on the channel $F_1$ and the relations \eqref{eq:fFromF1} and \eqref{eq:4Pt_AntiSelfCrossing}.

\paragraph{Leading order.}
At leading order, no Feynman diagrams contribute to the channel $F_1$, and thus we have
\begin{equation}
    F_1^{(0)} (x)
    =
    0\,.
    \label{eq:1111_F1_LO}
\end{equation}
As explained in Section \ref{subsubsec:Crossing_FourPointFunctions}, the function $f(x)$ is chosen to be antisymmetric.
Therefore, at leading order, we readily obtain
\begin{equation}
    f^{(0)} (x)
    =
    0\,.
    \label{eq:1111_f_LO}
\end{equation}

\paragraph{Next-to-leading order.}
At next-to-leading order, the channel $F_1$ is determined by a single diagram, which we refer to as the X-diagram.
After removing the unit-normalization and the $R$-symmetry variables, this diagram evaluates to
\begin{equation}
    \NLOX
    =
    \frac{\lambda^3}{4} X_{1234}\,,
    \label{eq:1111_NLO_Diagram}
\end{equation}
where $X_{1234}$ is defined in \eqref{eq:X1234}.
The prefactor of $1/4$ comes from the symmetry factor and the trace of the diagram.

It is interesting to note that the channels $F_2$ and $F_3$ are more intricate to compute.
These channels involve boundary diagrams and contain functions of transcendentality weight $2$, while the channel $F_1$ has transcendentality weight $1$ only.
Therefore, it is a significant simplification to use the Ward identities from \eqref{eq:4Pt_SCWI} to derive the entire correlator from $F_1$ alone.

By expressing the integral in terms of Goncharov polylogarithms, the X-diagram results in the following for the channel $F_1$:
\begin{equation}
    F_1^{(1)} (x)
    =
    - \frac{1}{8\pi^2 x (1-x)} (x G(0,x) + (1-x) G(1, x))\,.
    \label{eq:1111_F1_NLO}
\end{equation}
From this, we can extract the corresponding function $f(x)$:
\begin{equation}
    f^{(1)} (x)
    =
    \frac{1}{8 \pi^2} \biggl( \frac{\pi^2}{6} + G(1,0,x) - G(0,1,x) \biggr)\,.
    \label{eq:1111_f_NLO}
\end{equation}

\subsubsection{Next-to-next-to-leading order}
\label{subsubsec:1111_NextToNextToLeadingOrder}

At next-to-next-to-leading order, the calculation becomes more intricate, involving both bulk and boundary diagrams. Table \ref{table:Diagrams1111NNLO} presents the relevant diagrams for calculating the channel $F_1$.
The complexity of this step is evident, but it would be even more challenging if we were required to compute the channels $F_2$ and $F_3$ directly, as those would introduce diagrams with multiple integrals along the Wilson line.

\begin{table}[t!]
    \centering
    \caption{The relevant Feynman diagrams for the computation of $F_1(x)$ at next-to-next-to-leading order.
    The horizontal double line separates {\normalfont bulk} from {\normalfont boundary} diagrams.
    In the last row, the colored dots along the Wilson line indicate where the gluon can connect.
    Explicit expressions can be found in Appendix \ref{app:FeynmanDiagramsOf1111}.}
    \begin{tabular}{lc}
        \hline
        Self-energy & \DefectSSSSTwoLoopsSelfEnergyOne\ \DefectSSSSTwoLoopsSelfEnergyTwo\ \DefectSSSSTwoLoopsSelfEnergyThree\ \DefectSSSSTwoLoopsSelfEnergyFour\ \\[2ex]
        \hline
        XX & \DefectSSSSTwoLoopsXXOne\ \DefectSSSSTwoLoopsXXTwo\ \\[2ex]
        \hline
        XH & \DefectSSSSTwoLoopsXHOne\ \DefectSSSSTwoLoopsXHTwo\ \DefectSSSSTwoLoopsXHThree\ \DefectSSSSTwoLoopsXHFour\ \\[2ex]
        \hline
        Spider & \DefectSSSSTwoLoopsSpider\ \\[2ex]
        \hline \hline
        XY & \DefectSSSSTwoLoopsXYOne\ \DefectSSSSTwoLoopsXYTwo\ \DefectSSSSTwoLoopsXYThree\ \DefectSSSSTwoLoopsXYFour\ \\[2ex]
        \hline
    \end{tabular}
    \label{table:Diagrams1111NNLO}
\end{table}

The diagrams can be computed explicitly, as described in Appendix \ref{app:FeynmanDiagramsOf1111}.
The resulting expression for the channel $F_1$ at NNLO is
\begin{equation}
    \begin{split}
    F_1^{(2)} (x)
    &=
    - \frac{1}{64 \pi^4 x (1-x)}
    \biggl(
    \frac{\pi^2}{3} (x G(0,x) + G(1,x)) \\
    &\phantom{=\ }
    +
    x (G(0,0,1,x) - G(1,1,0,x) + G(1,0,1,x) - G(0,1,0,x) + 3 \zeta_3) \\
    & \phantom{=\ }
    +
    G(1,0,1,x)
    +
    G(1,1,0,x)
    -
    2 (G(0,1,1,x) + G(1,0,0,x) + G(0,1,0,x))
    \biggr)\,.
    \end{split}
    \label{eq:1111_F1_NNLO}
\end{equation}
From this expression, we can deduce the corresponding function $f(x)$, which takes the form:
    \begin{equation}
        \begin{split}
            f^{(2)} (x)
            &=
            \frac{1}{64 \pi^4}
            \biggl(
            \frac{\pi^4}{15}
            +
            3 \zeta_3 G(1,x)
            +
            \frac{\pi^2}{3} (G(1,0,x) -G(0,1,x) + G(1,1,x)) \\
            &\phantom{=\ }
            +
            2( G(1,1,0,1,x) - G(0,0,1,0,x) + G(0,0,1,1,x) - G(1,1,0,0,x) \\
            &\phantom{=\ } + G(0,1,0,0,x) - G(1,0,1,1,x) )
            +
            G(1,0,1,0,x) - G(0,1,0,1,x) \\
            &\phantom{=\ } + G(1,0,0,1,x) - G(0,1,1,0,x)
            \biggr)\,.
        \end{split}
        \label{eq:1111_f_NNLO}
    \end{equation}
Notice that all terms have homogeneous transcendentality, and that the coefficients of the Goncharov polylogarithms and zeta functions are simple rational numbers.
This structure was exploited in prior works using a bootstrap approach, where an Ansatz was constructed and solved for these coefficients \cite{Cavaglia:2022qpg}.
In the present context, the Feynman diagrams serve to calculate the coefficients directly.
It should be pointed out that the coefficients of such an Ansatz are fully fixed by knowing $F_1$ only up to the following order:
\begin{equation}
    F_1^{(2)} (x)
    =
    \frac{1}{64 \pi^4}
    \biggl(
    - \log^2 x - \frac{\pi^2 - 12}{3} \log x + \frac{\pi^2}{3} - 3(2 + \zeta_3) - \frac{3}{2} x \log^2 x + \Om(x \log x)
    \biggr)\,.
    \label{eq:1111_F1_FirstOrdersInx}
\end{equation}
These \textit{four} coefficients could have been determined numerically if the Feynman diagrams of Table \ref{table:Diagrams1111NNLO} would have been too complicated to calculate analytically.
Understanding the necessary number of terms in the expansion for next-to-next-to-next-to-leading order would provide insight into extending these methods further.
The Ansatz approach also plays a role in developing the bootstrap algorithm discussed in Section \ref{sec:HigherPointFunctions}, which will be used to derive higher-point results.

\subsection{$\vev{1122}$}
\label{subsec:1122}

We now examine the four-point functions of two elementary fields $\Op_1$ and two composite operators $\Op_2$.
It is convenient to decompose the correlator into its fully connected and factorized components, as follows:
\begin{equation}
    \Am_{1122}
    =
    \Am_{1122}^{\text{(fact)}}
    +
    \Am_{1122}^{\text{(conn)}}\,
    \label{eq:1122_DisconnectedAndConnected}
\end{equation}
where by factorized we mean diagrams that are given products of lower-point defect Feynman diagrams.
For the channel of interest $F_1$, defined via \eqref{eq:4Pt_NaturalBasis}, the factorized part corresponds to the correlator $\vev{1111}$ discussed in Section \ref{subsec:1111}, though with necessary subtractions for overcounted terms:
\begin{equation}
    \left.
    \Am_{1122}^{\text{(fact)}}
    \right|_{(13)(24)}
    =
    \frac{n_1^2}{n_2}
    \left.
    \Am_{1111}
    \right|_{(13)(24)}
    -
    \text{(overcounted)}\,.
    \label{eq:1122_DisconnectedFrom1111}
\end{equation}
The overcounted terms correspond to boundary diagrams and are further discussed in Section \ref{subsubsec:1122_NextToNextToLeadingOrder}.
Up to next-to-leading order, there is no overcounting, and thus the second term in \eqref{eq:1122_DisconnectedFrom1111} can be discarded.

\subsubsection{Leading and next-to-leading orders}
\label{subsubsec:1122_LeadingAndNextToLeadingOrders}

\paragraph{Leading order.}
At leading order, there is no contribution to $F_1$ due to planarity.
Given our chosen normalization, we have
\begin{equation}
    f^{(0)} (x)
    =
    0\,.
    \label{eq:1122_f_LO}
\end{equation}

\paragraph{Next-to-leading order.}
At next-to-leading order, there is no fully connected contribution, so the correlator is identical to $\vev{1111}$.
The channel $F_1$ is therefore given by \eqref{eq:1111_F1_NLO}, and the solution to the Ward identities is
\begin{equation}
    f^{(1)} (x)
    =
    \frac{1}{8 \pi^2} \biggl( \frac{\pi^2}{6} + G(1,0,x) - G(0,1,x) \biggr)\,.
    \label{eq:1122_f_NLO}
\end{equation}

\subsubsection{Next-to-next-to-leading order}
\label{subsubsec:1122_NextToNextToLeadingOrder}

At next-to-next-to-leading order, new diagrams contribute to $F_1$ in addition to the factorized terms corresponding to $\vev{1111}$.
These diagrams are shown in Table \ref{table:Diagrams1122NNLO}.
We have four fully connected bulk diagrams, but we must also account for the overcounting discussed in \eqref{eq:1122_DisconnectedFrom1111}.
From Table \ref{table:Diagrams1111NNLO}, it is clear that three factorized boundary diagrams need to be subtracted from $\vev{1111}$, as depicted in the third row of Table \ref{table:Diagrams1122NNLO}.

\begin{table}[t!]
    \centering
    \caption{The diagrams contributing at next-to-next-to-leading order to the channel $F_1$ of the correlator $\vev{1122}$.
    Here we include only the connected contributions, which are called XX and XH, and the overcounted terms (XY) which have to be subtracted from the product of $\vev{1111}$ and $\vev{11}$ in order to obtain the factorized part.
    Explicit expressions can be found in Appendix \ref{app:FeynmanDiagramsOf1122}.}
    \begin{tabular}{lc}
        \hline
        XX & \NNLOXXOne \quad \NNLOXXTwo \\[2ex]
        \hline
        XH & \NNLOXHOne \quad \NNLOXHTwo \\[2ex]
        \hline \hline
        XY (subtractions) & \NNLOXYSubtractOne\ \NNLOSubtractFactorOne \quad \NNLOXYSubtractTwo\ \NNLOSubtractFactorOne \quad \NLOX\ \NNLOSubtractFactorTwo \\[2ex]
        \hline
    \end{tabular}
    \label{table:Diagrams1122NNLO}
\end{table}

These diagrams are straightforward to compute using standard techniques, and the results are summarized in Appendix \ref{app:FeynmanDiagramsOf1122}.
The corresponding channel $F_1$ is given by
\begin{equation}
    \begin{split}
    F_1^{(2)} (x)
    &=
    \frac{1}{64 \pi^4 x (1-x)}
    \biggl(
    - \frac{\pi^2}{6} (4 x G(0,x) + (x-2) G(1,x)) \\
    &\phantom{=\ }
    +
    \frac{x}{2} (2 G(0,1,0,x) - G(1,0,1,x) - 2 G(0,0,1,x) + G(1,1,0,x) - 3 \zeta_3)\\
    &\phantom{=\ }
    +
    2 (G(1,0,0,x) + G(0,1,1,x))
    -
    2 G(0,1,0,x) - G(1,0,1,x)
    -
    G(1,1,0,x)
    \biggr)\,.
    \end{split}
    \label{eq:1122_F1_NNLO}
\end{equation}
The solution to the Ward identities is
\begin{equation}
    \begin{split}
    f^{(2)} (x)
    &=
    \frac{1}{64 \pi^4}
    \biggl(
    \frac{17 \pi^4}{180}
    +
    \frac{3}{2} \zeta_3 G(1,x)
    +
    \frac{\pi^2}{6} ( 4 G(1,0,x) - 2 G(0,1,x) + G(1,1,x) ) \\
    &\phantom{=\ }
    -
    2 (G(1,0,1,1,x) + G(1,1,0,0,x) + G(0,0,1,0,x) - G(0,0,1,1,x) \\
    &\phantom{=\ }
    - G(0,1,0,0,x))
    -
    G(0,1,0,1,x) - G(0,1,1,0,x) + G(1,0,0,1,x) \\
    &\phantom{=\ }
    +
    G(1,0,1,0,x)
    +
    \frac{3}{2} G(1,1,0,1,x)
    +
    \frac{1}{2} G(1,1,1,0,x)
    \biggr)\,.
    \end{split}
    \label{eq:1122_f_NNLO}
\end{equation}
Note that this expression is similar to \eqref{eq:1111_f_NNLO}, in that it contains no rational functions of $x$ -- only Goncharov polylogarithms and $\zeta$ functions.

This result can be partially verified by pinching $\vev{1122}$ to $\vev{132}$ and $\vev{222}$, as described in \eqref{eq:1122_PinchingTo132}-\eqref{eq:1122_PinchingTo222}.
We find a perfect match with the localization data provided in \eqref{eq:lambda123}-\eqref{eq:lambda222}.

\subsection{$\vev{1212}$}
\label{subsec:1212}

We conclude this section by calculating the correlator $\vev{1212}$.
This correlator is fully fixed up to next-to-next-to-leading order simply by the train track integral \eqref{eq:TrainTrack_Definition}.

\subsubsection{Leading and next-to-leading orders}
\label{subsubsec:1212_LeadingAndNextToLeadingOrders}

\paragraph{Leading order.}
At leading order, the function $F_1$ vanishes, as with the other correlators in this section.
Thus, we have
\begin{equation}
    f^{(0)} (x)
    =
    0\,.
    \label{eq:1212_f_LO}
\end{equation}

\paragraph{Next-to-leading order.}
At next-to-leading order, the intertwining of operators $\Op_1$ and $\Op_2$ makes it impossible to draw a planar diagram for the channel $F_1$.
Therefore, we obtain
\begin{equation}
    F_1^{(1)} (x)
    =
    0\,.
    \label{eq:1212_F1_NLO}
\end{equation}
This means that the solution to the Ward identities is constant at this order.
The anti-self-crossing condition \eqref{eq:4Pt_AntiSelfCrossing} fixes it to be zero:
\begin{equation}
    f^{(1)} (x)
    =
    0\,.
    \label{eq:1212_f_NLO}
\end{equation}

\subsubsection{Next-to-next-to-leading order}
\label{subsubsec:1212_NextToNextToLeadingOrder}

At next-to-next-to-leading order, the correlator is fully governed by a single integral, which corresponds to the pinching limit of the train track integral to the kite integral, as discussed in \eqref{eq:KiteIntegral}:
\begin{equation}
    \NNLOFourPtTrainTrack\
    =
    - \frac{\lambda^5}{8} I_{24} K_{13,24}\,.
    \label{eq:1212_NNLO_TrainTrackDiagram}
\end{equation}
We do not keep track of the $R$-symmetry variables, and the numerical prefactor accounts for a symmetry factor and for the trace.
The channel $F_1$ is then
\begin{equation}
    \begin{split}
    F_1^{(2)} (x)
    &=
    \frac{1}{64 \pi^4 x (1-x)}
    \biggl(
    - \frac{\pi^2}{6} (4 G(0,x) + (2-x) G(1,x)) \\
    &\phantom{=\ }
    +
    x \biggl(\frac{1}{2} G(1, 1, 0, x) - G(0, 0, 1, x)
    + G(0, 1, 0, x) - \frac{1}{2} G(1, 0, 1, x)
    - \frac{3}{2} \zeta_3\biggr) \\
    &\phantom{=\ }
    + 2 G(1, 0, 0, x) - G(1, 1, 0, x)
    - 2 G(0, 1, 0, x) - G(1, 0, 1, x)
    + 2 G(0, 1, 1, x)
    \biggr)\,,
    \end{split}
    \label{eq:1212_F1_NNLO}
\end{equation}
which gives the solution to the Ward identities as
\begin{equation}
    \begin{split}
    f^{(2)} (x)
    &=
    \frac{1}{64 \pi^4}
    \biggl(
    -
    \frac{7 \pi^4}{180}
    +
    \frac{1}{2}
    (G(0,0,0,1,x) + G(0,0,1,0,x) + G(0,1,0,1,x) \\
    &\phantom{=\ }+ G(0,1,1,0,x) - G(1,0,0,1,x) - G(1,0,1,0,x) - G(1,1,0,1,x) \\
    &\phantom{=\ }- G(1,1,1,0,x)
    +
    2 G(1,0,1,1,x) - 2 G(0,1,0,0,x) \\
    &\phantom{=\ }+ 2 G(1,1,0,0,x) - 2 G(0,0,1,1,x))
    \biggr)\,.
    \end{split}
    \label{eq:1212_f_NNLO}
\end{equation}
Once again, no rational function appears in this solution.
As mentioned before, this observation plays a crucial role in the next section for deriving the correlators $\vev{11112}$ and $\vev{111111}$.

\section{Higher-point functions}
\label{sec:HigherPointFunctions}

Building upon the results of the previous section, we now introduce a bootstrap method for determining the higher-point functions $\vev{11112}$ and $\vev{111111}$ up to next-to-next-to-leading order.
Specifically, we derive these correlators with minimal reliance on Feynman diagrams, utilizing either vanishing contributions or channels composed of a single train track diagram.
The key ingredient is the construction of a suitable Ansatz for multipoint correlators, based on symbols and Goncharov polylogarithms.
The remaining input consists of protected data.

\subsection{$\vev{11112}$}
\label{subsec:11112}

\begin{figure}
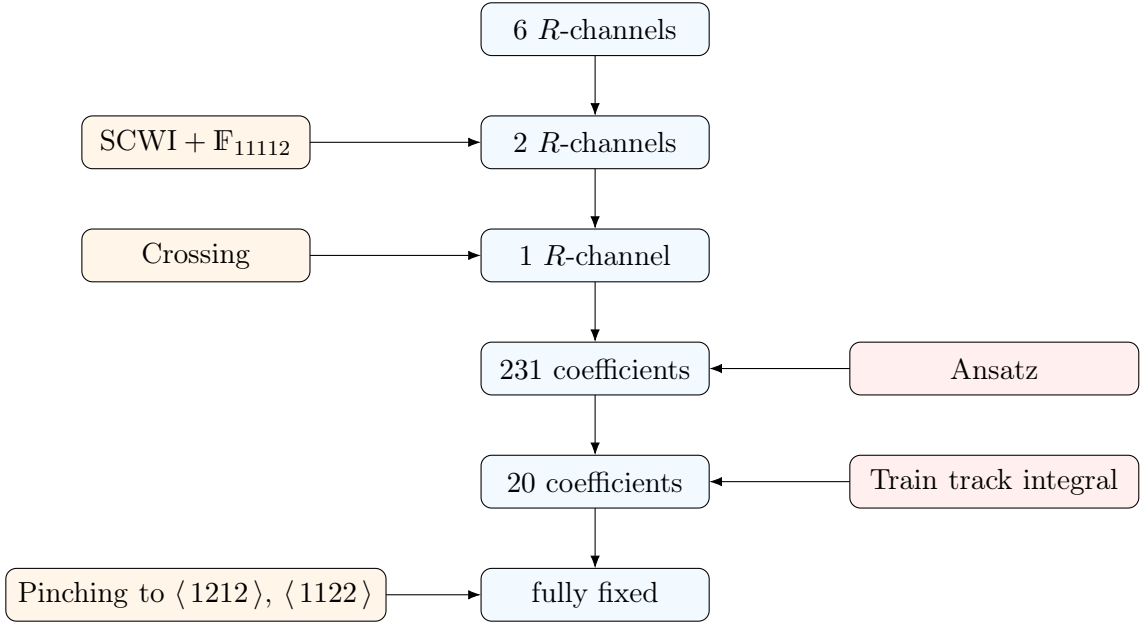

    \centering
    \MethodFivePoint
    \caption{
    Illustration of the steps involved in the bootstrap method used to derive the correlator $\vev{11112}$ at next-to-next-to-leading order.
    Superconformal Ward identities and crossing symmetry determine five of the six $R$-channels, in the basis $\tilde{R}_j$.
    An educated Ansatz is then formulated based on homogeneous transcendentality, limiting the number of open coefficients to $231$.
    By inputting the train track integral \eqref{eq:TrainTrack_Result}, which completely determines the channel $F_1$, all but $20$ coefficients are fixed.
    The remaining coefficients are resolved using the pinching constraints discussed in Section \ref{subsubsec:Pinching_FivePointFunctions}.}
    \label{fig:11112_Illustration}
\end{figure}

We now turn to the five-point function $\vev{11112}$, introduced earlier in Section \ref{subsubsec:Correlators_FivePointFunctions}.
Below, we outline a method to derive this correlator up to next-to-next-to-leading order without the need to explicitly compute all the relevant diagrams.
The key idea is to leverage the non-perturbative constraints discussed in Section \ref{sec:NonPerturbativeConstraints}, which reduce the number of functions to determine from $6$ to $1$.
Next, we construct an Ansatz for the perturbative order of interest and incorporate the channel $F_1$ from diagram computations.
At next-to-leading order, the channel $F_1$ vanishes, while at next-to-next-to-leading order, it is determined by the train track integral.
This procedure is explained in further detail below and is summarized in Figure \ref{fig:11112_Illustration} for the next-to-next-to-leading order calculation.

\subsubsection{Leading and next-to-leading orders}
\label{subsubsec:11112_LeadingAndNextToLeadingOrders}

\paragraph{Leading order.}
At leading order, the correlator is determined up to a constant by the fact that the channel $F_1$, as defined by \eqref{eq:11112_NaturalBasis}, vanishes:
\begin{equation}
    F_1^{(0)} (x_1, x_2)
    =
    0\,.
    \label{eq:11112_F1_LO}
\end{equation}
With our chosen normalization for the solution of the Ward identities \eqref{eq:11112_SolutionSCWI}, we find
\begin{equation}
    f^{(0)} (x_1, x_2)
    =
    1\,.
    \label{eq:11112_f_LO}
\end{equation}

\paragraph{Next-to-leading order.}
The next-to-leading order is more interesting and, although it is already known \cite{Barrat:2021tpn}, we use it here to demonstrate our bootstrap method.

The first step is to construct a suitable perturbative Ansatz for the function $f(x_1,x_2)$, defined via \eqref{eq:11112_SolutionSCWI} and \eqref{eq:11112_f1f2_Crossing}, following a perturbative expansion akin to \eqref{eq:f_PerturbativeExpansion}.
At any given order in perturbation theory, we assume the following properties for the Ansatz:
\begin{enumerate}
    \item $f^{(\ell)} (x_1, x_2)$ contains no rational functions, consisting solely of Goncharov polylogarithms and $\zeta$-functions;\footnote{It is likely that the function space would need to be expanded beyond a certain order in perturbation theory.
    For instance, multiple $\zeta$-values are certainly expected to appear.}
    \item $f^{(\ell)} (x_1, x_2)$ is homogeneous in transcendentality;
    \item $f^{(\ell)} (x_1, x_2)$ has transcendentality weight $2\ell$.
\end{enumerate}

At next-to-leading order, this implies that the Ansatz must include a constant and Goncharov polylogarithms of at most transcendentality $2$.
The condition of homogeneous transcendentality immediately cancels terms of transcendentality $1$, as they would multiply $\zeta_1$, which diverges.

To construct the appropriate Goncharov polylogarithms, we write the Ansatz in symbols, with the elements
\begin{equation}
    \xi_i
    =
    \lbrace
    x_1, 1-x_1, x_2, 1-x_2, x_{12}
    \rbrace\,.
    \label{eq:11112_SymbolBasis}
\end{equation}
The Ansatz then takes the form
\begin{equation}
    f^{(1)} (x_1, x_2)
    =
    c_0
    +
    \sum_{i \neq j} c_{ij}\, \xi_i \otimes \xi_j\,,
    \label{eq:11112_AnsatzInSymbols}
\end{equation}
where terms with $i=j$ are excluded since they would correspond to second-order anomalous dimensions in a conformal block expansion, which are not expected to appear.
In other words, we exclude $\log^2$ terms in the OPE limits $x_i \to 0, (1-x_i) \to 0, x_{ij} \to 0$.

The Ansatz in \eqref{eq:11112_AnsatzInSymbols} is not quite yet in a usable form, as it is not finite for every choice of coefficients.
Requiring finiteness (also known as integrability condition), we fix $6$ out of the $21$ free coefficients, yielding the following Ansatz:
\begin{equation}
    \begin{split}
        16 \pi^2 f^{(1)} (x)
        &=
        c_0
        +
        c_9 G(1, x_1) G(1, x_2)
        +
        G(0, x_1) ((c_2 + c_{11}) G(0, x_2) + c_7 G(1, x_2)) \\
        &\phantom{=\ }
        +
        c_{10} G(1, x_2) G(x_2, x_1)
        +
        G(0, x_2) ((c_6 + c_{13}) G(1, x_1) + c_4 G(x_2, x_1)) \\
        &\phantom{=\ }
        +
        (c_4 + c_{12}) G(0, 0, x_2)
        +
        c_5 G(0, 1, x_1)
        +
        (c_8 + c_{10}) G(0, 1, x_2) \\
        &\phantom{=\ }
        +
        c_{11} G(0, x_2, x_1)
        +
        c_1 G(1, 0, x_1)
        +
        (c_3 + c_{14}) G(1, 0, x_2) \\
        &\phantom{=\ }
        +
        c_{13} G(1, x_2, x_1)
        +
        (-c_4 + c_{11} + c_{12}) G(x_2, 0, x_1) \\
        &\phantom{=\ }
        +
        (-c_{10} + c_{13} + c_{14}) G(x_2, 1, x_1)\,,
    \end{split}
    \label{eq:11112_Ansatz}
\end{equation}
where we have relabeled the coefficients from $c_0$ to $c_{14}$ for clarity.
A review of the integrability condition for symbols is found in Appendix \ref{app:SymbolsAndGoncharovPolylogarithms}.

With the Ansatz constructed, we can now input data from the $R$-channels to fix as many coefficients as possible.
For instance, it is straightforward to see that the channel $F_1$ remains zero at this order:
\begin{equation}
    F_1^{(1)} (x)
    =
    0\,.
    \label{eq:11112_F1_NLO}
\end{equation}
Comparing this result to \eqref{eq:11112_Ansatz} via the Ward identities fixes $13$ more coefficients.
The final two coefficients are fixed by applying the pinching conditions for $\vev{11112} \to \vev{1122}$ and $\vev{11112} \to \vev{1212}$, as given in \eqref{eq:11112_PinchingTo1122}–\eqref{eq:11112_PinchingTo1212}.

The final result is
\begin{equation}
    \begin{split}
    f^{(1)} (x_1, x_2)
    &=
    \frac{1}{8 \pi^2}
    \biggl(
    \frac{\pi^2}{2}
    +
    G(0, x_2) G(1, x_1)
    -
    G(1, x_1) G(1, x_2)
    +
    G(1, x_2) G(x_2, x_1) \\
    &\phantom{=\ }
    +
    G(0, 1, x_2)
    -
    G(1, 0, x_2)
    +
    G(1, x_2, x_1)
    -
    G(x_2, 1, x_1)
    \biggr)\,.
    \end{split}
    \label{eq:11112_f_NLO}
\end{equation}
This result can be compared with the explicit calculation performed in \cite{Barrat:2021tpn} using diagrammatic recursion relations, and we observe perfect agreement between the two methods.

\subsubsection{Next-to-next-to-leading order}
\label{subsubsec:11112_NextToNextToLeadingOrder}

We apply the same bootstrap method to compute the five-point function $\vev{11112}$ at next-to-next-to-leading order.
The Ansatz must now include terms of transcendentality up to weight $4$, while terms of transcendentality $3$ are excluded by the homogeneous transcendentality condition.
This results in an Ansatz expressed in symbols with initially $651$ coefficients. However, imposing the finiteness condition immediately fixes $420$ of these coefficients, leaving $231$ free coefficients to determine.

On the input side, it is crucial to note that the channel $F_1$ is now determined by a single diagram: the train track integral discussed in Section \ref{subsec:OneIntegralToRuleThemAll}.
This is given by
\begin{align}
    \NNLOFivePt
    =
    - \frac{\lambda^5}{8}
    B_{125,345}\,,
    \label{eq:11112_Diagram_NNLO}
\end{align}
where, as usual, we do not include the $R$-symmetry variables.
The integral can be evaluated analytically, and in terms of Goncharov polylogarithms it reads
\begin{equation}
    \begin{split}
        F_1^{(2)} (x_1, x_2)
        &=
        \frac{1}{128 \pi^4 x_1 (1-x_2)}
        \bigl(
        2 (G(0, x_2) G(1, 0, x_1) - 
        G(1, x_2) G(1, 0, x_1)) \\
        &\phantom{=\ } - 
        G(x_2, x_1) (G(0, 1, x_2) + G(1, 0, x_2)) + 
        G(1, x_1) (-2 G(0, 0, x_2) \\
        &\phantom{=\ } + G(0, 1, x_2) + G(1, 0, x_2)) - 
        G(0, x_2) G(1, x_2, x_1) + 2 G(1, x_2) G(x_2, 0, x_1) \\
        &\phantom{=\ } + 
        G(1, 0, x_2, x_1) + G(1, x_2, 0, x_1) - G(x_2, 0, 1, x_1) - 
        G(x_2, 1, 0, x_1))
        \bigr)\,.
    \end{split}
    \label{eq:11112_F1_NNLO}
\end{equation}
This result, along with the pinching conditions, is sufficient to completely fix the Ansatz.
Although the full result for $f^{{(2)}} (x_1, x_2)$ is lengthy and is provided in the ancillary \textsc{Mathematica} notebook, we present a couple of terms here for illustrative purposes:
\begin{equation}
    f^{(2)} (x_1, x_2)
    =
    \frac{1}{64 \pi^4}
    \bigl(
    G(x_2, x_2, 1, 0, x_1)
    -
    2 G(x_2, x_2, 1, x_2, x_1)
    +
    \ldots
    \bigr)\,.
\end{equation}

\subsection{$\vev{111111}$}
\label{subsec:111111}

We now compute the six-point function $\vev{111111}$ using the same method as in the previous section, but adapted to the case of three spacetime cross-ratios.
In particular, the basis of symbols is now
\begin{equation}
    \xi_i
    =
    \lbrace
    x_1, x_2, x_3,
    1-x_1, 1-x_2, 1-x_3,
    x_{12}, x_{13}, x_{23}
    \rbrace\,.
\end{equation}
In the following, we only list the results since the steps are fully equivalent to the five-point case.

\begin{figure}
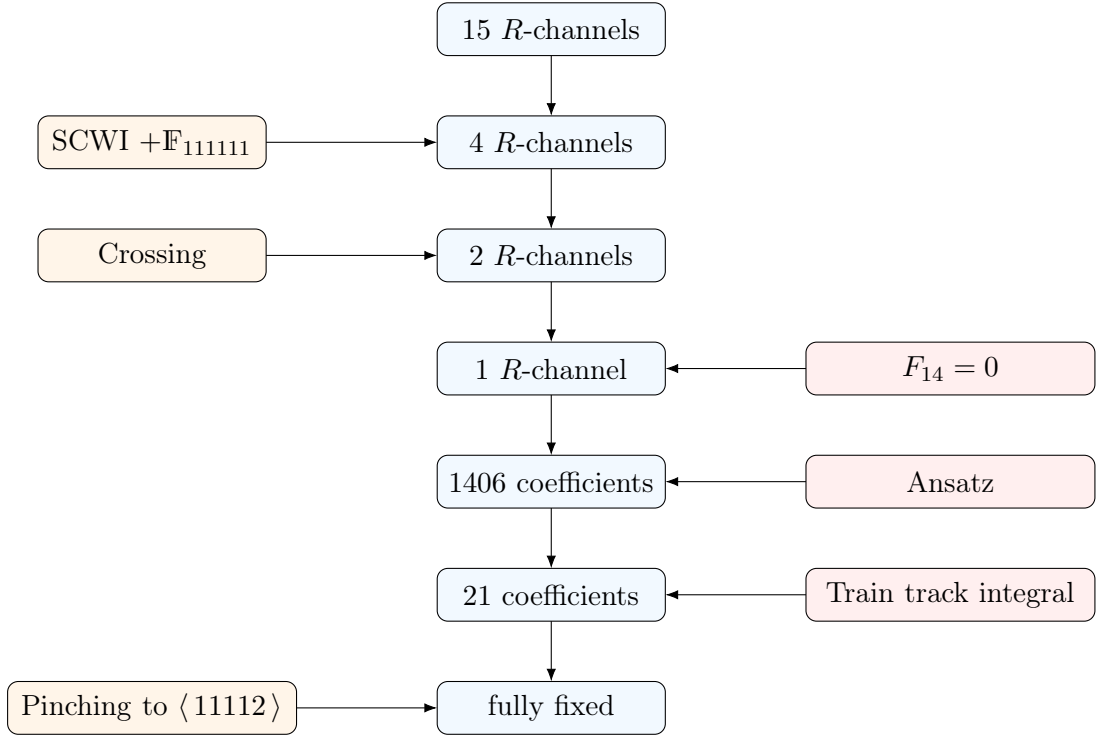

    \centering
    \MethodSixPoint
    \caption{Illustration of the method used for bootstrapping the six-point function $\vev{111111}$ at next-to-next-to-leading order.}
    \label{fig:111111_Illustration}
\end{figure}

\subsubsection{Leading and next-to-leading orders}
\label{subsubsec:111111_LeadingAndNextToLeadingOrders}

\paragraph{Leading order.}
At leading order, the functions $f_{1,2}(x_1,x_2,x_3)$ are simply given by
\begin{equation}
	\begin{split}
    f_1^{(0)} (x_1,x_2,x_3)
    &=
    1\,, \\
    f_2^{(0)} (x_1,x_2,x_3)
    &=
    0\,.
    \end{split}
    \label{eq:111111_f_LO}
\end{equation}
$f_2$ has been chosen such that it is in a one-to-one correspondence with the channel $F_{14}$ of the natural basis.
This choice is convenient for our calculations, as it is easy to see that $F_{14}=0$ up to next-to-next-to-leading order.

\paragraph{Next-to-leading order.}
At next-to-leading order, we obtain
\begin{equation}
    \begin{split}
        f_1^{(1)} (x_1,x_2,x_3)
        &=
        \frac{1}{12 \pi^2}
        \biggl(
        G(0,1,x_1) - G(1,0,x_1)
        +
        2\bigl(
        G(1,x_1) G(1,x_2) \\
        &\phantom{=\ }
        - G(x_2,x_1) G(1,x_2) - G(1,x_1) G(1,x_3) - G(0,x_2) (G(1,x_1)-G(x_3,x_1)) \\
        &\phantom{=\ }
        + G(1,x_3) G(x_3,x_1) + G(x_2,x_1) G(x_3,x_2) + G(1,x_3) G(x_3,x_1) \\
        &\phantom{=\ }
        + G(x_2,x_1) G(x_3,x_2)
         +G(0,x_3) ( G(1,x_1) - G(x_3,x_1) + G(x_3,x_2)) \\
        &\phantom{=\ }
        - G(x_3,x_1) G(x_3,x_2)
        - G(0,1,x_2) - G(0,1,x_3) + G(0,x_3,x_2) \\
        &\phantom{=\ }
        + G(1,0,x_2) + G(1,0,x_3) - G(1,x_2,x_1) + G(1,x_3,x_1)
        + G(x_2,1,x_1) \\
        &\phantom{=\ } - G(x_2,x_3,x_1) - G(x_3,0,x_2) - G(x_3,1,x_1) + G(x_3,x_2,x_1)
        \bigr) \\
        &\phantom{=\ }
        +
        3\bigl(
        G(1,x_3) G(1,x_2) - G(1,x_3) G(x_3,x_2) - G(0,x_3) G(1,x_2) \\
        &\phantom{=\ }
        - G(1,x_3,x_2) + G(x_3,1,x_2)
        \bigr)
        +
        \frac{2 \pi^2}{3}
        \biggr)\,, \\
        f_2^{(1)} (x_1,x_2,x_3)
        &=
        0\,,
    \end{split}
\end{equation}
which is in agreement with the results of \cite{Barrat:2021tpn}.

\subsubsection{Next-to-next-to-leading order}
\label{subsubsec:111111_NextToNextToLeadingOrder}

At next-to-next-to-leading order, the correlator is controlled by the train track integral introduced in Section \ref{subsec:OneIntegralToRuleThemAll}.
The corresponding diagram is
\begin{equation}
    \NNLOSixPt
    =
    \frac{\lambda^5}{8} B_{156,234}\,.
\end{equation}
This is the only diagram appearing in the channel $F_4$.
The determination of $f_1$ is completely analogous to the calculation of $f$ for the correlator $\vev{11112}$, the only difference being technical since we deal with a high number of open coefficients.
Before imposing the integrability condition, the Ansatz consists of $6643$ coefficients.
Finiteness reduces this number to $1406$ unknowns.
In order to deal with such a high number of terms, we create a linear system of equations in \textsc{Mathematica} to match the Ansatz to $f_4$, that we iteratively solve by using row reduction.
Inputting the train-track fixes all the coefficients except for $21$ of them.
The pinching limit \eqref{eq:111111_PinchingTo11112} consists of six relations.
One of them suffices to fix the remaining open coefficients, while the other five serve as checks of our final result.
Since the expressions are lengthy, we provide the solution of the Ward identities $f_1(x_1, x_2, x_3)$ in the ancillary notebook, where the $R-$channels in the natural basis are also given explicitly.
To illustrate the results, we show here some of the contributing terms:
\begin{equation}
	\begin{split}
		f_1^{(2)} (x_1, x_2, x_3)
		&=
		\frac{1}{64 \pi^4}
		\biggl(
		G(x_2, x_3, 1, 0, x_1)
		- \frac{8}{3}
		G(x_3, x_3, x_3, x_2, x_1)
		+
		\ldots
		\biggr)\,, \\
		f_2^{(2)} (x_1,x_2,x_3) &= 0\,.
	\end{split}
\end{equation}
The complete algorithm is summarized in Figure \ref{fig:111111_Illustration}.

\subsection{Checks}
\label{subsec:Checks}

\begin{figure}
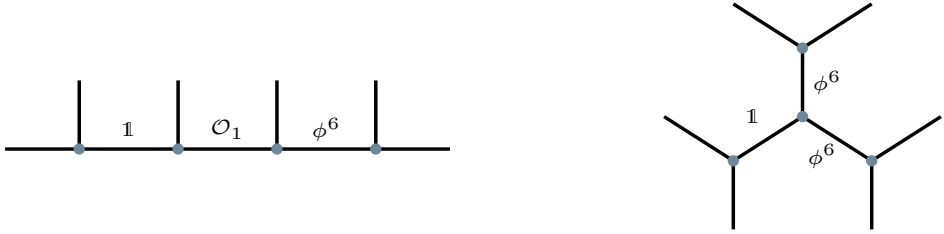

\centering
\begin{subfigure}{.55\textwidth}
  \centering
  \CombChannel
\end{subfigure}%
\begin{subfigure}{.45\textwidth}
  \centering
  \SnowflakeChannel
\end{subfigure}
\caption{The two OPE configurations used for performing checks on our result for the six-point function $\vev{111111}$ are displayed here.
The left figure presents the comb channel and the leading non-trivial exchange of operators.
The right figure shows the leading non trivial contribution in the snowflake channel.}
\label{fig:CombsAndSnowflakes}
\end{figure}

We now proceed to some elementary checks for the results of this section.
Specifically, we wish to compare the lowest OPE coefficients for each correlator to the existing literature.
For the six-point function, we can use the two different channels, namely the comb and snowflake configurations depicted in Figure \ref{fig:CombsAndSnowflakes}.\footnote{We thank Andrea Cavagli{\`a} and Marco Meineri for suggesting these checks.}

In the case of the correlator $\vev{11112}$, the easiest check that one can perform is the coefficient
\begin{equation}
    F_2 (\eta_1, \eta_2)
    =
    \lambda_{112} + \Om(\eta_1, \eta_2)\,,
    \label{eq:11112_BlockExpansion}
\end{equation}
where $\eta_i$ refers to the cross-ratios of the comb channel used in Section 4.2.2 of \cite{Barrat:2024nod}.
This OPE coefficient is protected and given exactly in \eqref{eq:lambda112}.
Perturbatively, it reads
\begin{equation}
    \lambda_{112} = 1 + \frac{\lambda}{48} - \frac{29\lambda^2}{23040} + \Om(\lambda^3)\,,
    \label{eq:lambda112}
\end{equation}
and we find a perfect match order by order with our correlator.
A strongest check would be to compare the OPE coefficients of unprotected operators.
The simplest case would then be the combination $\lambda_{11\phi^6}^2 \lambda_{2 \phi^6\phi^6}$.
However, to the best of our knowledge, this coefficient is presently not known up to next-to-next-to-leading order.

The check is less trivial for the six-point function $\vev{111111}$.
Using the variables $\eta_i$, our correlator takes the form
\begin{equation}
    F^{(2)}_2 (\eta_1, \eta_2, \eta_3)
    =
    \frac{\pi^2 - 18 + 6 \log \eta_1}{192 \pi^4} \lambda^2\, \eta_1
    +
    \Om(\eta_1^2, \eta_2, \eta_3)\,,
    \label{eq:111111_BlockExpansion}
\end{equation}
from which we can read the OPE coefficient and the anomalous dimension of the unprotected operator $\phi^6$:
\begin{equation}
	\begin{split}
		\left. \lambda_{11\phi^6}^2 \eta_1 \right|_{\Om(\lambda^2)}
    		&=
    		\frac{\pi^2 - 18}{192 \pi^4}\,, \\
    		\gamma^{(1)}_{\phi^6}
    		&=
    		\frac{1}{4 \pi^2}\,.
	\end{split}
	\label{eq:111111_lambda116}
\end{equation}
These values match the literature (see, e.g., \cite{Cavaglia:2021bnz}).
A similar check can be performed in another OPE limit, which corresponds to the snowflake configuration.
Using the variables $z_i$ of \cite{Barrat:2022eim,Peveri:2023qip,Barrat:2024nod} and the conformal blocks of \cite{Fortin:2023xqq}, the highest-weight channel is then given by
\begin{equation}
    F^{(2)}_2 (z_1, z_2, z_3)
    =
    \frac{\pi^2 - 18 + 6 \log z_1 z_2}{192 \pi^4} \lambda^2\, z_1 z_2
    +
    \Om(z_1^2, z_2^2, z_3)\,,
    \label{eq:111111_BlockExpansion2}
\end{equation}
which again matches \eqref{eq:111111_lambda116}.

Note that an expansion in superconformal blocks would give access to the three-point function $\vev{\phi^6 \phi^6 \phi^6}$ at next-to-next-to-leading order, a quantity which is presently unknown to the best of our knowledge.
The superblocks are currently not available, and we postpone this analysis to future work.

636\section{Conclusions}
\label{sec:conclusions}

In this paper, we explored the multipoint correlators of scalar half-BPS insertions along the Maldacena-Wilson line in $\Nm=4$ SYM, developing a perturbative bootstrap approach up to next-to-next-to-leading order.
Specifically, we focused on the five-point function $\vev{11112}$ and the six-point function $\vev{111111}$, while deriving new results for $\vev{1122}$ and $\vev{1212}$ as well.
Our method combined both perturbative and non-perturbative insights into these correlators.
On the non-perturbative front, we employed the superconformal Ward identities introduced in \cite{Liendo:2016ymz,Bliard:2024und,Barrat:2024ta}, applying crossing symmetry constraints to reduce the correlators to a low number of functions of the spacetime cross-ratios.
These functions were further constrained by pinching operators down to lower-point functions. On the perturbative side, we found that, for $\vev{11112}$ and $\vev{111111}$, the correlator was governed by a single integral, referred to as the six-point train track.
This integral was evaluated in \cite{Rodrigues:2024znq} for all relevant orderings.
By incorporating this result into a carefully constructed Ansatz, based on transcendentality properties, we succeeded in fully determining the correlators.

There are several promising avenues for further exploration following this work.
While our focus was on higher-point correlators, the techniques we developed here can also be applied to obtain four-point functions involving higher-weight operators.
A natural next step would be to construct a recursion relation, similar to the one presented in \cite{Barrat:2021tpn}, for four-point functions at next-to-next-to-leading order.
This would extend the results of \cite{Kiryu:2018phb} for arbitrary correlators $\vev{\Delta_1 \Delta_2 \Delta_3 \Delta_4}$ to the next-to-next-to-leading order, which could be particularly valuable for bootstrap applications.

Another interesting extension would be to conduct a numerical analysis of the correlator $\vev{111111}$, building upon the methods outlined in \cite{Antunes:2023kyz}.
This approach could potentially be merged with the bootstrability techniques applied so far to four-point functions \cite{Cavaglia:2023mmu}, thereby providing access to bounds on four-point functions and to a broader range of CFT data.
Such a framework would serve as an excellent testing ground for methods that may be applicable to higher-dimensional systems.
To carry out this analysis, it will be crucial to determine the superconformal blocks in the comb channel, a task that is currently being addressed in ongoing work \cite{Barrat:2024ta2}.
This work seeks to rederive and extend the strong-coupling results of \cite{Giombi:2023zte} for $\vev{111111}$, employing the techniques introduced in \cite{Barrat:2024ta2}.
Additionally, integrated correlators in the gist of \cite{Drukker:2022pxk,Cavaglia:2022qpg,Pufu:2023vwo,Fiol:2023cml,Billo:2023ncz,Billo:2024kri,Dempsey:2024vkf,Drukker:2024ta} were shown to significantly improve the numerical bootstrap results.
It would certainly also be useful to revisit the solution of the Ward identities for six-point functions in order to understand if more constraints can be used to reduce the number of functions to one after imposing crossing symmetry.
An interesting quantity that could be reachable is the three-point function $\vev{\phi^6 \phi^6 \phi^6}$, thanks to the interplay between comb and snowflake channels in six-point functions.

This work underscores the significant role of train track integrals in multipoint correlators of defect operators.
We anticipate that, at next-to-next-to-next-to-leading order, the eight-point function $\vev{11111111}$ will similarly be governed by lower-point functions and the integral 
\begin{equation}
    \NNNLOEightPt\,.
    \label{eq:EightPointIntegral}
\end{equation}
While obtaining the lower-point functions might pose challenges, it would still be valuable to further investigate the class of integrals \eqref{eq:EightPointIntegral}.
They have been studied in the literature and are conjectured to be integrable \cite{Chicherin:2017frs,Kazakov:2023nyu,Loebbert:2024fsj,Duhr:2024hjf}.
In cases where the external points are not aligned, elliptic behavior is expected.
Drawing from our experience with the six-point function, it is plausible that all $n$-point train track integrals, in the collinear limit, could be expressible in terms of Goncharov polylogarithms.
One promising approach to study such integrals could involve the Wilson-line defect CFT of the fishnet theory \cite{Gromov:2021ahm}, where the vertices significantly limit the number of contributing diagrams for a given configuration.
By combining techniques from the conformal bootstrap and integrability, it has been possible to derive exact correlators within this theory \cite{Grabner:2017pgm,Gromov:2018hut}.
It would be fascinating to explore to what extent these insights can be extended to the defect CFT framework.

Recently, the magnetic line defect CFT in the $\mathrm{O}(N)$ model has garnered significant interest, both from the perspective of large $N$ expansions and the $\veps$-expansion.
A similar defect CFT has been explored in Yukawa theories \cite{Giombi:2022vnz}, with correlators studied in \cite{Barrat:2023ivo}.
Yukawa CFTs are thought to exhibit emergent supersymmetry under specific configurations \cite{Fei:2016sgs}.
In gauge theories across dimensions, conformal Wilson lines have been systematically classified in \cite{Aharony:2023amq}.
These models hold interesting applications in condensed-matter physics, and the study of multipoint correlators in this context could yield valuable insights.
In the framework of the $\veps$-expansion, it is likely that many of the structures observed in this paper will persist, even in the absence of supersymmetry.

\acknowledgments

We are particularly grateful to Carlos Bercini, Lorenzo Bianchi, Gabriel Bliard, Andrea Cavaglià, Luke Corcoran, Burkhard Eden, Pietro Ferrero, Valentina Forini, Aleix Gimenez-Grau, Nikolay Gromov, Moritz Kade, Ziwen Kong, Carlo Meneghelli, Sophie Müller, Jan Plefka, and Yingxuan Xu for useful discussions.
DA is funded, and GP has been funded, by the Deutsche Forschungsgemeinschaft (DFG, German Research Foundation) -- Projektnummer 417533893/GRK2575 ``Rethinking Quantum Field Theory''.
JB is supported by ERC-2021-CoG - BrokenSymmetries 101044226, and has benefited from the German Research Foundation DFG under Germany’s Excellence Strategy – EXC 2121 Quantum Universe – 390833306.

\appendix

\section{Symbols and Goncharov polylogarithms}
\label{app:SymbolsAndGoncharovPolylogarithms}

In this appendix, we provide a brief overview of symbols and their connection to Goncharov polylogarithms, which are key elements in formulating the Ansatz in Section \ref{sec:HigherPointFunctions}.
Much of the content presented here is based on \cite{Duhr:2019tlz}.

\subsection{Symbols}
\label{subsec:Symbols}

Symbols are formally defined via the Hopf algebra structure of multiple polylogarithms.
Specifically, the symbol of an expression represents the highest iteration of the coproduct, modulo $i \pi$.
This feature is particularly useful for building an Ansatz, as symbols contain essential information about the arguments of the Goncharov polylogarithms.
For example, the symbol of the dilogarithm $\text{Li}_2(x)$ is given by
\begin{equation}
    \Sm ( \text{Li}_2 (x) )
    =
    - (1-x) \otimes x\,.
    \label{eq:Symbols_Example}
\end{equation}
The \textsc{Mathematica} package \textsc{PolyLogTools} is particularly effective for manipulating symbols and Goncharov polylogarithms.

Symbols obey the following properties:
\begin{align}
    A \otimes ( \pm 1 ) \otimes B &= 0\,, \label{eq:Symbols_Preperty1} \\
    A \otimes (x \cdot y) \otimes B
    &=
    A \otimes x \otimes B + A \otimes y \otimes B\,.
    \label{eq:Symbols_Property2}
\end{align}

In Section \ref{sec:HigherPointFunctions}, we demonstrate how expressing correlators through symbols facilitates the construction of an Ansatz.
However, the resulting expression may not always be finite.
Finiteness is achieved by enforcing the \textit{integrability condition}.
For an expression written as
\begin{equation}
    S
    =
    \sum c_{i_1 \ldots i_n} A_{i_1} \otimes \ldots \otimes A_{i_n}\,,
    \label{eq:Symbols_Ansatz}
\end{equation}
where the $c_{i_1 \ldots i_n}$ are numerical coefficients, $S$ is finite if it satisfies
\begin{equation}
    \sum c_{i_1 \ldots i_n} A_{i_1} \otimes \ldots \otimes A_{i_{p-1}} \otimes A_{i_{p+2}} \otimes \ldots \otimes A_{i_n} d\log A_{i_p} \wedge d\log A_{i_{p+1}} = 0\,,
    \label{eq:Symbols_IntegrabilityCondition}
\end{equation}
for all pairs $(i_p, i_{p+1})$ with $1 \leq p < n$.
This condition is efficiently implemented in \textsc{PolyLogTools}.

Additionally, symbols obey the \textit{shuffle algebra} relation:
\begin{equation}
    \Sm (x \cdot y)
    =
    \Sm (x) \shuffle\,\Sm(y)\,,
    \label{eq:Symbols_ShuffleAlgebra}
\end{equation}
which serves as a useful property for simplifying expressions.

\subsection{Goncharov polylogarithms}
\label{subsec:GoncharovPolylogarithms}

Goncharov polylogarithms are a crucial component of our Ansatz.
Here, we define Goncharov polylogarithms and outline their key properties.

One definition of Goncharov polylogarithms is given by the iterated integral
\begin{equation}
    G(a_1, \ldots, a_n, x)
    =
    \int_0^x \frac{dt}{t-a_1} G(a_2, \ldots, a_n, t)\,,
    \label{eq:Goncharov_Definition}
\end{equation}
with $n \geq 0$, and the initial condition
\begin{equation}
    G(x)
    =
    1\,.
    \label{eq:Goncharov_InitialCondition}
\end{equation}
The transcendentality weight of a Goncharov polylogarithm defined in this manner is $n$.
In the special case where all arguments are zero except for the last one, Goncharov polylogarithms reduce to
\begin{equation}
    G(\underbrace{0, \ldots, 0}_{n \text{ times}}, x)
    =
    \frac{1}{n!} \log^n x\,.
    \label{eq:Goncharov_SpecialCase1}
\end{equation}
Another notable special case is
\begin{equation}
    G(0, \ldots, 0, a, x)
    =
    - \text{Li}_n \biggl( \frac{x}{a} \biggr)\,,
    \label{eq:Goncharov_SpecialCase2}
\end{equation}
where $\text{Li}_n$ denotes the $n$th dilogarithm.
Further relations between Goncharov polylogarithms and harmonic polylogarithms (HPLs) can be found in \cite{Duhr:2019tlz}.

Goncharov polylogarithms also obey a shuffle algebra inherited from the symbols.
This relation is expressed as
\begin{equation}
    G(\vec{a}, x) G(\vec{b}, x)
    =
    \sum_{\vec{c} = \vec{a} \shuffle \vec{b}} G(\vec{c}, x)\,.
    \label{eq:Goncharov_ShuffleAlgebra}
\end{equation}

\section{Integrals}
\label{app:Integrals}

This appendix is dedicated to the integrals encountered throughout this work.
We consider here bulk integrals, i.e., integrals that do not involve the Wilson line as an internal vertex.
Integrals along the defect are one-dimensional and easy to perform with the \textsc{PolyLogTools} package \cite{Duhr:2019tlz} by using the analytical expressions of this section.

The master integral for massless scalar four-point integrals at next-to-leading order is the well-known X-integral \cite{Usyukina:1994iw,Usyukina:1994eg}\footnote{See also \cite{Beisert:2002bb,Drukker:2008pi} for the modern notation.}
\begin{align}
    X_{1234}
    =
    \XIntegral\
    =
    \int d^4 x_5\, I_{15} I_{25} I_{35} I_{45}
    =
    \frac{I_{12} I_{34}}{16\pi^2}\ z \zb\, D( z, \zb)\,,
    \label{eq:X1234}
\end{align}
with the Bloch-Wigner function \cite{bloch1978applications}
\begin{equation}
    D(z, \zb)
    =
    \frac{1}{z - \zb} \left( 2 \Li_2 (z) - 2 \Li_2 (\zb) + \log z \zb\, \log \frac{1-z}{1-\zb} \right)\,,
    \label{eq:BlochWigner}
\end{equation}
where the cross-ratios $z$ and $\zb$ are defined through
\begin{equation}
    z \zb = \frac{x_{12}^2 x_{34}^2}{x_{13}^2 x_{24}^2}\,.
    \qquad
    (1-z)(1-\zb) = \frac{x_{14}^2 x_{23}^2}{x_{13}^2 x_{24}^2}\,.
    \label{eq:4dCrossRatios}
\end{equation}
Note that the Bloch-Wigner function is crossing symmetric:
\begin{equation}
    D(z, \zb) = D(1-z\,, 1-\zb)\,.
    \label{eq:BWIsCrossingSymmetric}
\end{equation}
From the X-integral, it is possible to obtain the three-point Y-integral defined as
\begin{align}
    Y_{123}
    =
    \YIntegral\
    &=
    \int d^4 x_4\, I_{14} I_{24} I_{34} \notag \\
    &=
    \lim\limits_{\tau_4 \to \infty} I_{34}^{-1} X_{1234}
    =
    \frac{I_{12}}{16\pi^2} z \zb\, D( z, \zb)\,,
    \label{eq:Y123}
\end{align}
where now the cross-ratios $z, \zb$ are defined as the limit $\tau_4 \to \infty$ of the four-point variables defined above:
\begin{equation}
    z \zb
    =
    \frac{I_{13}}{I_{12}}\,,
    \qquad
    (1-z)(1-\zb)
    =
    \frac{I_{13}}{I_{23}}\,.
    \label{eq:CrossRatiosY}
\end{equation}
The Y-integral becomes $\log$-divergent when two points are pinched together:
\begin{equation}
    Y_{112} = Y_{122} = \IntegralYOneTwoTwo\ = - \frac{I_{12}}{16\pi^2} \left( \log \frac{I_{12}}{I_{11}} - 2 \right)\,.
    \label{eq:Y112}
\end{equation} 
The pinching of two points in the X-integral can be related to Y-integrals:
\begin{equation}
    X_{1233}
    =
    \IntegralXOneTwoThreeThree\
    =
    \frac{1}{2} ( I_{13} Y_{223} + I_{23} Y_{113} ) - \frac{I_{13} I_{23}}{32 \pi^2} \log \frac{I_{13} I_{23}}{I_{12}^2}\,.
    \label{eq:X1233}
\end{equation}
For completion, we also consider the limit where the external points coincide pairwise.
This is equivalent to a two-point integral with doubled propagators. In this case, we have
\begin{equation}
    X_{1122}
    =
    \IntegralXOneOneTwoTwo\
    =
    2 I_{12} Y_{112} - \frac{I_{12}^2}{8 \pi^2}\,.
    \label{eq:X1122}
\end{equation}

We often encounter derivatives of the $Y$-integral, for which the following identities hold:
\begin{equation}
    \begin{split}
    \partial_{1 \mu} Y_{123} &= - (\partial_{2 \mu} + \partial_{3 \mu}) Y_{123}\,, \\[.6em]
    \partial_1^2\, Y_{123} &= - I_{12} I_{13}\,, \\
    \left( \partial_1 \cdot \partial_2 \right) Y_{123} &= \frac{1}{2} \left( I_{12} I_{13} + I_{12} I_{23} - I_{13} I_{23} \right)\,.
    \end{split}
    \label{eq:YIdentities}
\end{equation}
All these identities are elementary to prove, using integration by parts and the scalar Green's equation
\begin{equation}
    \pd_1^2 I_{12}
    =
    - \delta^{(4)} (x_{12})\,.
    \label{eq:GreensEquation}
\end{equation}

As mentioned in the main text, the H-integral, defined through
\begin{equation}
    H_{12,34}
    =
    \int d^4 x_5\, I_{15} I_{25}\, Y_{345}\,,
    \label{eq:H1234}
\end{equation}
is \textit{not} conformal, and as far as we can tell, no exact solution has been obtained yet.
In this work, it appears in two contexts: associated with derivatives, or in the one-dimensional limit.
In the first case, using the identities given in \eqref{eq:YIdentities}, it is straightforward to obtain the following relations:
\begin{equation}
    \begin{split}
    \partial_1^2 H_{12,34}
    &=
    - I_{12}\, Y_{134}\,, \\
    \left( \partial_1 \cdot \partial_2 \right) H_{12,34}
    &=
    \frac{1}{2} \left[ I_{12} ( Y_{134} + Y_{234} ) - X_{1234} \right]\,.
    \end{split}
    \label{eq:H_Identities}
\end{equation}
There is no identity known for $\left( \partial_1 \cdot \partial_3 \right) H_{12,34}$.
For the second case, in which the points $x_k$ are aligned, the integral $H_{12,34}$ can be obtained analytically from the train track integral, as explained in Section \ref{subsec:OneIntegralToRuleThemAll} (see in particular \eqref{eq:H_From_B} for the result itself).

We also encounter divergent H-integrals, which correspond to the case in which two points on the same side coincide.
It is easy to derive analytical results for the specific cases of interest in the conformal frame \eqref{eq:ConformalFrame}:
\begin{align}
    H_{23,11} &= \frac{1}{8 \pi^2} (1 - \log \veps) Y_{123} - \frac{1}{16 \pi^2} A_1\,, \label{eq:H2311} \\
    H_{13,22} &= \frac{1}{8 \pi^2} (1 - \log \veps) Y_{123} - \frac{1}{16 \pi^2} A_2\,, \label{eq:H1322} \\
    H_{12,33} &= \frac{1}{8 \pi^2} (1 - \log \veps) Y_{123} - \frac{1}{16 \pi^2} A_3\,, \label{eq:H1233}
\end{align}
where the integral $A_k$ is defined as
\begin{equation}
    A_k = \int d^4 x_5\, I_{15} I_{25} I_{35} \log x_{k5}^2\,.
    \label{eq:Ak}
\end{equation}
In the conformal frame \eqref{eq:ConformalFrame}, the relevant results for this integral are
\begin{align}
    A_1 &= \log x\, Y_{123}\,, \label{eq:A1} \\
    A_2 &= (\log x + \log (1-x)) Y_{123}\,, \label{eq:A2} \\
    A_3 &= \log (1-x)\, Y_{123}\,. \label{eq:A3}
\end{align}

The identities \eqref{eq:H_Identities} can be used to determine integrals that arise in the computation of Feynman diagrams.
For instance, diagrams with two scalar-scalar-gluon vertices give rise to the expression
\begin{equation}
    F_{12,34}
    =
    \frac{\pd_{12} \cdot \pd_{34}}{I_{12} I_{34}} H_{12,34}\,,
    \label{eq:F1234_Definition}
\end{equation}
where we have used the shorthand notation $\pd_{ij}^\mu := \pd_i^\mu - \pd_j^\mu$.
This integral can be elegantly expressed in terms of X- and Y-integrals as \cite{Beisert:2002bb}
\begin{equation}
    \begin{split}
    F_{12,34}
    =\ &
    \frac{X_{1234}}{I_{13}I_{24}} - \frac{X_{1234}}{I_{14}I_{23}} + \left( \frac{1}{I_{14}} - \frac{1}{I_{13}} \right) Y_{134} + \left( \frac{1}{I_{23}} - \frac{1}{I_{24}} \right) Y_{234} \\
    & + \left( \frac{1}{I_{23}} - \frac{1}{I_{13}} \right) Y_{123} + \left( \frac{1}{I_{14}} - \frac{1}{I_{24}} \right) Y_{124}\,.
    \end{split}
    \label{eq:F1234_Result}
\end{equation}
Note that the F-integral also reduces to a simple expression when two external points coincide:
\begin{equation}
    \begin{split}
    F_{13,23} =\ \IntegralFOneThreeTwoThree\ =\ &  \frac{1}{2}
    \left(\frac{Y_{113}}{I_{13}} + \frac{Y_{223}}{I_{23}} \right) + Y_{123} \left( \frac{1}{I_{13}} + \frac{1}{I_{23}} - \frac{2}{I_{12}} \right) \\
    &+ \frac{1}{32 \pi^2} \log \frac{I_{13} I_{23}}{I_{12}^2}\,.
    \end{split}
    \label{eq:F1323}
\end{equation}

Another important family of integrals is
\begin{equation}
    K_{ij}
    =
    \int d^4 x_5\, I_{15} I_{25} I_{35}\, Y_{ij5}\,.
    \label{eq:Kij_Definition}
\end{equation}
In the conformal frame \eqref{eq:ConformalFrame}, if one of the points $(i,j)$ is $\tau_4$, the integral gives
\begin{equation}
    K_{i4}
    =
    \frac{I_{34}}{16 \pi^2} ( (2 + \log \tau_4^2) Y_{123} - A_i)\,.
    \label{eq:Ki4}
\end{equation}
If both points are $\tau_4$, we have
\begin{equation}
    K_{44}
    =
    - \frac{1}{8 \pi^2} I_{34} Y_{123} (\log \veps - 1 - \log \tau_4)\,.
    \label{eq:K44}
\end{equation}

For fermionic integrals, there exists a special star-triangle identity \cite{d1971theoretical,Vasiliev:1981yc,Baxter:1997tn}:
\begin{equation}
    \VertexFermionFermionScalar
    =
    \spd_1 \spd_3\, Y_{123}
    =
    - 4 \pi^2 \sx_{12} \sx_{23} I_{12} I_{13} I_{23}\,,
    \label{eq:StarTriangle}
\end{equation}
which is used to compute integrals with a Yukawa coupling (for instance the diagram given in \eqref{eq:1111_Spider}).

\section{Feynman diagrams of $\vev{1111}$}
\label{app:FeynmanDiagramsOf1111}

In this appendix, we gather the results for the diagrams relevant for the computation of the correlator $\vev{1111}$ at next-to-next-to-leading order, for which the results are presented in Section \ref{subsec:1111}.
The diagrams contributing to the channel $F_1$ are gathered in Table \ref{table:Diagrams1111NNLO}, and can be split into \textit{bulk} and \textit{boundary} diagrams.

It is important to understand that all the diagrams are evaluated in the conformal frame
\begin{equation}
    (\tau_1, \tau_2, \tau_3, \tau_4)
    =
    (0, x, 1, \infty)\,.
    \label{eq:ConformalFrame}
\end{equation}
In particular, the limit $\tau_4 \to \infty$ provides an important simplification for the calculations.
However, the diagrams are not conformal on their own; it should thus be understood that the results for each individual diagram is only correct for the conformal frame.
At the end of the calculation, we expect that summing up all the diagrams cancels the divergences and the spurious $\log \tau_4$ terms, and that the final result is conformal.

In all the expressions above, we ignore the trivial $R$-symmetry factors for readability.

\subsection{Bulk diagrams}
\label{subsec:BulkDiagrams}

We begin by calculating the bulk diagrams.
The results are given in terms of known integrals, for which the results can be found in Appendix \ref{app:Integrals}.
More details about the manipulations of the integrals can be found in \cite{Peveri:2023qip,Barrat:2024nod}.

\subsubsection{Self-energy diagrams}
\label{subsubsec:SelfEnergyDiagrams}

The first line in Table \ref{table:Diagrams1111NNLO} consists of self-energy diagrams.
The manipulations in order to obtain the diagrams are elementary, and we obtain
\begin{align}
    &\DefectSSSSTwoLoopsSelfEnergyOne\
    +
    \DefectSSSSTwoLoopsSelfEnergyTwo\ \notag \\
    & \quad
    +
    \DefectSSSSTwoLoopsSelfEnergyThree\
    +
    \DefectSSSSTwoLoopsSelfEnergyFour\
    =
    - \frac{\lambda^4}{2} (I_{34} H_{23,11} + I_{34} H_{13,22} + I_{34} H_{12,33} + K_{44})\,.
\end{align}
The results for the $H$- and $K$-integrals are given in \eqref{eq:H2311}-\eqref{eq:H1233} and \eqref{eq:K44}, respectively.
Each of these terms is divergent.
Notice also that $K_{44}$ contains a term $\log \tau_4$, which ultimately must cancel with terms in other diagrams for the correlator to be conformal.

\subsubsection{XX diagrams}
\label{subsubsec:XXDiagrams}

The diagrams that contain two $X$-vertices are more intricate.
Taking into account the trace as well as the symmetry factors, their sum can be shown to be equal to
\begin{align}
    \DefectSSSSTwoLoopsXXOne\
    +
    \DefectSSSSTwoLoopsXXTwo
    &=
    - \frac{\lambda^4}{4} \int d^4 x_5\, ( I_{35} I_{45}\, X_{1255} + I_{15} I_{45}\, X_{2355} ) \notag \\
    &\sim
    - \frac{\lambda^4}{8}
    \biggl(
    I_{34} H_{23,11}
    +
    I_{34} H_{13,22}
    +
    I_{34} H_{12,33}
    +
    K_{44} \notag \\
    &\phantom{=\ }
    -
    \frac{1}{16 \pi^2} \log \tau_4^2\, I_{34} Y_{123}
    \biggr)\,.
    \label{eq:1111_XX}
\end{align}

\subsubsection{XH diagrams}
\label{subsubsec:XHDiagrams}

The diagrams with one $X$ and two $Y$ vertices, which we refer to as XH-diagrams, can be evaluated in the same way as the other diagrams.
Summing them up, we find
\begin{align}
    &\DefectSSSSTwoLoopsXHOne\
    +
    \DefectSSSSTwoLoopsXHTwo\ \notag \\
    &\qquad
    +
    \DefectSSSSTwoLoopsXHThree\
    +
    \DefectSSSSTwoLoopsXHFour
    =
    \frac{\lambda^4}{8}
    \biggl(
    I_{34} H_{23,11}
    +
    I_{34} H_{13,22}
    +
    I_{34} H_{12,33}
    +
    K_{44} \notag \\
    &\phantom{\qquad
    +
    \DefectSSSSTwoLoopsXHThree\
    +
    \DefectSSSSTwoLoopsXHFour =\ }
    -
    H_{12,13}
    -
    H_{13,23}
    -
    2 H_{12,23} \notag \\
    &\phantom{\qquad
    +
    \DefectSSSSTwoLoopsXHThree\
    +
    \DefectSSSSTwoLoopsXHFour =\ } 
    +
    \frac{1}{16 \pi^2} (4 - \log x (1-x) + \log \tau_4^2) I_{34} Y_{123}
    \biggr)\,.
    \label{eq:1111_XH}
\end{align}

\subsubsection{Spider diagram}
\label{subsubsec:SpiderDiagram}

We call spider diagram the diagram with a fermionic loop.
By using the fermionic star-triangle identity \eqref{eq:StarTriangle} and the trace identity
\begin{equation}
    \tr \slashed{x}_1 \slashed{x}_2 \slashed{x}_3 \slashed{x}_4
    =
    16
    [
    (x_1 \cdot x_2) (x_3 \cdot x_4)
    -
    (x_1 \cdot x_3) (x_2 \cdot x_4)
    +
    (x_1 \cdot x_4) (x_2 \cdot x_3)
    ]\,,
    \label{eq:TraceIdentity}
\end{equation}
the diagram can be shown to be equal to
\begin{align}
    \DefectSSSSTwoLoopsSpider
    &=
    \frac{\lambda^4}{4} \tr \int d^4 x_5 \int d^4 x_6\, I_{25} I_{46}\, \spd_6 \spd_5 Y_{156}\, \spd_5 \spd_6 Y_{356} \notag \\
    &=
    \frac{\lambda^4}{4}
    \biggl(
    I_{34} H_{23,11}
    +
    I_{34} H_{13,22}
    +
    I_{34} H_{12,33}
    +
    K_{44} \notag \\
    &\phantom{=\ }
    -
    2 H_{12,23}
    -
    \frac{1}{16 \pi^2} \log \tau_4^2\, I_{34} Y_{123}
    \biggr)\,,
    \label{eq:1111_Spider}
\end{align}
in the conformal frame \eqref{eq:ConformalFrame}.
In order to obtain this result, we also used the result for the kite integral given in \eqref{eq:KiteIntegral}.

\subsection{Boundary diagrams}
\label{subsec:BoundaryDiagrams}

We now treat the boundary diagrams of Table \ref{table:Diagrams1111NNLO}.
They all consist of one $X$- and one $Y$-vertex, with the insertion of a gluon field coming from the expansion of the Wilson line.
The results are expressed in terms of divergent integrals and Goncharov polylogarithms.

\subsubsection{First XY diagram}
\label{subsubsec:FirstXYDiagram}

The first such diagram is given by the two slices
\begin{align}
    \DefectSSSSTwoLoopsXYOne
    &\sim
    \frac{\lambda^4}{8} \tau_{13}^2 \tau_{24}^2 \int_{-\infty}^{\tau_2} d\tau_6\, \veps(1 3 6) \int d^4 x_5\, I_{25} I_{35} I_{45}\, \pd_{15} Y_{156} \notag \\
    &\sim
    \frac{\lambda^4}{4} I_{34} H_{23,11}
    -
    \frac{\lambda^4}{8} I_{34} H_{12,23} \notag \\
    &\phantom{=\ }
    +
    \frac{I_{34}}{2048 \pi^6}
    \biggl(
    (G(0,0,1,x)+(x - (1-x)) (G(0,1,0,x) + G(1,1,0,x)) \notag \\
    &\phantom{=\ }
    -
    2 x (G(0,1,1,x) + G(1,0,0,x)) + G(1,0,1,x)
    +
    \frac{\pi^2}{3} (1-x) G(1,x)
    \biggr)\,.
    \label{eq:1111_XY1}
\end{align}

\subsubsection{Second XY diagram}
\label{subsubsec:SecondXYDiagram}

For the second diagram, we obtain
\begin{align}
    \DefectSSSSTwoLoopsXYTwo
    &=
    \frac{\lambda^4}{8} \tau_{13}^2 \tau_{24}^2 \int_{\tau_1}^{\tau_3} d\tau_6\, \veps(2 4 6) \int d^4 x_5\, I_{15} I_{35} I_{45}\, \pd_{25} Y_{256} \notag \\
    &\sim
    \frac{\lambda^4}{4} I_{34} H_{12,33}
    -
    \frac{\lambda^4}{8} I_{34} H_{12,23} \notag \\
    &\phantom{=\ }
    +
    \frac{I_{34}}{2048 \pi^6}
    \biggl(
    x (2 (G(1,0,0,x)- G(0,1,1,x)) + G(1,0,1,x) + G(1,1,0,x)\notag \\
    &\phantom{=\ } + 3 \zeta_3)
    +
    (1-x) G(0,0,1,x) - (1+x) G(0,1,0,x)\notag \\
    &\phantom{=\ }
    +
    2 (G(0,1,1,x)-G(1,0,1,x))
    -
    \frac{\pi^2}{3} G(1,x)
    \biggr)\,.
    \label{eq:1111_XY2}
\end{align}

\subsubsection{Third XY diagram}
\label{subsubsec:ThirdXYDiagram}

The third XY-diagram yields
\begin{align}
    \DefectSSSSTwoLoopsXYThree
    &=
    \frac{\lambda^4}{8} \tau_{13}^2 \tau_{24}^2 \int_{\tau_2}^{\tau_4} d\tau_6\, \veps(1 3 6) \int d^4 x_5\, I_{15} I_{25} I_{45}\, \pd_{35} Y_{356} \notag \\
    &\sim
    \frac{\lambda^4}{4} I_{34} H_{13,22}
    -
    \frac{\lambda^4}{8} I_{34} (H_{12,23} + H_{13,23}) \notag \\
    &\phantom{=\ }
    -
    \frac{I_{34}}{2048 \pi^6}
    \biggl(
    2 x (G(1,0,1,x) + 3 \zeta_3) + 2 (1-x) (G(0,1,1,x) + G(1,0,0,x)) \notag \\
    &\phantom{=\ }
    -
    G(0,1,0,x)
    -
    G(1,0,1,x)
    -
    G(1,1,0,x)
    +
    (x - (1-x)) G(0,0,1,x) \notag \\
    &\phantom{=\ }
    +
    \frac{\pi^2}{3} x G(1,x)
    \biggr)\,.
    \label{eq:1111_XY3}
\end{align}

\subsubsection{Fourth XY diagram}
\label{subsubsec:FourthXYDiagram}

The last diagram is found to be
\begin{align}
    \DefectSSSSTwoLoopsXYFour
    &\sim
    \frac{\lambda^4}{8} \tau_{13}^2 \tau_{24}^2 \int_{-\infty}^{\tau_1} d\tau_6\, \veps(2 4 6) \int d^4 x_5\, I_{15} I_{25} I_{35}\, \pd_{45} Y_{456} \notag \\
    &\sim
    \frac{\lambda^4}{4} K_{44}
    -
    \frac{\lambda^4}{8} (K_{14} + K_{34}) \notag \\
    &\phantom{=\ }
    +
    \frac{I_{34}}{2048 \pi^6}
    \biggl(
    2 x G(0,0,x) + G(0,1,x) + G(1,0,x) + 2 (1-x) G(1,1,x) \notag \\
    &\phantom{=\ }
    +
    \biggl(\frac{2\pi^2}{3}-4\biggr) (x G(0,x) + (1-x) G(1,x))
    \biggr)\,.
    \label{eq:1111_XY4}
\end{align}

\section{Feynman diagrams of $\vev{1122}$}
\label{app:FeynmanDiagramsOf1122}

We now list the results for the diagrams of the channel $F_1$ of the correlator $\vev{1122}$ at next-to-next-to-leading order.
The relevant diagrams, after subtracting the factorized pieces, are summarized in Table \ref{table:Diagrams1122NNLO}, and can be classified into bulk and boundary diagrams, similarly to $\vev{1111}$.
As before, the final expressions are given in the conformal frame \eqref{eq:ConformalFrame}.

\subsection{Bulk diagrams}
\label{subsec:1122_BulkDiagrams}

We begin by calculating the bulk diagrams.

\subsubsection{XX-diagrams}
\label{subsubsec:1122_XXDiagrams}

The XX-diagrams can be calculated to give
\begin{align}
    \NNLOXXOne\
    +
    \NNLOXXTwo\
    &\sim
    \frac{\lambda^5}{4 \pi^2} (1 - \log \veps + \log \tau_4) I_{24} I_{34} Y_{123}\,,
\end{align}
where analytical expressions for all the integrals can be found in Appendix \ref{app:Integrals}.

\subsubsection{XH-diagrams}
\label{subsubsec:1122_XHDiagrams}

The XH-integrals can be evaluated as well, and they yield
\begin{align}
    \NNLOXHOne\
    +
    \NNLOXHOne\
    &\sim
    \frac{\lambda^5}{8 \pi^2} (3 - 2 \log \veps + 2 \log \tau_4) I_{24} I_{34} Y_{123}\,.
\end{align}

\subsection{Boundary diagrams}
\label{subsec:BoundaryDiagrams}

Boundary diagrams are more intricate, but they can also be evaluated analytically.
The first diagram gives
\begin{align}
    \NNLOXYSubtractOne\ \NNLOSubtractFactorOne
    &\sim
    \frac{\lambda^5}{2048 \pi^{6}}
    \log \veps\, I_{24} I_{34} Y_{123}
    +
    \frac{\lambda^5}{4096 \pi^{6}} \frac{I_{24} I_{34}}{x (1-x)}
    \biggl(
    \frac{x}{6} \bigl(3 (G(0,1,x)+G(1,0,x) \notag \\
    &\phantom{=\ }
    -
    2 G(1,1,x)+2 G(1,0,1,x)-2 G(1,1,0,x)+6 \zeta_3) \notag \\
    &\phantom{=\ }
    +
    2 \left(3-2 \pi^2\right) G(0,x)+2 \left(\pi^2-3\right)
    G(1,x)\bigr)+G(1,x)+G(1,1,x)
    \biggr)\,,
\end{align}
while the second evaluates to
\begin{align}
    \NNLOXYSubtractTwo\ \NNLOSubtractFactorOne
    &\sim
    - \frac{\lambda^5}{128 \pi^{2}}
    \biggl(2 + \frac{2\pi^2}{3} - \log \veps + 2 \log \tau_4 \biggr) I_{24} I_{34} Y_{123} \notag \\
    &\phantom{=\ }
    -
    \frac{\lambda^5}{8192 \pi^{6}} \frac{I_{24} I_{34}}{x (1-x)}
    \left(
    x (G(0,1,x)+G(1,0,x)) + 2 (1-x) G(1,1,x)
    \right)\,.
\end{align}
The last diagram is found to give
\begin{align}
    \NLOX\ \NNLOSubtractFactorTwo
    &\sim
    \frac{\lambda^5}{64 \pi^{2}}
    \biggl(
    \left(\frac{2 \pi ^2}{3}-1\right)
    +
    \log \veps - 2 \log \tau_4
    \biggr) I_{24} I_{34} Y_{123}\,.
\end{align}

\bibliography{./auxi/biblio.bib}
\bibliographystyle{./auxi/JHEP}

\end{document}